\documentclass[linenumbers,fleqn,10pt]{wlscirep}
\usepackage[utf8]{inputenc}
\usepackage[T1]{fontenc}
\usepackage{graphicx}
\usepackage{xcolor}
\usepackage{amssymb}

\usepackage{comment}
\usepackage{mwe}
\usepackage{amsmath}
\usepackage{pbox}
\usepackage{float}
\usepackage{booktabs}
\usepackage{placeins}
\usepackage{bm}
\usepackage{subcaption}
\usepackage{cases}
\usepackage{url}
\usepackage{multirow}
\usepackage{tabularx}
\usepackage[ruled,vlined]{algorithm2e}
\providecommand{\SetAlgoLined}{\SetLine}
\SetKwInput{KwData}{Input}
\SetKwInput{KwResult}{Output}

\title{Matrix Completion of World Trade}

\author[1]{Giorgio Gnecco}
\author[1,*]{Federico  Nutarelli}
\author[1]{Massimo Riccaboni}
\affil[1]{IMT School for Advanced Studies, AXES Research Unit, Lucca, 55100, Italy}

\affil[*]{federico.nutarelli@imtlucca.it}


\begin{abstract}
This work applies Matrix Completion (MC) -- a class of machine-learning methods commonly used in the context of recommendation systems -- to analyze economic complexity. 
MC is applied to reconstruct 
the Revealed Comparative Advantage (RCA) matrix, whose elements express the relative advantage of countries in given classes of products, as evidenced by yearly trade flows. A high-accuracy binary classifier is derived from the application of MC, with the aim of discriminating between elements of the RCA matrix that are, respectively, higher or lower than one. We introduce a novel Matrix cOmpletion iNdex of Economic complexitY (MONEY) based on MC, which is related to the predictability of countries' RCA (the lower the predictability, the higher the complexity). 
Differently from previously-developed indices of economic complexity, the MONEY index takes into account the various singular vectors of the matrix reconstructed by MC, whereas other indices are based only on one/two eigenvectors of a suitable symmetric matrix, derived from the RCA matrix. Finally, MC is compared with a state-of-the-art economic complexity index (GENEPY). 
We show that the false positive rate per country of a binary classifier constructed starting from the average entry-wise output of MC can be used as a proxy of GENEPY.
\end{abstract}

\begin{document}

\setlength{\parskip}{0pt}

\maketitle

%
\thispagestyle{empty}

\vspace{-0.5cm}


\section{Introduction}\label{sec:Introduction}

Since the early 2000s, building metrics for measuring economic complexity has been a set goal. Starting from the Economic Complexity Index (ECI) developed by Hidalgo and Hausmann (2009) \cite{HidalgoHausmann2009}, it has become clear how most traditional economic growth theories often shrank internal socio-economic dynamics of countries through strict assumptions, restricting the analysis to a small subset of pre-determined factors. Unlike traditional growth theories, economic complexity measures are based on a data-driven approach and are generally agnostic about the determinants of countries' competitiveness. For instance, the ECI seeks to explain the knowledge accumulated by a country and expressed in all the economic activities present in that country.\\
More and more refined measures of economic complexity have become available in the last few years. In a recent review, Hidalgo (2021)\cite{Hidalgo2021} identifies two main streams of the literature on economic complexity: the first involves metrics of so-called relatedness, whereas the second concerns economic complexity metrics, which apply dimensionality reduction techniques based, e.g., on Singular Value Decomposition (SVD). Metrics of relatedness measure the affinity between an activity and a location, while methods related to dimensionality reduction search for the best combination of factors explaining the structure of a given specialization matrix.\\
According to the principle of relatedness, the probability that a location $c$ (e.g., a country) enters or exits an economic activity $p$ (e.g., a sector) is influenced by the presence of related activities in that location. This poses, however, deeper questions about the role played by similar countries in determining the likelihood that the location $c$ enters the economic activity $p$. Furthermore, while the principle of relatedness attempts to model the probability of entering an activity $p$ by the location $c$, it does not provide hints about whether $c$ will enter $p$ successfully or not. Besides, there is a strong connection between the concept of production function -- a function connecting economic inputs to outputs -- and economic complexity via the SVD factorization of a suitable specialization matrix ${\bf R}$. 
Therefore, SVD is used to learn the singular vectors (factors) that best explain the structure of ${\bf R}$. The ECI index is closely related to the leading singular vectors of that specialization matrix (Hidalgo, 2021)\cite{Hidalgo2021}, i.e. the truncated SVD of the matrix. These are also the leading eigenvectors of the product of the specialization matrix with its transpose. Usually, scholars select one of the first two eigenvectors (i.e., the ones associated with the two largest eigenvalues) because it carries out the maximum amount of information. Recently, Sciarra et al. (2020)\cite{Sciarraetal2020} combined information coming from the first two eigenvectors into a unique index called GENeralised Economic comPlexitY (GENEPY). Nevertheless, it is worth noticing that, by doing this, the other eigenvectors are neglected and, together with them, further information which could potentially better explain economic complexity. Therefore, it looks reasonable to explore a suitable way to carefully select some other most informative eigenvectors beyond the first two.\\
This paper exploits a class of machine-learning methods called Matrix Completion (MC) to consider the information provided by a suitable number of eigenvectors of the specialization matrix. The main idea is to adopt MC to infer information about the Relative Comparative Advantage (RCA), or disadvantage, of a country in a given trade category of products. Such information is collected, for each year, in a matrix, ${\bf RCA} \in \mathbb{R}^{C \times P}$, where $C$ is the number of countries considered, and $P$ is the number of products examined (at a given aggregation level). In formulas, one has
\begin{equation}\label{eq:RCA}
RCA_{c,p}:=\frac{\frac{D_{c,p}}{\sum_{p'=1}^P D_{c,p'}}}{\frac{\sum_{c'=1}^C D_{c',p}}{\sum_{c'=1}^C \sum_{p'=1}^P D_{c',p'}}}\,,
\end{equation}
where $D_{c,p}$ is the return in international dollars of the exports of the product $p$ by country $c$.\footnote{In case one among $D_{c,p}$, $D_{c,p'}$, $D_{c',p}$, and $D_{c',p'}$ in Eq. (\ref{eq:RCA}) is not available, one gets $RCA_{c,p}=NaN$. In this case, as a pre-processing step, that $RCA_{c,p}$ value can be replaced by $0$.}\\
In the paper, MC is applied several times (starting from different training subsets of suitably discretized RCA values associated with several countries and products, excluding originally $NaN$ values) to estimate the expected RCA values of pairs of countries $c$ and products $p$ that have not been used in the training phase. To fulfill this task, the adopted MC technique is based on a soft-thresholded SVD, which selects each time -- via a suitable regularization technique --  the subset of most informative singular values and corresponding singular vectors. 
The predictions provided by MC are then exploited to construct two surrogate incidence matrices, one of which is used to compute a novel index of economic complexity, and the other one is used as an input to the GENEPY algorithm (Sciarra et al., 2020)\cite{Sciarraetal2020}. 


The work contributes to the literature on economic complexity in three ways: (i) it applies for the first time MC to assess the complexity of countries; (ii) it defines a novel index of economic complexity based on MC; (iii) it builds up a comparison with a state-of-the-art index of economic complexity (GENEPY), revealing a high correlation between the output of GENEPY when it is applied to the original incidence matrix and the false positive rate of a binary classifier derived by the repeated application of MC.
The results of our analysis show that MC performs well in estimating the RCA of countries. Supported by the high quality predictions of MC, we propose a novel Matrix cOmpletion iNdex of Economic complexitY (MONEY) for countries, which exploits the accuracy of their RCA predictions derived from the repeated applications of MC. Such accuracy is expressed in terms of a suitably weighted Area Under the Curve (AUC), one for each country examined. The MONEY index ranks countries according to their predictability, taking into account also the complexity of the products. Specifically, the larger the AUC for a specific country and the larger the average with respect to a subset of the products of that country of the MC performance in estimating the discretized RCA values of country-product pairs, the less complex that country. Using MC to construct the proposed index helps to solve the shortcoming of GENEPY, i.e., the fact that, differently from MC, GENEPY takes into account only the information coming from two eigenvectors. 
Moreover, the GENEPY index computed using the MC surrogate incidence matrix reveals interesting discrepancies in terms of economic complexity with respect to the original GENEPY, i.e., the one calculated starting from the incidence matrix associated with the observed {\bf RCA} matrix. By considering multiple years of data (see the Supplemental), we find a strong and significant positive correlation between the false positive rate of the binary classifier derived from thresholding the average output of MC 
and the original GENEPY index.\\ 

\section{Predicting the revealed competitive advantage of countries: a matrix completion approach} \label{ourMC}

In this work, we apply Matrix Completion (MC) techniques to study economic complexity.
This class of machine-learning methods has been popularized by the so-called Netflix competition (see the Appendix for further details on MC and Hastie et al. 2015 \cite{HastieMC}, Alfakih et al. 2000 \cite{alfaik200}, and Cai et al. 2010 \cite{caietal2010} for some of its applications). This paper uses MC to estimate the expected revealed competitive advantage (RCA) of countries $c$ and products $p$. The specific MC method adopted in the paper consists in completing a partially observed matrix ${\bf A} \in \mathbb{R}^{C \times P}$ (which is derived from the ${\bf RCA}$ matrix in our case), by minimizing a suitable trade-off between the reconstruction error of the known portion of that matrix and a penalty term, which penalizes a high nuclear norm of the reconstructed (or completed) matrix. This is formulated via the following optimization problem\cite{Mazumder2010}:
\begin{equation}\label{eq:matrix_completion1}
\underset{{\bf Z} \in \mathbb{R}^{C \times P}}{\rm minimize} \left(\frac{1}{2} \sum_{(c,p) \in \Omega^{\rm tr}} \left(A_{c,p}-Z_{c,p} \right)^2 + \lambda \|{\bf Z}\|_*\right)\,,
\end{equation}
where $\Omega^{\rm tr}
$ 
is a training subset of pairs of indices $(c,p)$ corresponding to positions of known entries of the partially observed matrix ${\bf A} \in \mathbb{R}^{C \times P}$, ${\bf Z} \in \mathbb{R}^{C \times P}$ is the completed matrix (to be optimized), $\lambda \geq 0$ is a regularization constant (chosen by a suitable validation method), and $\|{\bf Z}\|_*$ is the nuclear norm of the matrix ${\bf Z}$, i.e., the sum of all its singular values. The reader if referred to the Appendix for further technical details on the optimization problem (\ref{eq:matrix_completion1}) and on the algorithm we used to solve it.\\
While MC has already found many applications in many fields (e.g., movie recommendation, sensor engineering, econometrics), to the best of our knowledge, this is the first time 
it is used to analyze economic complexity. More precisely, we applied MC to define a novel complexity index to be compared with state-of-the-art complexity indices. 


\noindent In our application of MC to economic complexity, the MC optimization problem (\ref{eq:matrix_completion1}) was solved several times by a specific algorithm previously developed for that purpose (named Soft Impute\cite{Mazumder2010}, see the Appendix), for different choices of the regularization parameter $\lambda$ and of the subset $\Omega^{\rm{tr}}$ (detailed later in this section).  Then, two MC surrogates $\overline{\bf M}^{(MC)}$ and ${\hat{\bf M}}^{(MC)}$ of the incidence matrix ${\bf M} \in \mathbb{R}^{C \times P}$ were generated\footnote{The reader is referred to the Appendix for details on how the incidence matrix ${\bf M}$ is defined, starting from the ${\bf RCA}$ matrix.}. On one hand, $\overline{\bf M}^{(MC)}$ was exploited to evaluate the performance of MC by changing a suitable threshold. This allowed to build up performance measures that compose the MONEY index (see Section \ref{MONEY} for details). On the other hand, ${\hat{\bf M}}^{(MC)}$ was used as input to the GENEPY algorithm, to construct a counterfactual ${\widehat{\rm GENEPY}}^{(MC)} \in \mathbb{R}^{C \times P}$ to be compared with the GENEPY index computed using the original incidence matrix ${\bf M}$ . In the following, we describe 
our approach of applying MC to the reconstruction of the ${\bf RCA}$ matrix for the case in which the products were aggregated at the $4$-digits level in the Harmonized System Codes 1992 (HS-1992). 
Consistently with the literature \cite{Sciarraetal2018}, we constructed the matrix ${\bf A}$ (one of the inputs to the optimization problem (\ref{eq:matrix_completion1})) by discretizing the elements of the ${\bf RCA}$ matrix. For the sake of brevity, we refer to the MC application to the definition of a measure of complexity of the countries. To get a measure of complexity of the products, it is enough to replace the matrix ${\bf A}$ with its transpose (see also the Supplemental for some related results). 
\vspace{-0.2cm}
\begin{enumerate}[itemsep=0pt, parsep=0pt, leftmargin=0.5cm]
\item For the matrix ${\bf A} \in \mathbb{R}^{C \times P}$ (where $C=119$ is the number of countries, and $P=1243$ is the number of products), the MC optimization problem (\ref{eq:matrix_completion1}) was solved $N=1000$ times by the Soft Impute algorithm, based on various choices for the training/validation/test sets (and, as already mentioned, for the regularization parameter $\lambda$).
\item For each such repetition $n=1,\ldots,N$, the sets above were constructed as follows. First, a (pseudo)random permutation of the rows of ${\bf A}$ was generated. Then, a subset $S_n$ of these rows was considered, by including in it the first row in the permutation and the successive $s \% \simeq 25 \%$ rows. In this way, the resulting number of elements of the set $S_n$ was $|S_n|=30$. Next, for each row in $S_n$, its elements belonging to all the groups except group "0" were obscured independently with probability $p_{\rm missing}=0.3$. The (indices ot the) remaining entries of the matrix ${\bf A}$ (excluding the ones belonging to the group "0") formed the training set (denoted by $\Omega^{\text{tr}_n}$). The obscured entries in one of the $|S_n|$ rows (say, row $h \in \{1,\ldots,|S_n|\}$) formed the test set (denoted by $\Omega^{\text{test}_{n,h}}$), whereas the obscured entries in the remaining $|S_n|-1$ rows formed the validation set (denoted by $\Omega^{\text{val}_{n,h}}$).
\item For each repetition $n$, the generation of the validation and test sets from the set $S_n$ was made $|S_n|$ times, each time with a different selection of the row $h$ associated with the test set (and, as a consequence, also of the $|S_n|-1$ rows associated with the validation set). Hence, the same training set was associated with $|S_n|$ different pairs of validation and test sets\footnote{The number of repetitions $N=1000$ and the percentage $s\% \simeq 25 \%$ were selected in order to associate each row with the test set a sufficiently large number of times, with high probability. In particular, with these choices, the average number of times each row was associated with the test set was about $250$.}. In this way, for each choice of $S_n$ and of the regularization parameter $\lambda$, the MC optimization problem  (\ref{eq:matrix_completion1}) was solved once instead of $|S_n|$ times, thus improving the computational efficiency. Finally, by construction, each time there was no overlap between the training, validation, and test sets.
\item To avoid overfitting, for each choice of the training set $\Omega^{\text{tr}_n}$, the optimization problem (\ref{eq:matrix_completion1}) was solved for $30$ choices $\lambda_k$ for $\lambda$, exponentially distributed as $\lambda_k = 2^{(k-1)/2}$ for $k=1,\dots,30$. The resulting completed and post-processed matrix was indicated as ${\bf Z}^{(n)}_{\lambda_k}$. Then, for each $\lambda_k$ and each of the $|S_n|$ selections of the validation sets associated with the same training set, the Root Mean Square Error (RMSE) of matrix reconstruction on that validation set was computed as
\begin{equation}
RMSE_{\lambda_k}^{{\rm val}_{n,h}}:=\sqrt{\frac{1}{|\Omega^{{\rm val}_{n,h}}|}\sum_{(c,p) \in \Omega^{{\rm val}_{n,h}}} \left(A_{c,p}-Z^{(n)}_{\lambda_k,c,p} \right)^2}\,,
\end{equation}
then the choice $\lambda_{k^\circ(n,h)}$ minimizing $RMSE_{\lambda_k}^{{\rm val}_{n,h}}$ for $k=1,\ldots,30$ was found. Finally, the RMSE of matrix reconstruction on the related test set was computed in correspondence of the so-obtained optimal value $\lambda_{k^\circ(n,h)}$ as
\begin{equation}
RMSE_{\lambda_{k^\circ(n,h)}}^{{\rm test}_{n,h}}:=\sqrt{\frac{1}{|\Omega^{{\rm test}_{n,h}}|}\sum_{(c,p) \in \Omega^{{\rm test}_{n,h}}} \left(A_{c,p}-Z^{(n)}_{\lambda_{k^\circ(n,h)},c,p} \right)^2}\,.
\end{equation}
\item \label{step:5} For each choice of $n$ and $h$, the MC predictions contained in the matrix ${\bf Z}^{(n)}_{\lambda_{k^\circ(n,h)}}$ were used to build a binary classifier. More precisely, each time an element $A_{c,p}$ of the matrix ${\bf A}$ was in the test set, such element was attributed to the class $0$ (corresponding to the case $0 \leq RCA<1$) when its MC prediction from ${\bf Z}^{(n)}_{\lambda_{k^\circ(n,h)}}$ was lower than $0$, otherwise it was attributed to the class 1 (corresponding to the case $RCA \geq 1$). Finally, the average classification of the element $A_{c,p}$ (with respect to all the test sets to which that element belonged) was indicated as $\overline{A}^{(MC)}_{c,p} \in [0,1]$, whereas its most frequent classification (either $0$ or $1$) was indicated as $\hat{A}^{(MC)}_{c,p}$. A random assignment between $0$ and $1$ was made to deal with ties. In the (unlikely) case the element $A_{c,p}$ appeared in none of the test sets\footnote{Due to the choice $p_{\rm missing}=0.3$, each element $A_{c,p}$ not associated with the group "0" appeared in the test set on average about 75 times. So, the probability that one such element appeared in none of the test sets was negligigle.}, both $\overline{A}^{(MC)}_{c,p}$ and $\hat{A}^{(MC)}_{c,p}$ were chosen to be equal to $0$.
\item \label{step:6}  A first MC surrogate $\overline{\bf M}^{(MC)} \in \mathbb{R}^{119 \times 1243}$ of the incidence matrix $\bf M$ was defined as follows:
\begin{equation}
\overline{M}^{(MC)}_{c,p}\doteq \begin{cases}
0\,,& {\,\,\rm if\,\,} RCA_{c,p}=NaN\,, \\
\overline{A}^{(MC)}_{c,p}\,,& {\,\,\rm otherwise\,.}
\end{cases}
\end{equation}
Similarly, a second MC surrogate $\hat{\bf M}^{(MC)} \in \mathbb{R}^{119 \times 1243}$ of the incidence matrix $\bf M$ was defined as follows:
\begin{equation}
\hat{M}^{(MC)}_{c,p}\doteq \begin{cases}
0\,,& {\,\,\rm if\,\,} RCA_{c,p}=NaN\,, \\
\hat{A}^{(MC)}_{c,p}\,,& {\,\,\rm otherwise\,.}
\end{cases}
\end{equation}
\item \label{step:7} Finally,  $\overline{\bf M}^{(MC)}$ was combined with several thresholds from 0 to 1 in the first part of the construction of the proposed MONEY index (see Section \ref{MONEY} for details). Instead,  $\hat{\bf M}^{(MC)}$ was provided as input to the GENEPY algorithm (replacing the original incidence matrix ${\bf M}$). In this way, a counterfactual GENEPY index, indicated as ${\widehat{\rm GENEPY}}^{(MC)}$, was generated.
\vspace{-0.2cm}
\end{enumerate}

\noindent In order to assess the prediction capability of the binary classifier associated with MC (see Step \ref{step:5} above), for each row (country) $c$ of ${\bf A}$, we also computed the false positive rate ${fpr}_c$ and the false negative rate ${fnr}_c$ as the average classification error frequency, respectively, of the true negative/true positive examples in all the test sets associated with that row (where the "negative class" refers to the class $0$ associated with $0 \leq RCA<1$, and the "positive class" to the class $1$ associated with $RCA \geq 1$). 


\section{The Matrix cOmpletion iNdex of Economic complexitY (MONEY)} \label{MONEY}
In this section, we introduce our proposed economic complexity index,  called  Matrix cOmpletion iNdex of Economic complexitY (MONEY), whose construction is based on MC.

The MONEY index is built starting from the matrix $\overline{\bf M}^{(MC)}$ introduced in Section \ref{ourMC}. It is based on constructing a binary classifier for each country by combining the corresponding row of $\overline{\bf M}^{(MC)}$ with a threshold, then  assessing the performance of the resulting MC classifications at the level of each country. 
First, for the binary classifier associated with each country, a Receiver Operating Characteristic (ROC) curve\footnote{We remind the reader that, for a binary classifier, the ROC curve expresses the trade-off between fall-out (false positive rate) and sensitivity (true positive rate) of that classifier, as a function of its threshold. It is recalled here that the  true positive rate is equal to 1 minus the false negative rate. In general, ROC curves closer to the top-left corner indicate a better performance. 
As a baseline (``Bench.''), a random guessing binary classifier is associated with a ROC curve with points lying along the diagonal indicated, e.g., in Fig. \ref{roc_global} (for which the true positive rate is equal to the false positive rate). The closer a ROC curve to the diagonal in the ROC space, the worse the performance of the associated binary classifier. It is worth reminding the reader that ROC curves do not depend on class frequencies. This makes them useful for evaluating classifiers predicting rare events as in the case of very high RCA values.} (denoted as $ROC_c$) is constructed, based on a country-dependent threshold. The corresponding Area Under the Curve (AUC)\footnote{We remind the reader that the AUC measures the area of the entire two-dimensional region underneath the entire ROC curve from $(0,0)$ to $(1,1)$. The AUC is exploited in the literature to provide an aggregate measure of performance across all possible classification thresholds. Formally, it represents  the probability that a classifier will rank a randomly chosen positive instance higher than a randomly chosen negative one, assuming that "positive" ranks higher than "negative" (Fawcett, 2006 \cite{Fawcett2006}).} is denoted as $AUC_c$. In more details, for each country $c$, the elements of the $c$-th row of the matrix $\overline{\bf M}^{(MC)}$ are compared with a threshold to construct the associated binary classifier. The  elements belonging to the same row of the original incidence matrix ${\bf M}$ are taken as ground truth. The discrimination threshold is varied from $0$ to $1$, using a step size equal to $0.01$. All the elements of $\overline{\bf M}^{(MC)}$ are used as dataset, except the ones having the same indices as the originally $NaN$ values in the {\bf RCA} matrix. This allows to form a binary classifier for each threshold and for each country. The idea now is to exploit the $AUC_c$ of the binary classifiers associated with the countries in order to provide a measure of complexity of such countries, based on the  predictability of the corresponding rows. Specifically, countries with lower $AUC_c$ may be considered as more complex, being harder for MC to predict their RCA entries. The $AUC_c$ alone, however, does not capture the reasons why MC performed poorly (or, vice versa, adequately). As an example, consider the three following hypothetical scenarios. Assume that MC performs poorly on a country $c$ by attributing $RCA \geq 1$ to a product $p$ when its true RCA was smaller than $1$, and assigned correctly a RCA smaller than $1$ to all the other countries for the same product $p$ (Scenario 1). Consider now the two following similar scenarios for which, for the same product $p$ and the same country $c$, MC performs poorly on the country $c$ by attributing $RCA \geq 1$ to the product $p$ when its true RCA was smaller than $1$, and it attributed  either correctly (Scenario 2) or incorrectly (Scenario 3) $RCA \geq 1$ to all the other countries for the same product. It is reasonable to suppose that, all other things being equal, 
the country $c$ to which MC assigned $RCA \geq 1$ for the product $p$ in Scenario 1 is more complex than the same country to which MC assigned $RCA \geq 1$ for the product $p$ in Scenarios 2 and 3. In fact, while in Scenario 2, MC could have been  driven to predict, for country $c$, a RCA of $p$ larger than or equal to 1 by the presence of several RCA entries larger than or equal to 1 for the other countries, this is not the case for Scenario 1.  Scenario 3 is  more unlikely to occur, since, as it is shown later in Section \ref{roc} and in the Supplemental, MC has typically a quite satisfying prediction capability in its specific application to the {\bf RCA} matrix. In this case, it is not possible to conclude that country $c$ is more complex than the other countries, since MC is wrongly attributing $RCA \geq 1$ to $p$, for all such countries.\\
The example above suggests us that, by adopting the $AUC_c$ alone as a complexity measure, country $c$ would be classified as equally complex in Scenarios 1, 2 and 3 (assuming the $AUC_c$ being equal in all these cases). In order to correct for this, we propose a refined complexity measure, based on weighting the $AUC_c$ for each country $c$. The rationale of the proposed complexity measure is that not only less predictable countries (according to MC) are more complex, but one should also take into account the product dimension when comparing the MC predictions obtained for different countries, controlling for the quality of each prediction. More precisely, it is proposed to associate a weight $w_{c}$ to each country $c$, which is constructed in such a way that the $AUC_c$'s of countries with an higher share of ``rare'' false positives are weighted less (since they are less predictable). In more details, the proposed complexity measure is constructed as follows.
\vspace{-0.2cm}
\begin{enumerate}[itemsep=0pt, parsep=0pt, leftmargin=0.5cm]
\item 
First, the MC analysis made for the countries is repeated for the products, still referring to the same year. This is obtained simply by replacing at the beginning of the analysis the $\bf{RCA}$ matrix with its transpose. Analogously, the matrices $\hat{\bf M}^{(MC)}$ and $\overline{\bf M}^{(MC)}$ are replaced by similarly constructed matrices $(\hat{\bf M}^\top)^{(MC)}$ and $(\overline{\bf M}^\top)^{(MC)}$. In particular, each element of the latter matrix represents the average MC prediction for the corresponding  product-country pair.
\item Then, a threshold $t$ is applied to the elements of the matrix $(\overline{\bf M}^\top)^{(MC)}$. For each value of that threshold, one constructs a matrix $\left(\overline{\bf M}_t^\top\right)^{(MC)} \in \{0,1\}^{P \times C}$, being each entry of it equal to $1$ whenever the corresponding element in the matrix $(\overline{\bf M}^\top)^{(MC)}$ is higher than or equal to $t$, otherwise being it equal to $0$.
\item At this point, for each product $p$ and each threshold $t$, one computes the quantity
\begin{equation}
ftot_{p,t}:=fpr_{p,t} \times \frac{N_p}{P_p + N_p}\,,
\end{equation}
being $fpr_{p,t}$ the false positive ratio for the classifications associated with that product (determined by the comparison between $\left(\overline{\bf M}_t^\top\right)^{(MC)}$ and ${\bf M}^\top$, restricted to the entries associated with that product) and $\frac{N_p}{N_p + P_p}$ the proportion of entries with true $RCA <1$ with respect to all the entries associated with that product (i.e., $119$). Besides, the average $\overline{ftot}_{p}$ of $ftot_{p,t}$ with respect to $t$ is computed.
\item Then, for each country $c$, the weight $w_c$ is defined as follows:
\begin{equation}
    w_c:=\frac{\sum_{p=1}^P  (\hat{M}^\top)^{(MC)}_{p,c} \times \overline{ftot}_p}{\sum_{p=1}^P (\hat{M}^\top)^{(MC)}_{p,c}}\,.
\end{equation}
In other words, for each country $c$, the weight $w_c$ is the average of $\overline{ftot}_{p}$  with respect to all the products $p$ for which one predicts $RCA\geq1$ through the surrogate incidence matrix $(\hat{\bf M}^\top)^{(MC)}$.
\item  Finally, the MONEY index for each country $c$ is  computed as: \begin{equation}
MONEY_c:=1-w_{c} \times AUC_{c}\,.
\end{equation}
\end{enumerate}

\section{Results}\label{sec:results}


\subsection{Global performance of matrix completion}\label{roc}
\begin{figure}[H]
        \centering
           \vspace{-0.2cm} \includegraphics[scale=0.6,trim=0.2cm 0.01cm 0cm 1.2cm,clip=true]{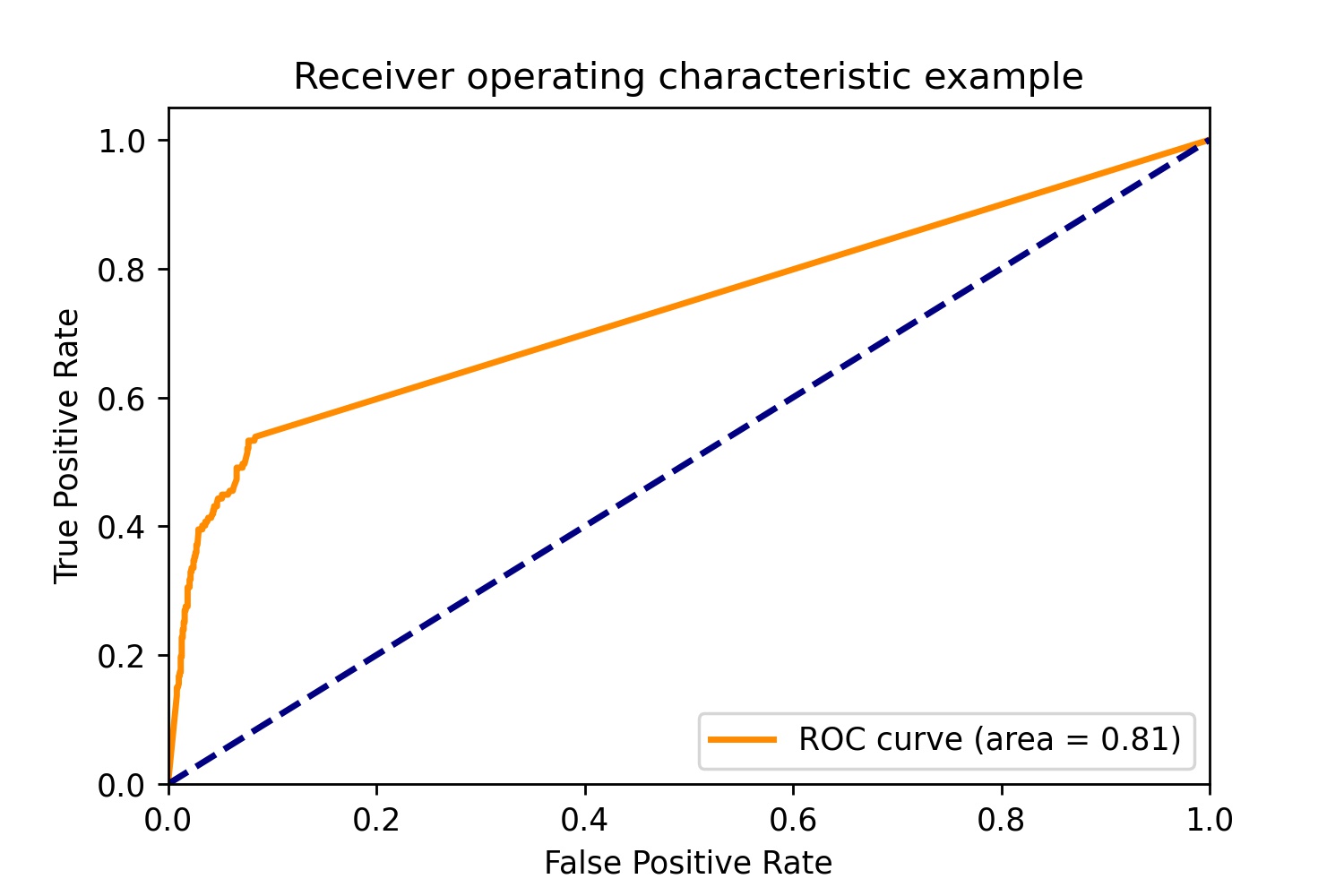}
           \vspace{-0.2cm}
            \caption{Global ROC curve constructed starting from the matrix $\overline{\bf M}^{(MC)}$, for the year 2018. ``Bench.'' stands for the line passing through the origin with slope $1$.}
            \label{roc_global}
        \end{figure}

\begin{figure}[H]
        \centering
           \vspace{-0.2cm} \includegraphics[scale=0.15,trim=0cm 0.4cm 0cm 0.5cm,clip=true]{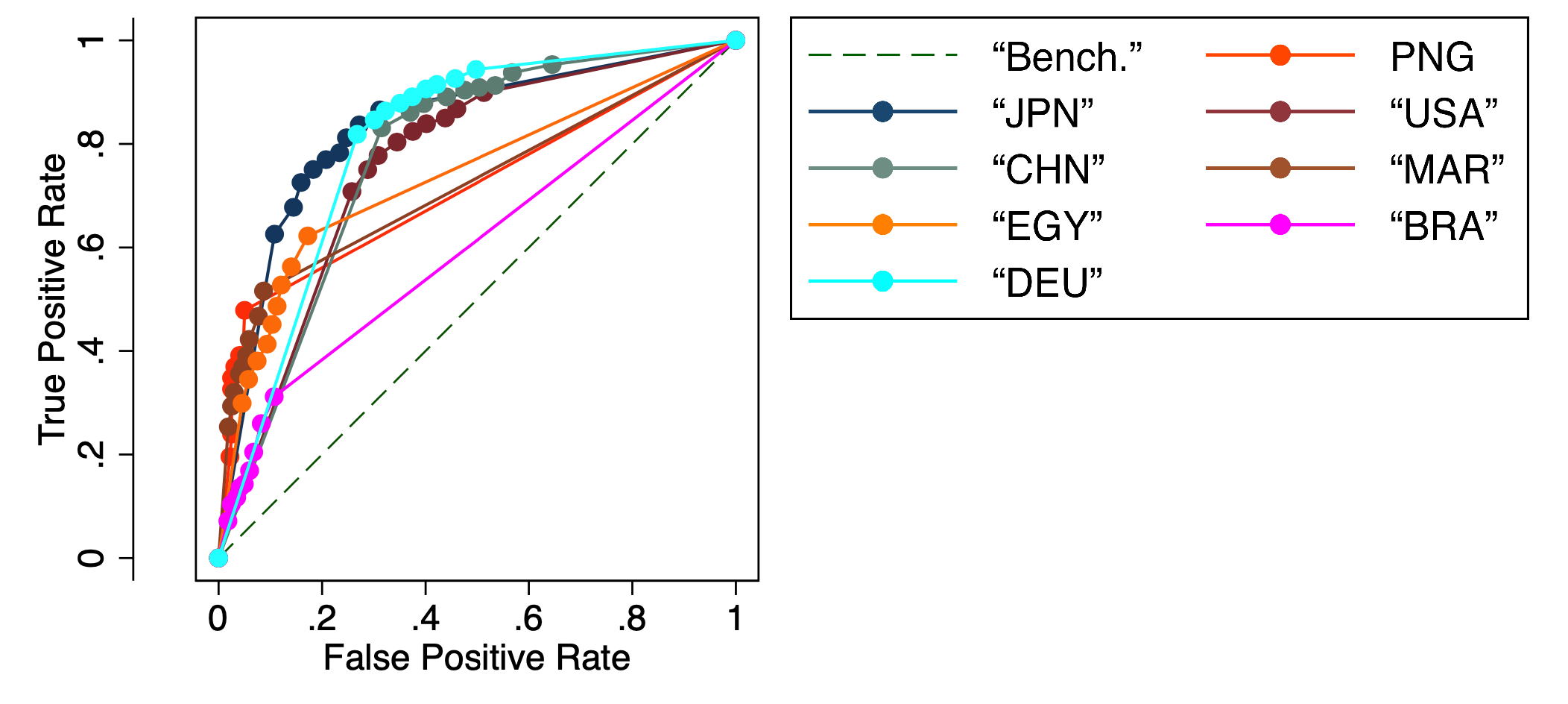}
           \vspace{-0.2cm}
            \caption{$ROC_c$ curves constructed starting from the matrix $\overline{\bf M}^{(MC)}$, for a sample of countries and for the year 2018. ``Bench.'' stands for the line passing through the origin with slope $1$.}
            \label{roc_new}
        \end{figure}

In the following, the diagnostic ability of MC is illustrated. Likewise in Section \ref{MONEY}, the matrix $\overline{\bf M}^{(MC)}$ was combined with a threshold to construct a binary classifier (in this case, however, differently from Section \ref{MONEY}, the threshold did not depend on the country). The discrimination threshold
was varied from $0$ to $1$, using a step size equal to $0.01$. All the elements of $\overline{\bf M}^{(MC)}$ were used as dataset, except the ones having the same indices as the originally \textit{NaN} values in the {\bf RCA} matrix. The ground truth was provided by the corresponding elements of the original incidence matrix ${\bf M}$. Fig. \ref{roc_global} shows the resulting ROC curve. Similarly, $ROC_c$ curves (see Section \ref{MONEY}) for a random sample of countries are displayed in Fig. \ref{roc_new}.

\begin{figure}[H]
        \centering
        \begin{subfigure}[b]{0.475\textwidth}
            \centering 
        \includegraphics[scale=0.4,trim=1.5cm 8cm 0cm 8.5cm,clip=true]{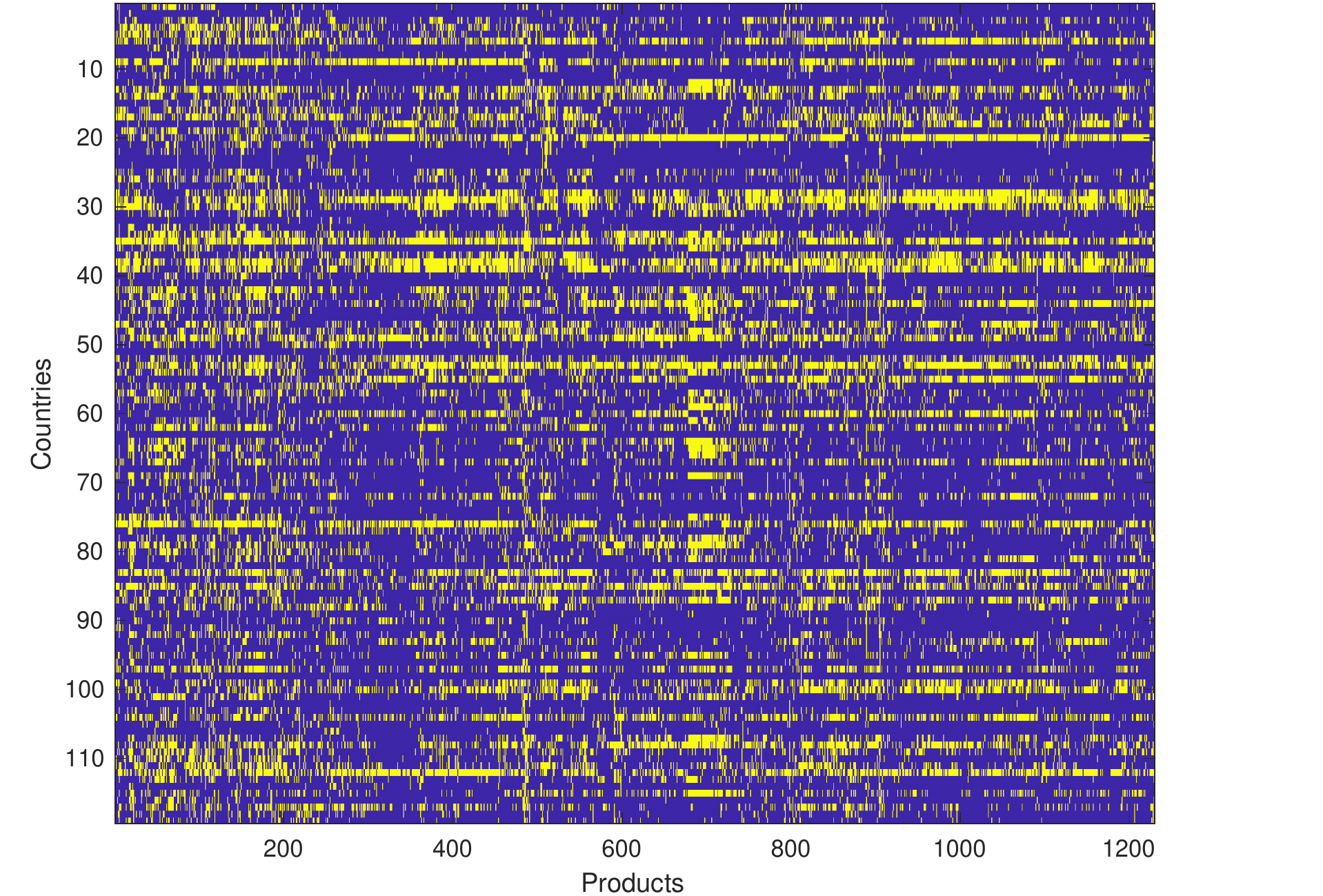}
            \caption{{\small Original incidence matrix ${\bf M}$ at the HS-4 level.}}
            \label{orighs4heat}
        \end{subfigure}
        \hfill
        \begin{subfigure}[b]{0.475\textwidth}  
            \centering 
        \includegraphics[scale=0.4,trim=1.5cm 8cm 0cm 8.5cm,clip=true]{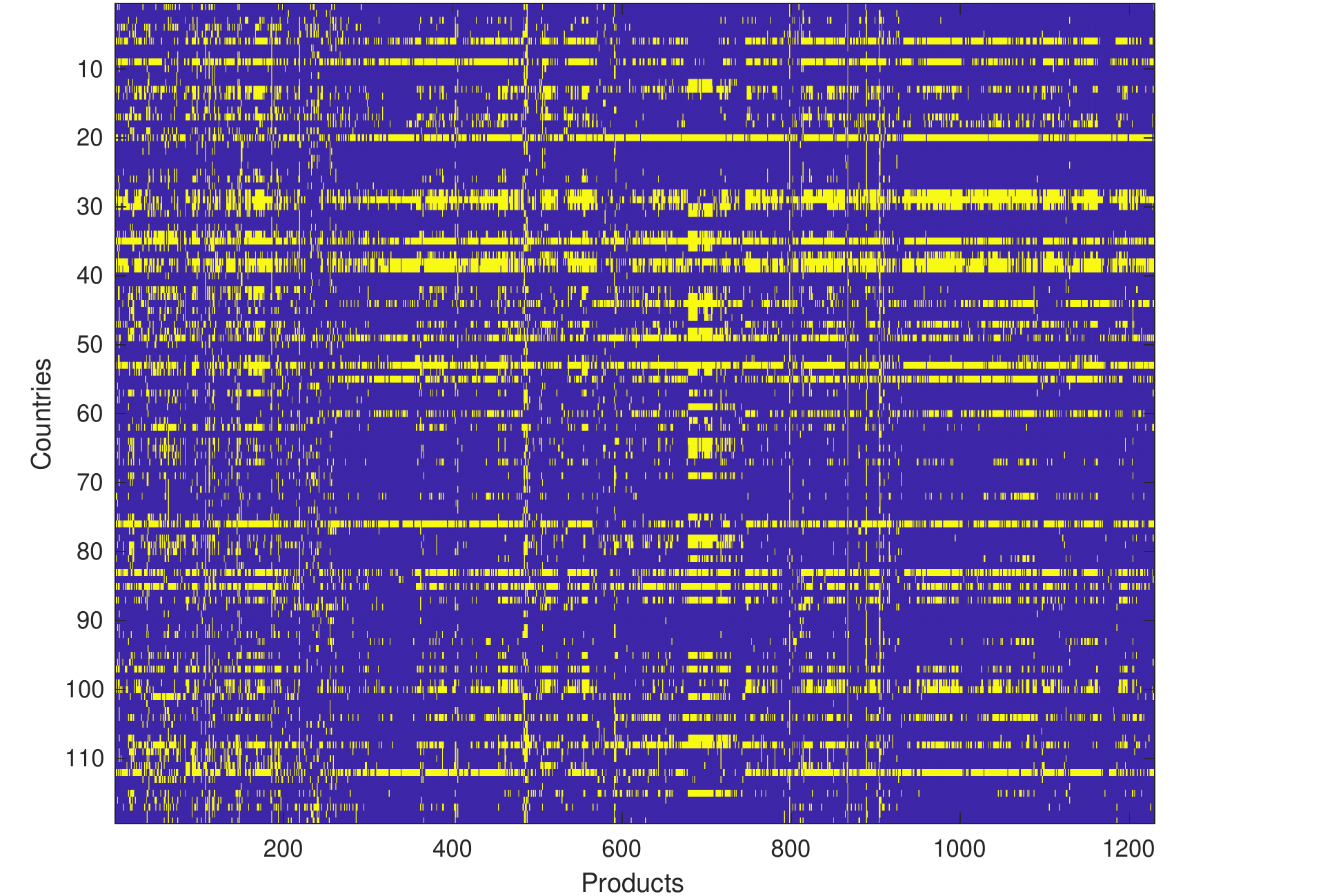}
            \caption{{\small Surrogate incidence matrix $\hat{\bf M}^{(MC)}$ at the HS-4 level.}}
            \label{predhs4heat}
        \end{subfigure}
        \vskip\baselineskip
        \caption{Original versus surrogate incidence matrix for the year 2018 at the HS-4 level.}
    \end{figure}

\vspace{-0.2cm}
\noindent As it is evident from Figs. \ref{roc_global} and \ref{roc_new}, MC performed quite well on average both globally and for developed countries such as Japan, United States and Germany. Its performance was poorer (though still above the baseline) for countries that either provided less information on their trade flows or whose trade flows were extremely volatile (i.e., they alternated between products with extremely high RCA values and products with very low RCA values). Specifically, ${fnr}_c$ was higher for the latter countries. Nonetheless, the average performance of MC over all the countries was high as depicted by the AUC reported in Fig. \ref{roc_global}, which turned out to be about 0.81 for the binary classifier described in Step \ref{step:5} of Section \ref{ourMC}.\\
As a further check, since  the positive and negative labels were unbalanced in the original dataset (specifically, entries with $RCA<1$ represented almost the 70\% of the entire dataset), we also applied the Balanced Accuracy (BACC) index\footnote{The BACC is a performance metric designed for binary classifiers in the case of unbalanced datasets. It is calculated as the average of the proportion of correctly classified elements of each class individually, and ranges from 0 (low balanced accuracy) to 1 (maximum balanced accuracy). Formally, it is equal to $\left({tpr} + {tnr} \right)/ 2$, where {\em t} stands for ``true''.}, which turned out to be 0.75. 
\noindent Figs.\ref{orighs4heat}-\ref{predhs4heat} display the original incidence matrix ${\bf M}$ as compared to the MC surrogate incidence matrix $\hat{\bf M}^{(MC)}$ obtained at the HS-4 level of product aggregation. The two matrices display similar but not identical entries. On one hand, their similarity confirms the good MC prediction performance at a global level. On the other hand, their differences could be attributed to the high complexity of specific country/product pairs being predicted. In other words, there may be a discrepancy between the actual RCA value of a country/product pair and its potential RCA value, predicted by MC on the basis of similar country/product pairs.
\vspace{-0.2cm}

\vspace{-0.2 cm}
\subsection{Results related to the MONEY index}\label{sec:money}
\begin{figure}[H]
        \centering
        \begin{subfigure}[b]{0.475\textwidth}
            \centering 
        \includegraphics[ width=8.4cm,height=4.7cm,scale=0.9]{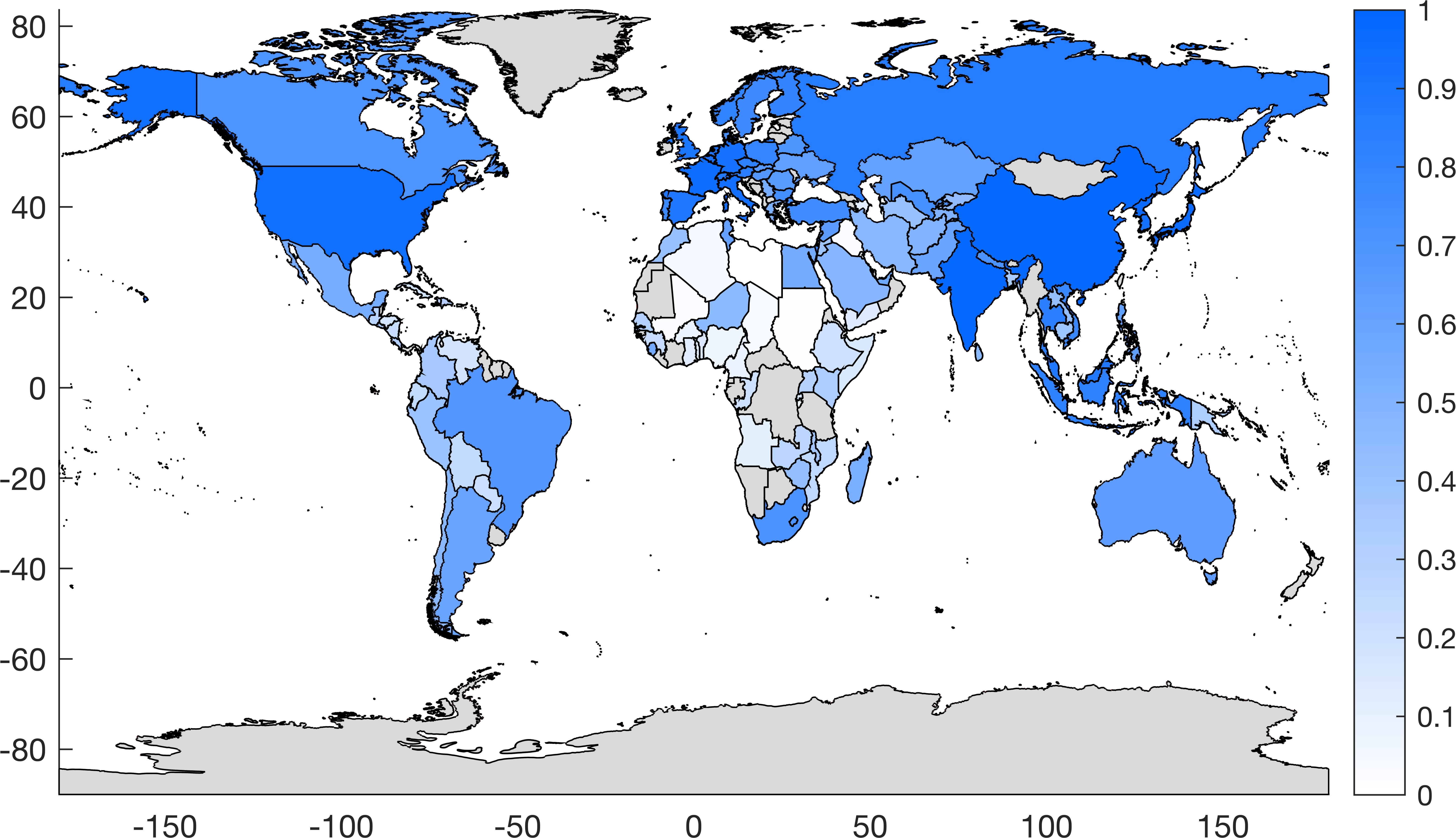}
            \caption{Countries colored according to the MONEY index. Countries in darker shades of blue are associated with a lower MONEY, hence they are considered more complex. Countries colored in grey are not considered in the analysis.}
            \label{money_rank}
        \end{subfigure}
        \hfill
        \begin{subfigure}[b]{0.475\textwidth}  
            \centering 
            \includegraphics[ width=8.4cm,height=4.7cm,scale=0.9]{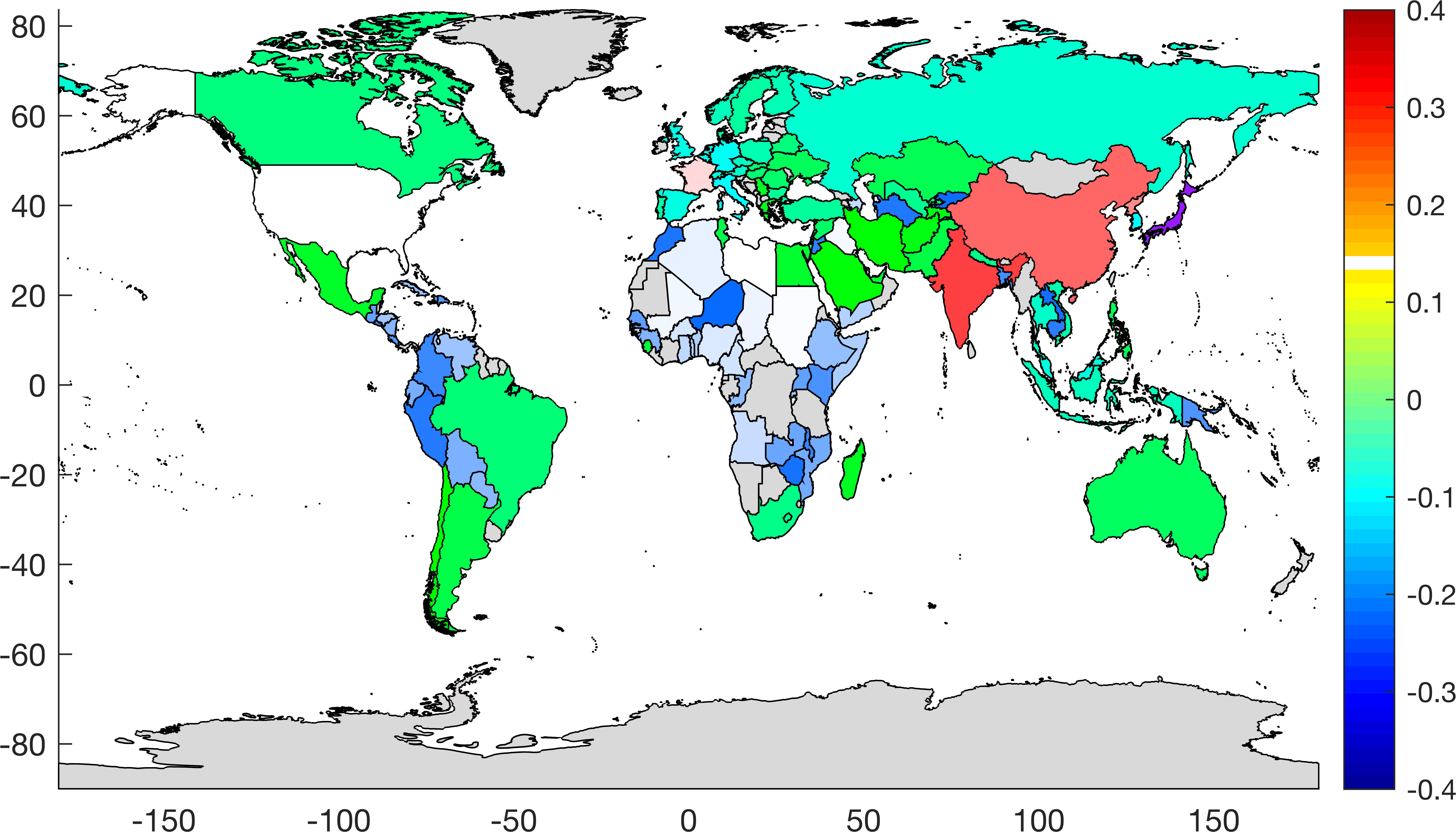}
            \caption{Countries colored according to the difference between the values of their MONEY and GENEPY indices. Countries in darker shades of blue are associated with a value of ${\rm MONEY}$<${\rm GENEPY}$, vice versa countries in red.}
            \label{genep_mon}
        \end{subfigure}
        \vskip\baselineskip
        \caption{Values of the MONEY index for the year 2018 at the HS-4 level of aggregation and their differences with respect to the corresponding values of the GENEPY index for the same year and the same level of aggregation.}
        \label{auc_w}
    \end{figure}

In this section we report the ranking of countries in terms of economic complexity as expressed by the MONEY index introduced in Section \ref{MONEY}. In particular, we represent the countries according to their MONEY index (Fig. \ref{money_rank}), then we compare the obtained ranking with the one expressed by GENEPY (Fig.\ref{genep_mon}). In Fig.\ref{money_rank}, countries are colored according to their MONEY values (normalized between 0 and 1), which are proportional to the shade of blue. In particular, the color map ranges from the least complex countries $c$ (colored in white) to the most complex ones (colored in dark blue).

\vspace{-0.2cm}
\vspace{-0.3cm}   
\noindent It is worth observing that both the GENEPY and the proposed MONEY index arise from the attempt to reconstruct (in a different way for each method) a matrix related to trade flows. In the case of GENEPY, the matrix is a proximity matrix ${\bf N}$ derived from the incidence matrix $\bf M$ (see the Appendix for the definition of the matrix $\bf N$), and its reconstruction is obtained as a nonlinear least-square estimate based on the components of the first two (normalized) eigenvectors of that matrix. Then, a successive evaluation on how the quality of the estimate changes by dropping specific components of such eigenvectors (the ones associated with a given country) is made. In our case, the matrix ${\bf A}$ is obtained as a discretization of the ${\bf RCA}$ matrix. Then, MC is applied several times to the matrix ${\bf A}$ to reconstruct a portion of that matrix which has been obscured, in the attempt to uncover a ``latent''  similarity between countries, which can be useful for the prediction of whether their RCA entries are lower than $1$, or higher than or equal to $1$. 
Another difference is that the matrix reconstruction on which GENEPY is based relies only on two eigenvectors of ${\bf N}$, whereas our method, being also based on MC, exploits a typically much larger number of left-singular/right-singular vectors to build the reconstructed matrix, for each application of MC. The choice of the number of such pairs is made automatically by the adopted validation procedure. Moreover, a final evaluation of the quality of the reconstruction is made, by considering several test sets, on which the $AUC_c$'s are based. A further quality assessment is provided by Tab. \ref{g20_tab}, which reports the number of G19+5\footnote{In the table we considered countries within G20. However, since G20 countries comprise EU (except France, Italy and Germany, which are accounted separately), that is an agglomerate of countries, we considered a group of 5 representative countries for EU, namely: Spain, Switzerland, Greece, Denmark and Hungary.} countries in the top 20, 30 and 40 positions, computed according to each among the GENEPY, ${\widehat{\rm GENEPY}}^{(MC)}$ and MONEY indices, then divided by 24. It is evident from the table that the largest ratio is obtained in correspondence of the proposed MONEY index. Additional robustness checks related to the MONEY index computed in years different than 2018 are reported in the Supplemental.

\begin{table}[H]
\setlength{\tabcolsep}{20pt}
\centering
\caption{Number of G19+5 countries in the top 20, 30 and 40 positions for the year 2018, computed according to each among the GENEPY, ${\widehat{\rm GENEPY}}^{(MC)}$ and MONEY indices, then divided by 24.}
\label{g20_tab}
\scalebox{0.7}{
\begin{tabular}{|c|c|c|c|}
\hline
\textbf{}                           & \multicolumn{3}{c|}{\bf \# of G19+5 countries in the top $x$ positions/24} \\
\hline
\textbf{Index}                           & \textbf{\begin{tabular}[c]{@{}l@{}}$x=20$\end{tabular}} & \textbf{\begin{tabular}[c]{@{}l@{}}$x=30$\end{tabular}} & \textbf{\begin{tabular}[c]{@{}l@{}}$x=40$\end{tabular}} \\ \hline
GENEPY                         &  42\%                                                                            & 58\%                                                                            & 75\%                                                                            \\ \hline
${\widehat{\rm GENEPY}}^{(MC)}$ & 55\%                                                                            & 64\%                                                                            & 71\%                                                                            \\ \hline
 MONEY                          & 55\%                                                                            & 66\%                                                                            & 79\%                                                                            \\ \hline

\end{tabular}}
\end{table}

\FloatBarrier \subsection{Differences in the GENEPY indices based on the original incidence matrix ${\bf M}$ and on $\hat{{\bf M}}^{(MC)}$}\label{sec:resultsGENEPY}
\begin{figure}[H]
        \centering
        \begin{subfigure}[b]{0.475\textwidth}
            \centering
          \hspace{-0.3cm}
           \includegraphics[scale=0.0175,trim=1.5cm 8.7cm 0.9cm 0cm,clip=true]{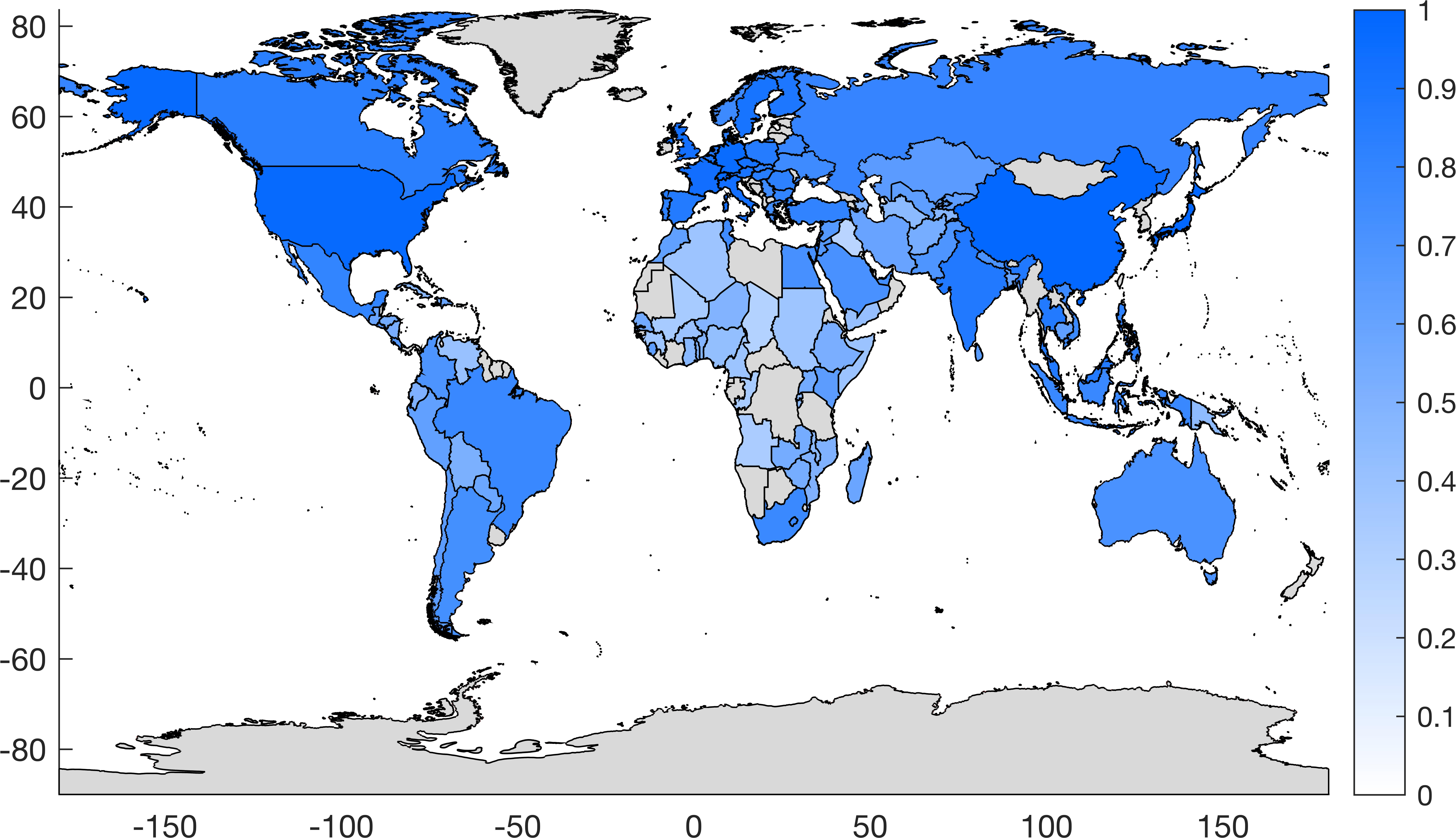}
            \caption{{\small Countries colored according to the GENEPY index computed starting from the original incidence matrix ${\bf M}$ at the HS-4 level of aggregation.}}
            \label{originalHS4}
        \end{subfigure}
        \hspace{0.3cm}
        \hfill
        \begin{subfigure}[b]{0.475\textwidth}  
            \centering 
           \includegraphics[scale=0.0175,trim=1.5cm 8.7cm 0.9cm 0cm,clip=true]{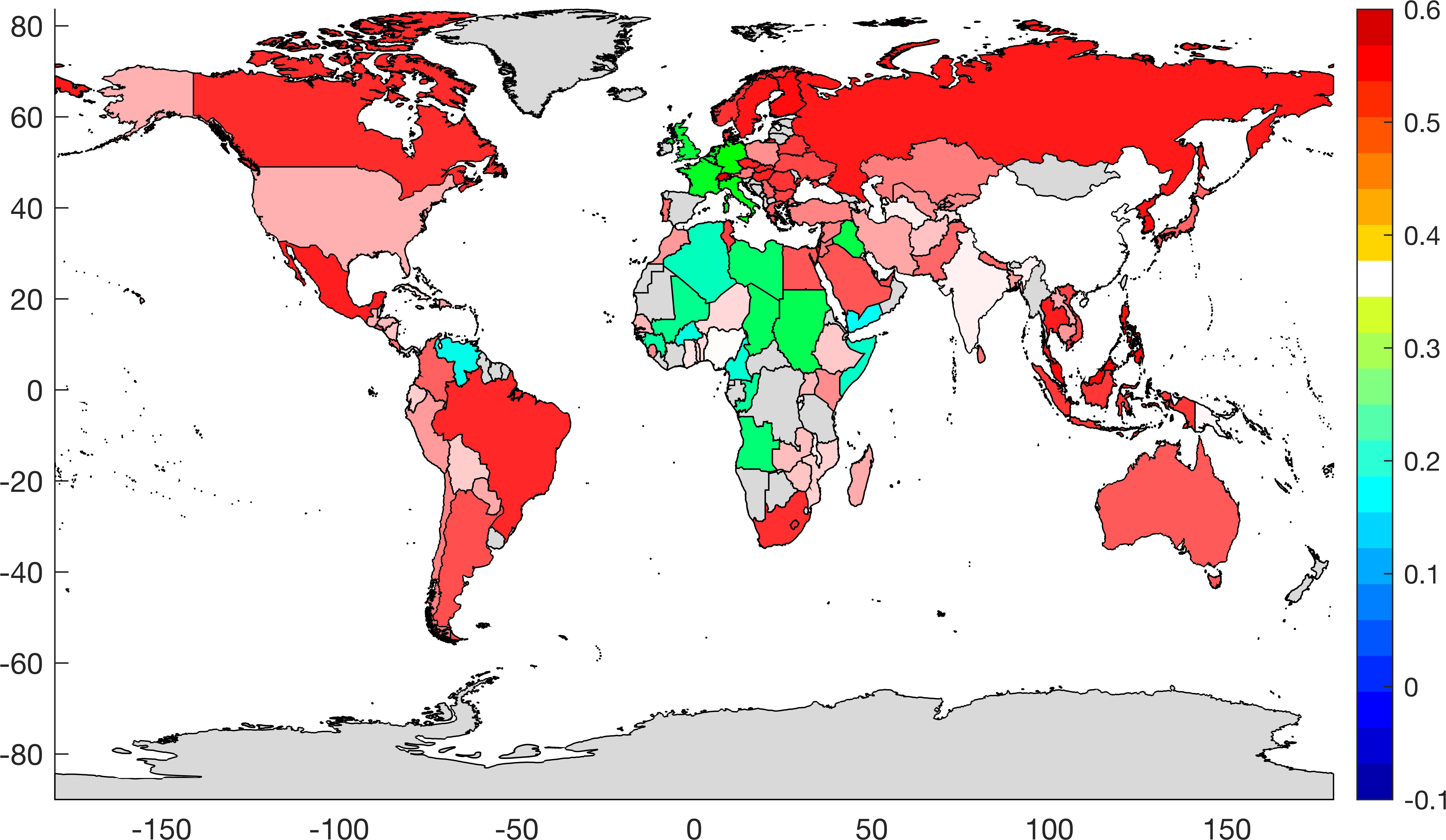}
            \caption{{\small Difference between the GENEPY index in (a) and the ${\widehat{\rm GENEPY}}^{(MC)}$ index based on its surrogate ${\hat{\bf M}}^{(MC)}$.}}
            \label{differenceHS4}
        \end{subfigure}
                \hfill
        \caption{Original  GENEPY values for countries for the year 2018 at the HS-4 level of aggregation,  and their comparison with their ${\widehat{\rm GENEPY}}^{(MC)}$ values. Countries colored in grey are not considered in the analysis.}
    \end{figure}

Fig. \ref{originalHS4} 
 represents the GENEPY index computed 
 based on the original incidence matrix ${\bf M}$. The interpretation is the same as in Fig.\ref{money_rank}. 
It is worth noticing that the GENEPY value computed based  on the incidence matrix ${\bf M}$ and the ${\widehat{\rm GENEPY}}^{(MC)}$ value based on its surrogate ${\hat{\bf M}}^{(MC)}$ are quite similar (see Fig. \ref{differenceHS4} for their difference). Hence, they provide analogous results in terms of the complexity of the countries, confirming the satisfactory prediction capability of MC for the specific learning task. Nevertheless, one can also notice that the two complexities differ in some countries. 
Such differences may be ascribed to surpluses/deficits of the actual complexities of such countries (i.e., the ones measured by GENEPY based on the original incidence matrix ${\bf M}$) with respect to the respective predicted complexities (i.e., the ones measured by ${\widehat{\rm GENEPY}}^{(MC)}$, which is based on the surrogate incidence matrix $\hat{\bf M}^{(MC)}$).\\
To 
quantify the correlation between the GENEPY rankings computed based on ${\bf M}$ and $\hat{\bf M}^{(MC)}$, respectively, we evaluated their Kendall rank correlation coefficient $\tau_k$. 
The statistical test produced  $\tau_k \simeq 0.8$ with a $p$-value near $0$, rejecting significantly the null hypothesis of independence between GENEPY and ${\widehat{\rm GENEPY}}^{(MC)}$.\\ 

\vspace{-0.6cm}
    
\vspace{-0.2cm}
\noindent It is worth noticing that, with a few exceptions (China, France, Italy, UK and Germany) the more complex the country according to GENEPY, the higher the difference between GENEPY and ${\widehat{\rm GENEPY}}^{(MC)}$. Finally, Fig. \ref{fposths4_2018} displays the false positive rate ${fpr}_c$ for each country considered in the analysis, which turned out to produce a ranking of countries quite similar to the one generated by GENEPY ($\tau_k=0.75$).
 \begin{figure}[H]{0.47}  
        \centering 
          \includegraphics[scale=0.02,trim=1.4cm 8.7cm 0cm 0cm,clip=true]{HQ_images/false_positive_HQ.jpg}
           \caption{{\small False positive rate ${fpr}_c$}, reported proportionally to the shade of blue for the year 2018 and the product aggregation level HS-4.}
            \label{fposths4_2018}
\end{figure}

 \section{Discussion}
 In the present work, we applied Matrix Completion (MC) to investigate in various ways the economic complexity of countries. First, we assessed a quite high accuracy of the MC predictions, when MC was applied to reconstruct the Revealed Comparative Advantage (RCA) matrix, which is at the basis of the construction of several existing economic complexity indices (see the Appendix). Then, we proposed the Matrix cOmpletion iNdex of Economic complexitY (MONEY), based on the predictability of the RCA entries associated with different countries. As an additional contribution, we combined MC with a recently-developed economic complexity index (GENEPY), to assess the expected economic complexity of countries.  
 In the work, MC was exploited to infer the expected discretized RCA of a country $c$ in a certain class of goods or services $p$. The MC technique employed is based on a soft-thresholded SVD. This, combined with the MC validation phase, allows to select automatically a suitable number of singular vectors to be used to reconstruct the discretized ${\bf RCA}$ matrix. In this way, differently from previous economic complexity indices, the information extracted is not restricted to the first two singular vectors. \\
 The results of our analysis highlighted a generally quite good performance of MC in discerning country-product pairs with RCA values greater than or equal to the critical threshold of 1, denoting the competitiveness of $c$ in producing $p$. The outcomes were summarized by reporting the global ROC curve and comparing the heat-map of the true incidence matrix ${\bf M}$ and the one of its MC surrogate matrix $\hat{\bf M}^{(MC)}$, which was obtained from various applications of MC. Motivated by the high MC accuracy, we developed the MONEY index taking into account both the predictive performance of MC for each country (as measured by its $AUC_c$) and the product dimension. In other words, when constructing that index, each $AUC_c$ was weighted by the average of the $\overline{ftot}_p$'s with respect to a subset of products associated with the specific country.
 As a further step, we applied the GENEPY algorithm first to the incidence matrix derived directly from the original ${\bf RCA}$ matrix, then to the MC surrogate incidence matrix $\hat{\bf M}^{(MC)}$. 
 This allowed us to directly compare the values of the two GENEPY indices, thus assessing their potential discrepancies. On average, such discrepancies were higher for more complex countries according to the original GENEPY index. 

{\footnotesize
\section*{Appendix: methods}
\renewcommand{\thesection}{\Alph{section}}
\renewcommand{\thesubsection}{\Alph{section}.\arabic{subsection}}



\subsection*{Matrix completion via nuclear norm regularization of the reconstruction error}



Given a subset of observed entries of a matrix ${\bf A} \in \mathbb{R}^{C \times P}$, Matrix Completion (MC) works by finding a suitable low-rank approximation (say, with rank $R$) of ${\bf A}$, by assuming the following model:
\begin{equation}\label{eq:matrix_completion}
{\bf A}= {\bf C} {\bf G}^\top + {\bf W}\,,
\end{equation}
where ${\bf C} \in \mathbb{R}^{C \times R}$, ${\bf G} \in \mathbb{R}^{P \times R}$, whereas ${\bf W} \in \mathbb{R}^{C \times P}$ is a matrix of modeling errors. The rank-$R$ approximating matrix ${\bf C} {\bf G}^\top$ is found by solving a suitable optimization problem (provided by Eq. (\ref{eq:matrix_completion1}), in the case of the present article). 
Eq. (\ref{eq:matrix_completion}) can be written element-wise as 
$A_{c,p}=\sum_{r=1}^R C_{c,r} G_{p,r} + W_{c,p}$. 
A common interpretation of this equation 
is as follows (see, e.g., the application of MC to collaborative filtering for movie ratings \cite{Hastie2015}). The number $C_{c,r}$ can be interpreted as the degree of membership of row $c$ of matrix ${\bf A}$ to some ``latent'' cluster $r$ (for a total of $R$ such clusters), and $G_{p,r}$ as the prediction of an element in column $p$ of matrix ${\bf A}$, conditioned on its row $c$ belonging to cluster $r$. It is worth mentioning that such an interpretation holds regardless of the signs of the elements $C_{c,r}$ and $G_{p,r}$. As an example, in the case of collaborative filtering for movie ratings, $c$ denotes a specific person, $p$ a specific movie, whereas $r$ may be interpreted as a specific movie genre.

\noindent In this work, MC is formulated via the optimization problem (\ref{eq:matrix_completion1}).  
The objective function of this optimization problem is the sum of two terms: the first one refers to the reconstruction error of the known portion of the matrix, whereas the second one is a regularization term, which biases the reconstructed matrix to have a small nuclear norm. The regularization constant $\lambda$ controls the trade-off between fitting the known entries of the matrix ${\bf A}$ and achieving a small nuclear norm. The latter requirement is often related to getting a low rank of the optimal matrix ${\bf Z}^\circ$, which follows by geometric arguments similar to the ones typically adopted to justify how the classical LASSO (Least Absolute Shrinkage and Selection Operator) penalty term achieves effective feature selection in linear regression \cite{Tibshirani1996}.

\noindent The MC optimization problem (\ref{eq:matrix_completion1}) can be also written as
\begin{equation}\label{eq:matrix_completion2}
\underset{{\bf Z} \in \mathbb{R}^{C \times P}}{\rm minimize} \left(\frac{1}{2} \|{\bf P}_{\Omega^{\rm tr}}({\bf A})-{\bf P}_{\Omega^{\rm tr}}({\bf Z})\|_F^2 + \lambda \|{\bf Z}\|_*\right)\,,
\end{equation}
where, for a matrix ${\bf Y} \in \mathbb{R}^{C \times P}$, $(P_{\Omega^{\rm tr}}({\bf Y}))_{c,p}:= Y_{c,p}$ if $(c,p) \in \Omega^{\rm tr}$, otherwise it is equal to $0$. Here, $P_{\Omega^{\rm tr}}({\bf Y})$ represents the projection of ${\bf Y}$ onto the set of positions of observed entries of the matrix ${\bf A}$, and
$\|{\bf Y}\|_F$ denotes the Frobenius norm of ${\bf Y}$ (i.e., the square root of the summation of squares of all its entries).

\noindent The MC optimization problem (\ref{eq:matrix_completion2}) can be solved by applying the following Algorithm \ref{alg:1}, named Soft Impute \cite{Mazumder2010} (compared to the original version, here we have included a maximal number of iterations $N^{\rm it}$, which can be helpful to reduce the computational effort when one has to run the algorithm multiple times, e.g., for several choices of the training set $\Omega^{\rm tr}$ and of the regularization constant $\lambda$, as in the present work):

 \begin{algorithm}[H]\label{alg:1}
\SetAlgoLined
\KwData{Partially observed matrix ${\bf P}_{\Omega^{\rm tr}}({\bf A})$ , regularization constant $\lambda \geq 0$, tolerance $\varepsilon \geq 0$, maximal number of iterations $N^{\rm it}$}
\KwResult{Completed matrix ${\bf Z}_\lambda \in \mathbb{R}^{C \times P}$}
  \begin{enumerate}
  \item Initialize ${\bf Z}$ as ${\bf Z}^{\rm old}={\bf 0} \in \mathbb{R}^{C \times P}$
 \item Repeat for at most $N^{\rm it}$ iterations:
 \begin{enumerate}
 \item Set ${\bf Z}^{\rm new}\leftarrow {\bf S}_\lambda\left({\bf P}_{\Omega^{\rm tr}}({\bf A})+{\bf P}_{\Omega^{\rm tr}}^{\perp}({\bf Z}^{\rm old}) \right)$
 \item If $\frac{\|{\bf Z}^{\rm new}-{\bf Z}^{\rm old} \|_F^2}{\|{\bf Z}^{\rm old}\|_F^2} < \varepsilon$, exit\
 \item Set ${\bf Z}^{\rm old}\leftarrow{\bf Z}^{\rm new}$
 \end{enumerate}
 \item Set ${\bf Z}_\lambda\leftarrow{\bf Z}^{\rm new}$
  \end{enumerate}
 \caption{{\bf Soft Impute} \cite{Mazumder2010}}
\end{algorithm}

\noindent In Algorithm \ref{alg:1}, for a matrix ${\bf Y} \in \mathbb{R}^{C \times P}$, ${\bf P}_{\Omega^{\rm tr}}^{\perp}({\bf Y})$ represents the projection of ${\bf Y}$ onto the complement of $\Omega^{\rm tr}$, whereas 
${\bf S}_\lambda({\bf Y}):= {\bf U} \bm{\Sigma}_\lambda {\bf V}^\top,$ 
being
 ${\bf Y}={\bf U} \bm{\Sigma} {\bf V}^\top$ 
(with $\bm{\Sigma}={\rm diag} [\sigma_1,\ldots,\sigma_R]$) the singular value decomposition of ${\bf Y}$, and 
 $\bm{\Sigma}_\lambda:={\rm diag} [(\sigma_1-\lambda)_+,\ldots,(\sigma_R-\lambda)_+]$, 
with $t_+:=\max(t,0)$.

\noindent It is worth mentioning that a particularly efficient implementation of the operator ${\bf S}_{\lambda}(\cdot)$  
is possible (by means of the MATLAB function {\tt svt.m}\cite{LiZhou2017}), which is based on the determination of only the singular values $\sigma_i$ of ${\bf Y}$ that are higher than $\lambda$, and of their corresponding left-singular vectors ${\bf u}_i$ and right-singular vectors ${\bf v}_i$. Indeed, all the other singular values of ${\bf Y}$ are annihilated in $\bm{\Sigma}_\lambda$. 

\noindent A final remark has to be made about the trade-off between prediction capability and biasedness of MC. Biasedness in MC depends, among others issues, on the way the selection of unobserved entries is made \cite{Foucartetal2017,MaChen2019} (in the specific case of our  application of MC to the discretized {\bf RCA} matrix, only entries belonging to a suitable subset of rows of the matrix ${\bf A}$ are obscured). For some MC algorithms, de-biasing is possible \cite{Foucartetal2017}, and can even improve prediction capability. Nevertheless, in general biasedness can be beneficial to prediction capability, due to the well-known trade-off between bias and variance \cite{Hastieetal2009}. In the particular case of MC achieved via the Soft Impute algorithm, biasedness can be ascribed also to the presence of the regularization constant $\lambda$ (indeed, for both $\lambda \to 0$ and $\lambda \to +\infty$, the predictions of the optimal solution to the optimization problem (\ref{eq:matrix_completion2}) tend to $0$ for the unobserved entries), and to the fact that the Soft Impute algorithm is initialized by a matrix with all entries equal to $0$, and terminated at most after a given number of iterations.

\subsection*{Technical details on the construction of the matrix ${\bf A}$ and on the the application of the Soft Impute algorithm}
 This subsection details the construction of the matrix ${\bf A}$ for our specific problem. As a first step, we removed from the ${\bf RCA}$ matrix its rows associated with countries having less than 5 million inhabitants. Then, the remaining entries of the ${\bf RCA}$ matrix were encoded into 9 groups according to increasing percentiles in the distribution of RCA values. This pre-processing step was done in order to make the elements of the resulting matrix ${\bf A} \in \mathbb{R}^{119 \times 1243}$ 
 of the same order of magnitude. 
In particular, we defined 4 negative groups ("-4", "-3", "-2", "-1"), representing the case $0 \leq RCA <1$ (with the group "-4" being the one associated with the lowest values in the RCA distribution) and 4 positive groups ("1", "2", "3", "4"), representing the case $RCA \geq 1$ (with the group "4" being the one associated with the highest values in the RCA distribution)\footnote{As already reported in the Appendix, MC is biased towards low absolute values. To take this into account, we constructed groups that were symmetrically distributed around zero, as the final goal was to discriminate between RCA values respectively lower than $1$, and larger than or equal to $1$.}. Originally $NaN$ RCA values 
were included in the remaining group "0". 
In our  application of MC, the elements in this group "0" 
were included neither in the training set, nor in the validation/test set, since no ground truth was available for them. 
\\
For computational efficiency reasons, we combined the original MATLAB implementation of Soft Impute \cite{Mazumder2010} with the MATLAB function {\tt svt.m} \cite{LiZhou2017}. The tolerance of the algorithm was chosen as $\varepsilon=10^{-9}$. Its number of iterations was set to $N^{\rm it}=1500$. The regularization parameter $\lambda$ was sampled $30$ times uniformly on the closed interval $[-1,15]$ in a logaritmic scale with base $2$. A post-processing step was included in MC, thresholding to $-4$ any element (when present) whose MC reconstruction was lower than -4, and to $4$ any element (when present) whose MC reconstruction was higher than 4. 

\subsection*{Generalized economic complexity index and related economics complexity indices}

The GENeralised Economic comPlexitY  (GENEPY) index is a recently-introduced economic complexity index \cite{Sciarraetal2020}, which can be applied to assess the complexity of both countries and products. It is based on a multidimensional representation of their complexity, which makes it possible to combine, in a single index, the different features of some previously-developed one-dimensional economic complexity indices: the Fitness (F) for countries and Quality (Q) for products, both computed by the Fitness and Complexity (FC) algorithm \cite{Tacchellaetal2012}, and the Economic Complexity Index (ECI) for countries and Product Complexity Index (PCI) for products, both obtained by the earlier Method of Reflections (MR) \cite{HidalgoHausmann2009}. Each of the latter methods is typically able, indeed, to capture only a specific aspect of economic complexity: for instance, when applied to countries, FC is mainly related to the degree of diversification of the export basket of each country, while MR essentially captures the similarities in the export baskets of the different countries \cite{Sciarraetal2020}.\\
The GENEPY index arises from the first two (normalized) eigenvectors (with the eigenvalues ordered in a weakly decreasing way) of a suitable symmetric proximity matrix, which is derived from an incidence matrix ${\bf M} \in \mathbb{R}^{C \times P}$ obtained by thresholding and binarizing the matrix of Revealed Comparative Advantage (RCA) values. The two eigenvectors capture, respectively, information obtained by the FC method and the MR one.\\
The GENEPY index for countries is obtained in the following way (a similar construction holds for the GENEPY index for products).
 \vspace{-0.2cm}
\begin{enumerate}[itemsep=0pt, parsep=0pt, leftmargin=0.5cm]
\item First, for a specific year, the 
matrix ${\bf RCA} \in \mathbb{R}^{C \times P}$ of RCA values in that year is determined (see the Introduction for details). 

\item In order to extract topological information from the ${\bf RCA}$ matrix, an incidence matrix ${\bf M} \in \mathbb{R}^{C \times P}$ is generated, whose entries are defined as follows:
\begin{equation}
M_{c,p} := \begin{cases} 1 \,,& \mbox{\,if\,} RCA_{c,p} \geq 1\,, \\
0\,, & \mbox{otherwise}\,.
\end{cases}
\end{equation}
Then, its weighted version ${\bf W} \in \mathbb{R}^{C \times P}$ is considered, whose generic element is defined as $W_{c,p}:=\frac{M_{c,p}}{k_c k'_p}$, where $k_c:=\sum_{p=1}^P M_{c,p}$ is the degree of the country $c$ in the graph represented by the incidence matrix ${\bf M}$, and $k'_p:=\sum_{c=1}^C \frac{M_{c,p}}{k_c}$ represents the degree of the product $p$ corrected by how easily that product
is found within the subnetwork of countries.
\item The matrix ${\bf N \in \mathbb{R}^{C \times C}}$ is constructed, whose elements $N_{c,c^*}$ are defined as follows:
\begin{equation}\label{eq:matrixN}
N_{c,c^*}:=
\begin{cases} \sum_{p=1}^P W_{c,p} W_{c^*,p}\,,& \mbox{\,if\, } c \neq c^*,\\
0, &\, \mbox{otherwise}\,.
\end{cases}
\end{equation}
Due to the weighting involved in the construction of the matrix ${\bf W}$, the resulting matrix ${\bf N}$ is symmetric. Each entry $N_{c,c^*}$ of ${\bf N}$ represents the proximity of the two corresponding countries $c$ and $c^*$.
\item The (normalized) eigenvectors ${\bf x}_1,{\bf x}_2 \in \mathbb{R}^C$ associated with the two largest eigenvalues $\lambda_1 \geq \lambda_2 \geq 0$ of ${\bf N}$
are determined. Their components are denoted as $x_{c,1}$ and $x_{c,2}$, respectively, for $c=1,\ldots,C$.
\item Then, the GENEPY index of country $c$ for the specific year is defined as follows:
\begin{equation}\label{eq:GENEPY}
GENEPY_c:=\left(\sum_{i=1}^2 \lambda_i x_{c,i}^2 \right)^2 + 2 \sum_{i=1}^2 \lambda_i^2 x_{c,i}^2\,.
\end{equation}
The specific nonlinear transformation from $x_{c,1}$ and $x_{c,2}$ to 
$GENEPY_c$, which is used in Eq. (\ref{eq:GENEPY}), can be justified by rigorous statistical arguments, based on the use of the two (normalized) eigenvectors ${\bf x}_1$ and ${\bf x}_2$ to get a nonlinear least-square estimate of the matrix ${\bf N}$, and on the evaluation of how relevant $x_{c,i}$ and $x_{c,2}$ are to obtain that estimate \cite{Sciarraetal2018,Sciarraetal2020}.
\end{enumerate}
\vspace{-0.2cm}
It is worth mentioning the qualitative difference between the GENEPY index and the ones determined by the FC and MR methods, considering again the case in which they are all applied to countries.
\vspace{-0.2cm}
\begin{itemize}[itemsep=0pt, parsep=0pt, leftmargin=0.5cm]
\item The GENEPY index is highly related to a linearized version of the F index computed by the FC method\cite{Sciarraetal2020}, in which one searches for the (normalized) eigenvector associated with the largest eigenvalue of a slightly different matrix ${\bf N}_{F} \in \mathbb{R}^{C \times C}$ than the matrix ${\bf N}$. The specific matrix ${\bf N}_{F}$ is written as ${\bf N}_{F}:={\bf W} {\bf W}^\top$, where ${\bf W}$ is the same weighted incidence matrix considered in the context of the GENEPY index. The difference with respect to the case of the matrix ${\bf N}$ defined in Eq. (\ref{eq:matrixN}) is that its diagonal entries are not set to $0$ (in Eq. (\ref{eq:matrixN}), such a choice of the diagonal entries is done in order to make the resulting ${\bf N}$ be a proximity matrix). 
\item The ECI index, computed by MR, is based on searching for the (normalized) eigenvector associated with the second-largest eigenvalue of a slightly different matrix ${\bf N}_{ECI} \in \mathbb{R}^{C \times C}$ than the matrix ${\bf N}$ considered by GENEPY. The specific matrix ${\bf N}_{ECI}$ is written as ${\bf N}_{ECI}:={\bf W}_{ECI} {\bf W}_{ECI}^\top$, where the elements of ${\bf W}_{ECI} \in \mathbb{R}^{C \times P}$ are defined as $W_{ECI,c,p}:=\frac{M_{c,p}}{k_c k_p}$, being $k_p:=\sum_{c=1}^C M_{c,p}$ the degree of the product $p$ in the graph represented by the incidence matrix ${\bf M}$. The second-largest eigenvalue of ${\bf N}_{ECI}$ is considered, instead of its first-largest one, as one can show that the (normalized) eigevector associated with the latter is non-informative, for the specific matrix ${\bf N}_{ECI}$.
\end{itemize}
\vspace{-0.2cm}
Similar comments hold for the case of the FC and MR methods when they are applied to products (obtaining, respectively, the Q index and the PCI index).


\begin{thebibliography}{}
\bibitem{alfaik200}
Alfakih, A., \& Wolkowicz, H. (2000). Matrix completion problems. In Handbook of semidefinite programming (pp. 533--545). Springer.
\vspace{-0.2cm}
\bibitem{Balassa1965}
Balassa, B. (1965). Trade liberalisation ad revealed comparative advantage. Manchester School of Economics and Social Studies, 33, pp. 99--123.
\vspace{-0.2cm}
\bibitem{caietal2010}
Cai, J. F., Cand\`es, E. J., \& Shen, Z. (2010). A singular value thresholding algorithm for matrix completion. SIAM Journal on optimization, 20(4), pp. 1956--1982.
\vspace{-0.2cm}
\bibitem{Fawcett2006}
Fawcett, T., 2006. An introduction to ROC analysis. Pattern recognition letters, 27(8), pp.861-874.
\vspace{-0.2cm}
\bibitem{Foucartetal2017}
Foucart, S., Needell, D., Plan Y., \& Wootters, M. (2017). De-biasing low-rank projection for matrix completion. Proceedings of SPIE, 10394, Wavelets and Sparsity XVII, 1039417, 13 pages.
\vspace{-0.2cm}
\bibitem{Hastieetal2009}
Hastie, T., Tibshirani, R., \& Friedman, J. (2009). The elements of
statistical learning: data mining, inference, and prediction. Springer. 
\vspace{-0.2cm}
\bibitem{Hastie2015}
Hastie, T., Tibshirani, R., \& Wainwright, M. (2015). Statistical learning with sparsity: the Lasso and generalizations. 
CRC Press.
\vspace{-0.2cm}
\bibitem{HastieMC}
Hastie, T., Mazumder, R., Lee, J.D. and Zadeh, R., 2015. Matrix completion and low-rank SVD via fast alternating least squares. The Journal of Machine Learning Research, 16(1), pp. 3367--3402.
\vspace{-0.2cm}
\bibitem{HidalgoHausmann2009}
Hidalgo, C. A., \& Hausmann, R (2009). The building blocks of economic complexity. Proceedings of the National Academy of Sciences, 106, pp. 10570--10575.
\vspace{-0.2cm}
\bibitem{Hidalgo2021}
Hidalgo, C. A. (2021). Economic complexity theory and applications. Nature Reviews Physics, 3, pp. 92--113.
\vspace{-0.2cm}
\bibitem{LiZhou2017}
Li, C., \& Zhou, H. (2017). Svt: Singular value thresholding in MATLAB. Journal of Statistical Software, 81(2), DOI: 10.18637/jss.v081.c02.
\vspace{-0.2cm}
\bibitem{MaChen2019}
Ma, W., \& Chen, G. H. (2019). Missing not at random in matrix completion: the effectiveness of estimating missingness probabilities under a low nuclear norm assumption. In Advances in Neural Information Processing Systems 32, Proceedings of the 33$^{\rm rd}$ Conference on Neural Information Processing Systems (NeurIPS 2019), Vancouver, Canada.
\vspace{-0.2cm}
\bibitem{Mazumder2010}
Mazumder, R., Hastie, T., \& Tibshirani, R. (2010). Spectral regularization algorithms for learning large incomplete matrices. Journal of Machine Learning Research, 11, pp. 2287--2322.
\vspace{-0.2cm}
\bibitem{Sciarraetal2018}
Sciarra, C., Chiarotti, G., Laio, F., \& Ridolfi, L. (2018). A change of perspective in network centrality. Scientific Reports, 8, article no. 15269.
\vspace{-0.2cm}
\bibitem{Sciarraetal2020}
Sciarra, C., Chiarotti, G., Ridolfi, G., \& Laio, F. (2020). Reconciling contrasting views on economic
complexity. Nature Communications, 11, article no. 3352.
\vspace{-0.2cm}
\bibitem{Tacchellaetal2012}
Tacchella, A., Cristelli, M., Caldarelli, G., Gabrielli, A., \& Pietronero, L. (2012). A new metrics for countries' fitness and products' complexity. Scientific Reports, 2, article no. 723.
\vspace{-0.2cm}
\bibitem{Tibshirani1996}
Tibshirani, R. (1996). Regression shrinkage and selection via the Lasso. Journal of the Royal Statistical Society. Series B (Methodological), 58(1), pp. 267–-288.
\end{thebibliography}
\vspace{-0.3 cm}
\bibliographystyle{plain}



\section*{Author contributions statement}

M.R. conceived the idea of the work, G.G. and F.N. developed the proposed method and implemented it.  All the authors analysed the results and reviewed the manuscript. 


\section*{Additional information}

\textbf{Competing interests} The authors state that they have no competing interests. }


\clearpage
\newpage

\appendix

\section*{Supplemental}

\subsection*{MONEY index for the years 2005 and 2014}
\begin{figure}[H]
        \centering
        \begin{subfigure}[b]{0.475\textwidth}
            \centering 
            \includegraphics[width=8.2cm,height=4.5cm,trim=1cm 0.5cm 0cm 0.3cm,clip=true]{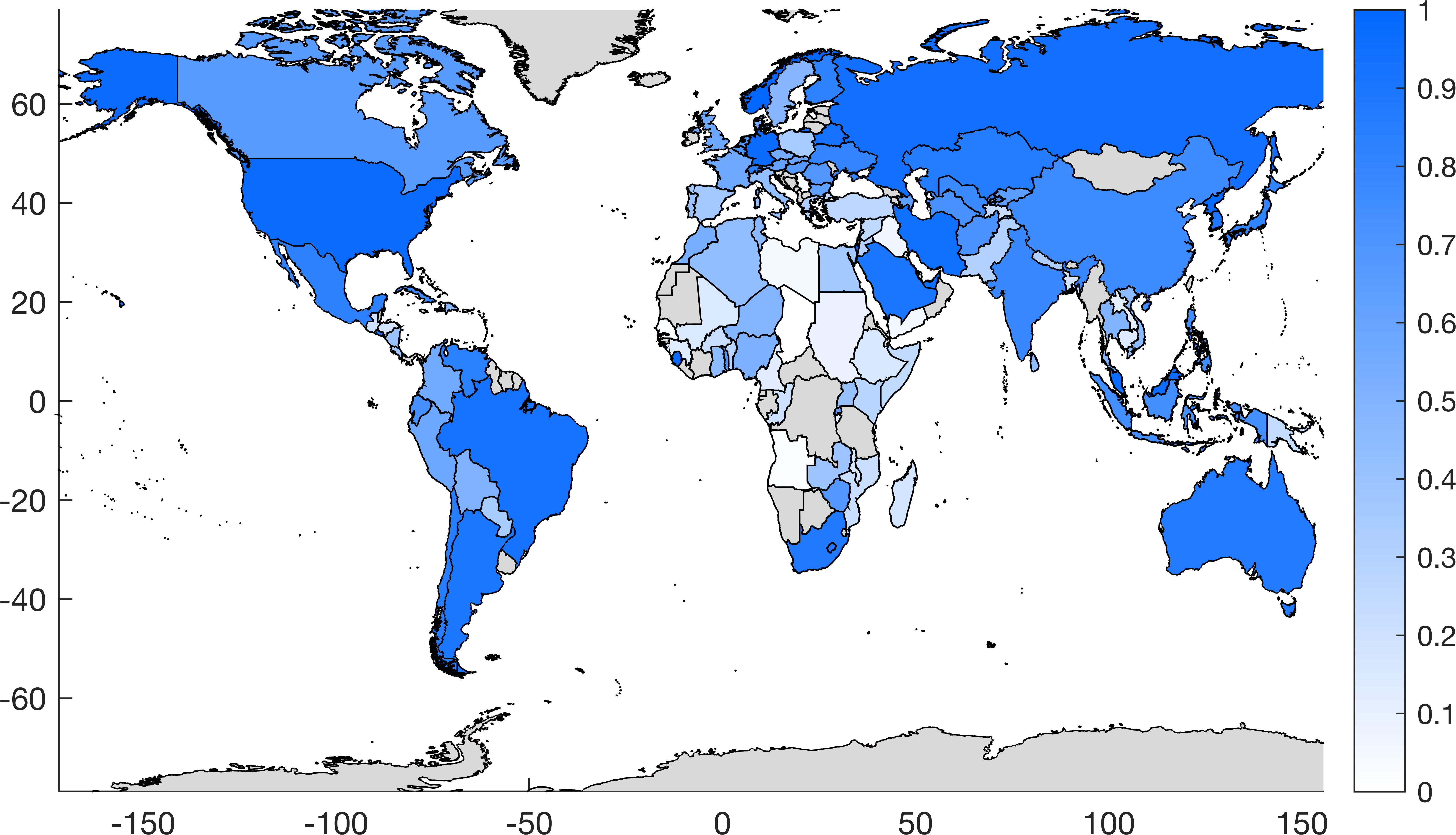}
            \caption{{\small Values of the MONEY index for the year 2005 at the HS-4 level of aggregation.}}
            \label{money_2005}
        \end{subfigure}
        \hfill
        \begin{subfigure}[b]{0.475\textwidth}  
            \centering 
               \includegraphics[width=8.2cm,height=4.5cm,trim=1cm 0.5cm 0cm 0.3cm,clip=true]{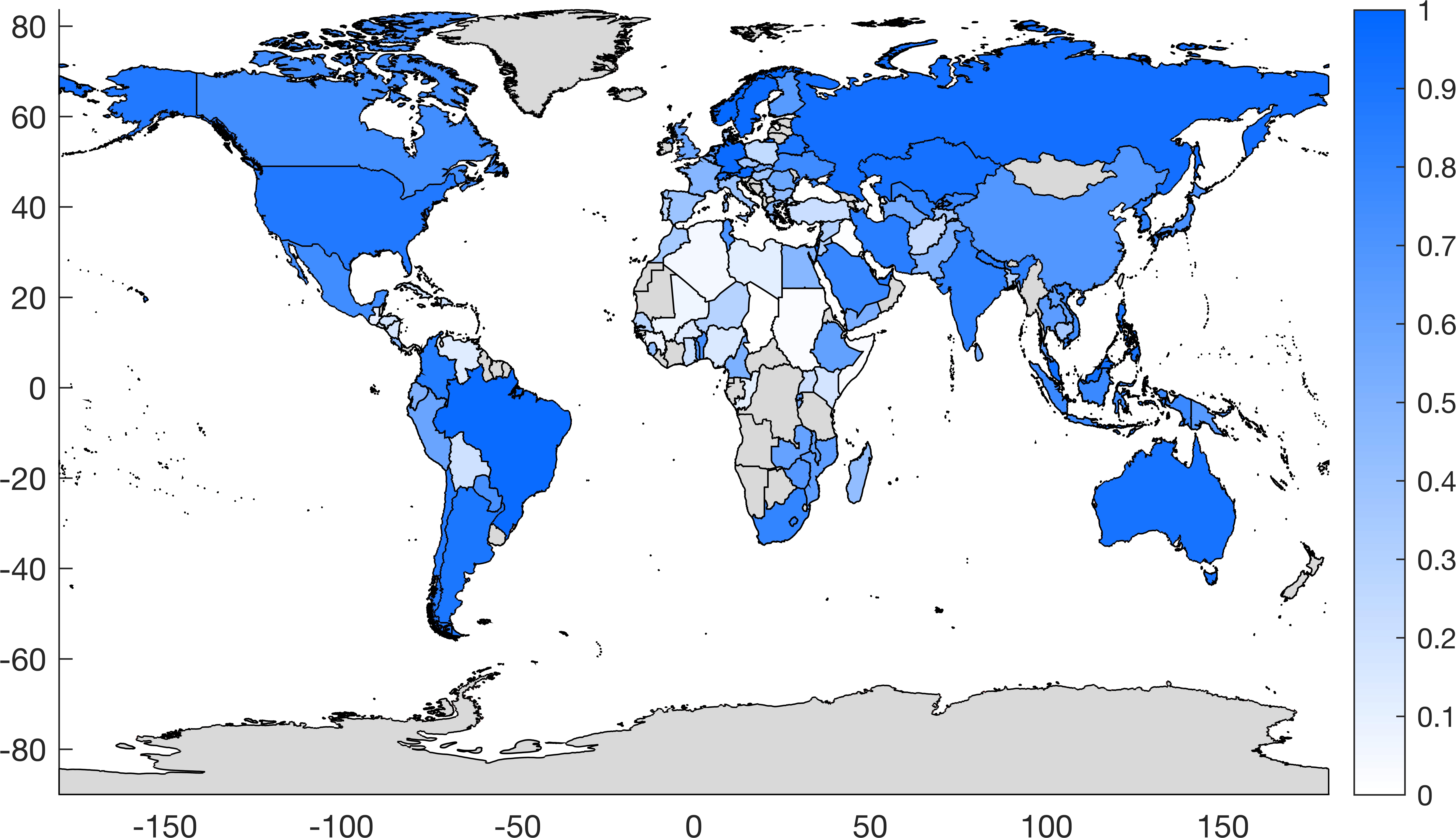}
            \caption{{\small Values of the MONEY index for the year 2014 at the HS-4 level of aggregation.}}
            \label{money_2014}
        \end{subfigure}
        \caption{Values of the MONEY index for the years 2005 and 2014 at the HS-4 level of aggregation.}
        \label{new}
\end{figure}

The present section provides robustness checks on the MONEY index. In particular, the whole analysis has been repeated for the years 2005 and 2014. For such years, the global AUC (see Section \ref{roc}) is respectively 0.76 for 2005 and 0.80 for 2014. For the same reason as reported in the main text -- that is, the fact that the two categories $RCA>1$ and $0 \leq RCA <1$ are unbalanced -- we also computed the BACC. The latter amounts to 0.74 for 2005 and to 0.75 for 2014. \\
Figures \ref{money_2005}-\ref{money_2014} display the MONEY index for the years 2005 and 2014.
    
\noindent The figures display similar results to those obtained in the main analysis. However, some differences emerge. Specifically, Russia and East Asia appear to be more complex in 2005 and 2014 than in 2018. Such countries were indeed rapidly developing in those years and their growth rate was higher than the one of EU, which was slowed down by the 2008 financial crisis. The crisis had an impact also on USA. Its MONEY index was in fact lower in 2014 as compared to 2005. In the successive years, yet, EU and USA recovered from the crisis, and their complexity, as measured by the MONEY index, raised accordingly in 2018.

\subsection*{Application of the second part of the analysis to countries for the year 2018, with products aggregated at the HS-2 level}

\begin{figure}[H]
        \begin{subfigure}[b]{0.475\textwidth}   
            \centering 
        \includegraphics[width=9.6cm,height=5cm,trim=2cm 8.5cm 0cm 8.3cm,clip=true]{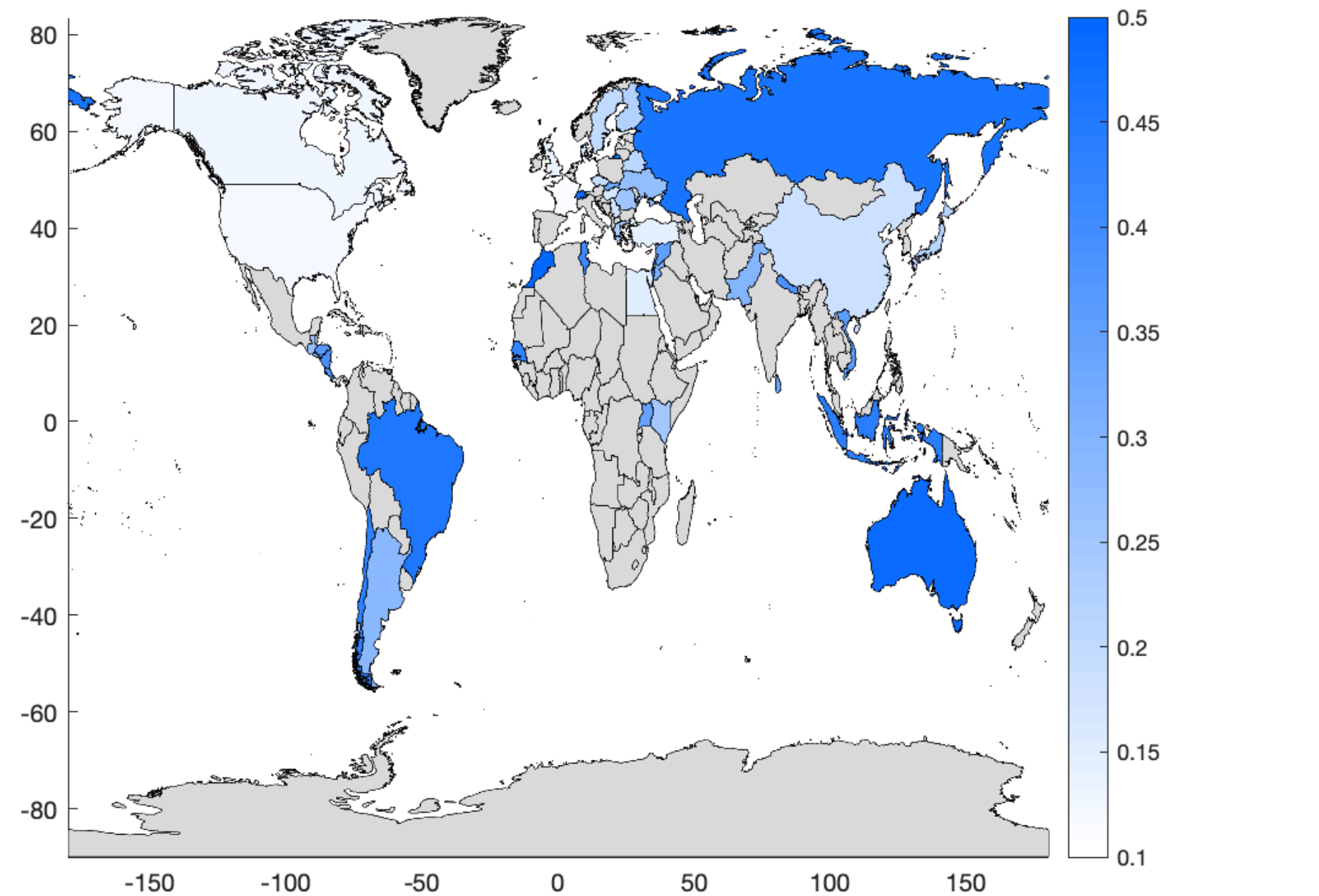}
            \caption{{\small False negative rate ${fnr}_c$ for countries $c$ having both average false negative and false positive rates lower than 0.5. In the figure, ${fnr}_c$ is reported proportionally to the shade of blue. Countries colored in grey are not considered in the analysis.}}    
        \label{signpostesworld}
        \end{subfigure}
        \hfill
        \begin{subfigure}[b]{0.475\textwidth}   
            \centering 
 \includegraphics[width=9.6cm,height=5cm,trim=2cm 8.5cm 0cm 8.3cm,clip=true]{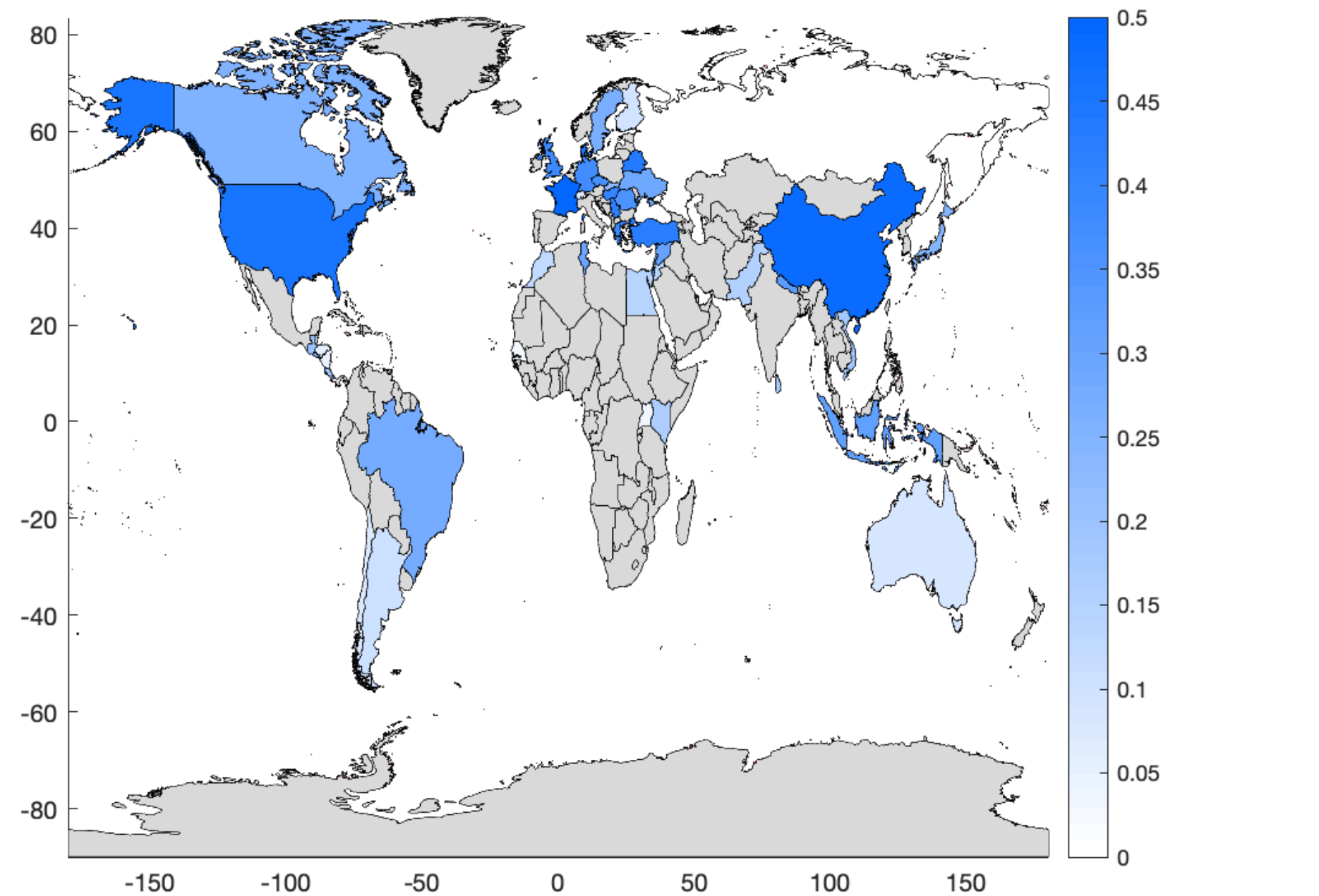}
            \caption{{\small False positive rate ${fpr}_c$ for countries $c$ having both average false negative and false positive rates lower than 0.5. In the figure, ${fpr}_c$ is reported proportionally to the shade of blue. Countries colored in grey are not considered in the analysis.}}    
     \label{signnegtesworld}
        \end{subfigure}
        \caption{False negative and false positive rates for a selection of countries, obtained by the  method reported in Step \ref{step:5} of Section \ref{ourMC} for the year 2018 and the HS-2 level of aggregation.}
        \label{nets}
    \end{figure}

Fig. \ref{new} reports, for the product aggregation level HS-2, results similar to those obtained in Section \ref{sec:resultsGENEPY} of the main text for the HS-4 level. For the sake of completeness, we report also results for the false negative rate. Fig. \ref{nets} provides similar results, restricting to the countries for which both the false negative rate and the false positive are lower than $0.5$. 

 \begin{figure}[H]
        \centering
        \begin{subfigure}[b]{0.475\textwidth}
            \centering 
        \includegraphics[width=9.6cm,height=5cm,trim=2cm 8.5cm 0cm 8.3cm,clip=true]{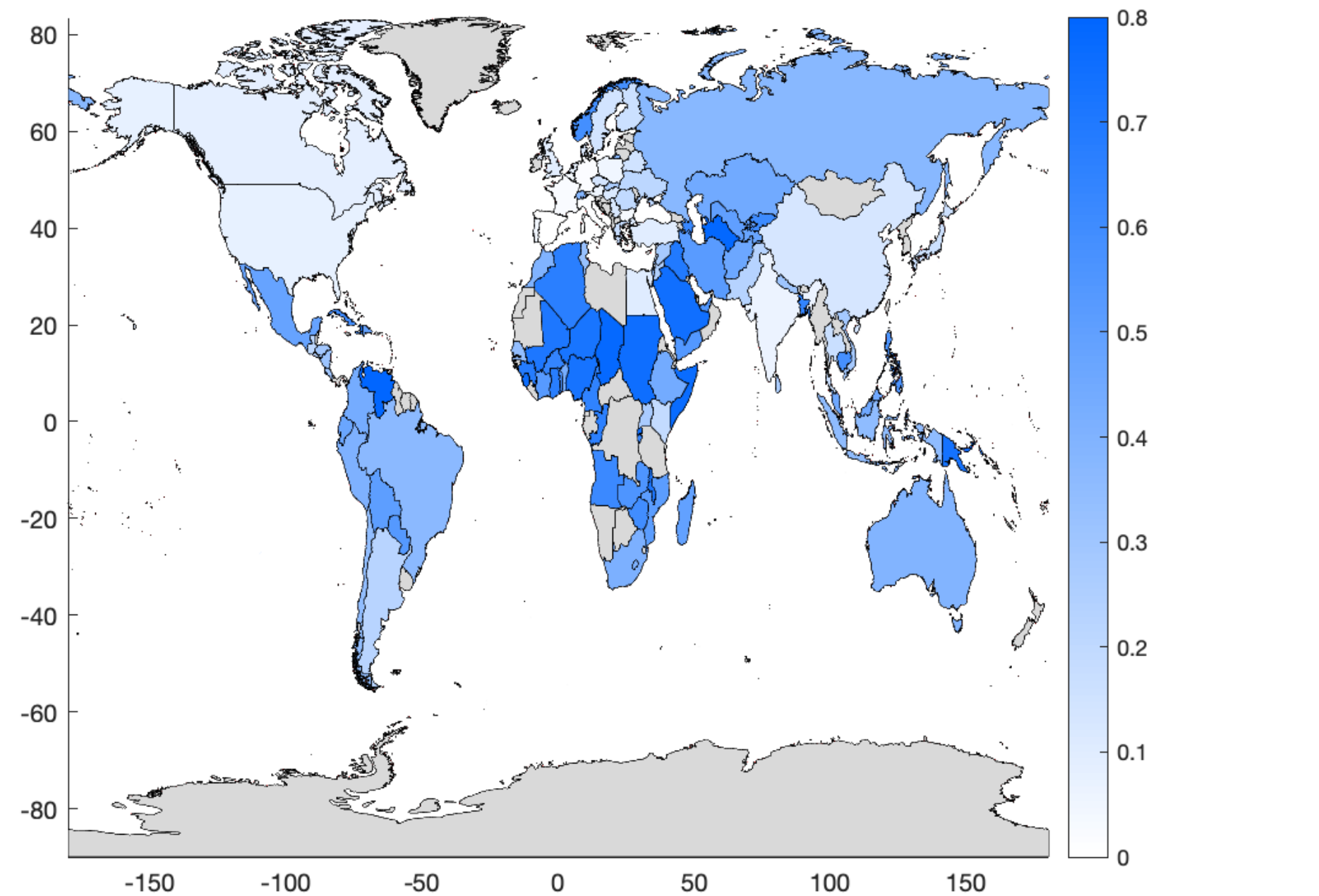}
            \caption{{\small False negative rate ${fnr}_c$}, \, reported proportionally to the shade of blue. Countries colored in grey are not considered in the analysis.}
            \label{postesworld}
        \end{subfigure}
        \hfill
        \begin{subfigure}[b]{0.475\textwidth}  
            \centering 
        \includegraphics[width=9.6cm,height=5cm,trim=2cm 8.5cm 0cm 8.3cm,clip=true]{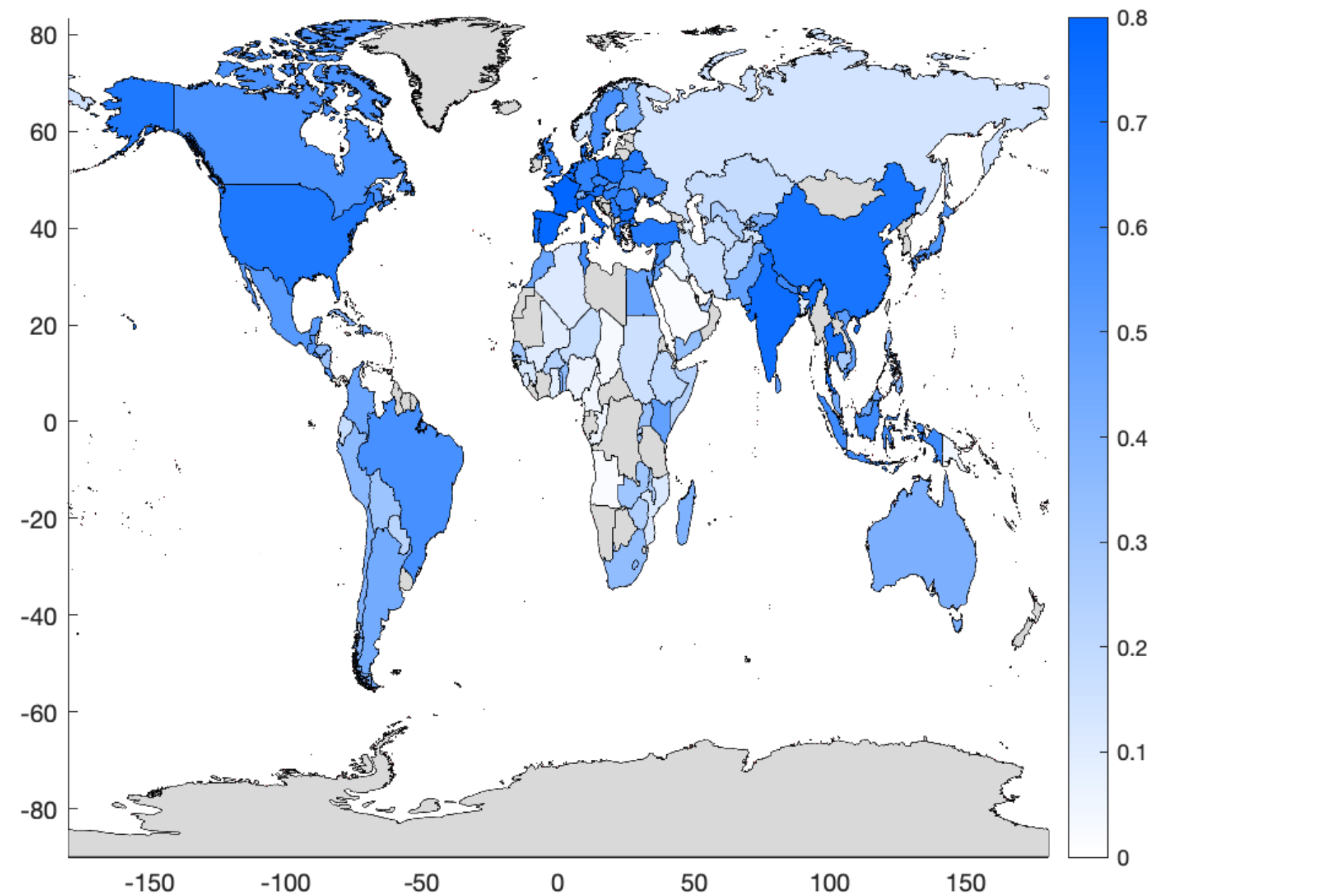}
            \caption{{\small False positive rate ${fpr}_c$}, reported proportionallly to the shade of blue. Countries colored in grey are not considered in the analysis.}
            \label{negtesworld}
        \end{subfigure}
        \caption{False negative and false positive rates for countries, obtained by the  method reported in Step \ref{step:5} of Section \ref{ourMC} for the year 2018 and the HS-2 level of aggregation.}
        \label{new}
\end{figure}

\noindent Additionally, Tab. \ref{genep_vs_ft_1} reports, for the product aggregation level HS-2, 
the Kendall correlation coefficients $\tau_k$ between the ranking produced using GENEPY against the ones produced using either ${fnr}_{c,hs-2}$ or ${fpr}_{c,hs-2}$. 

\begin{table}[H]
\centering
\begin{tabular}{|l|l|l|}
\hline
                            & \textbf{GENEPY ($\tau_k$)} & \textbf{GENEPY ($\text{$p$-value}$)} \\ \hline
\textbf{${fnr}_{c,hs-2}$} & 0.1230                      & 0.0575                    \\ \hline
\textbf{${fpr}_{c,hs-2}$} & 0.6476                    & 0.0000                     \\ \hline
\end{tabular}
\caption{Kendall rank correlation coefficients $\tau_k$ and corresponding $p$-values for the 2018 ranking of countries based on the HS-2 level of aggregation and  produced using GENEPY against the 2018 rankings produced respectively by ${fnr}_c$, and ${fpr}_c$.}
\label{genep_vs_ft_1}
\end{table}

\begin{figure}[H]
    \centering
    \begin{subfigure}[b]{0.475\textwidth}
            \centering
          \includegraphics[width=8.9cm,height=4.6cm,trim=2.4cm 8.5cm 0cm 8.3cm,clip=true]{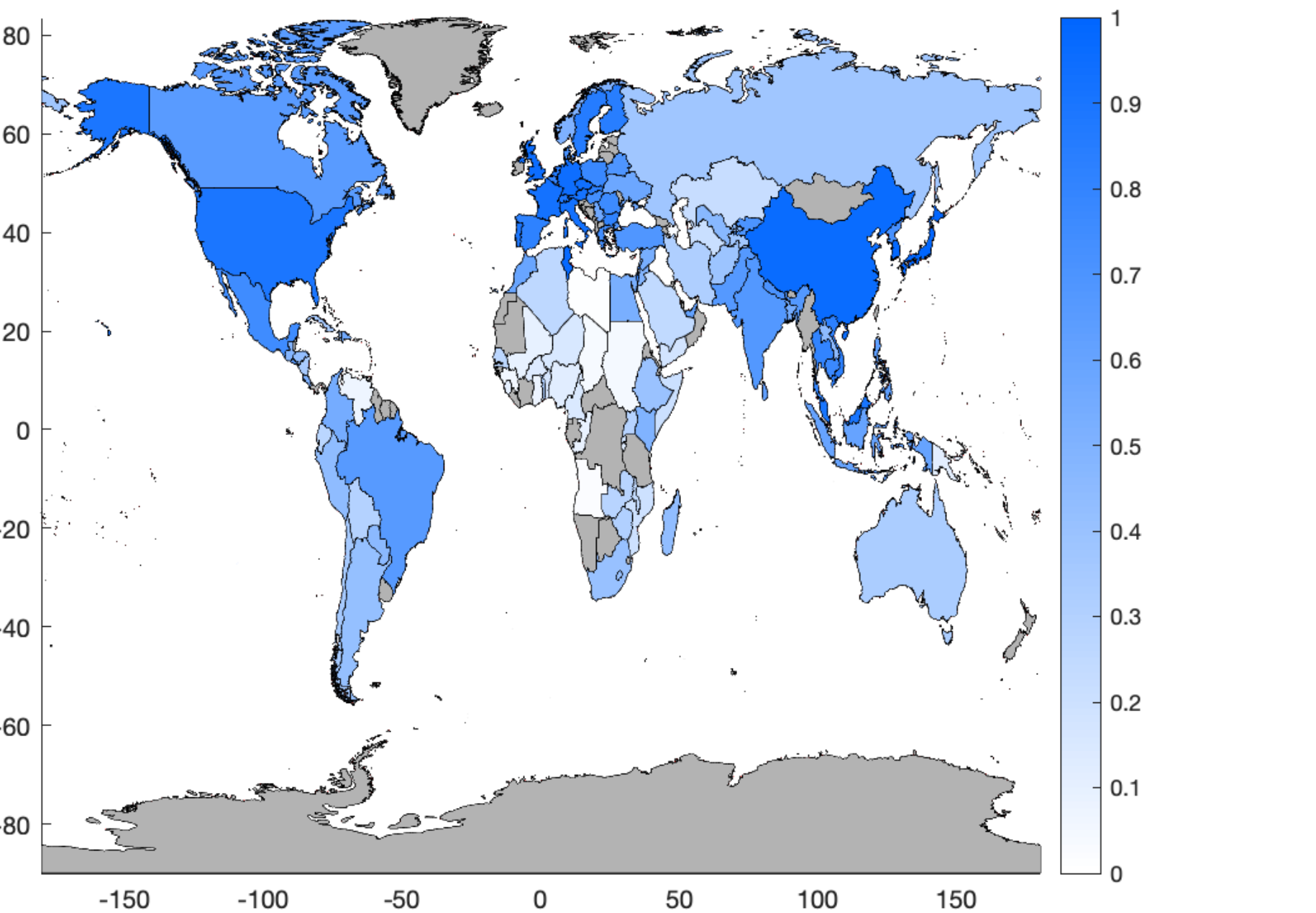}
            \caption{{\small Countries colored according to the GENEPY index computed starting from the original incidence matrix ${\bf M}$. Countries colored in grey are not considered in the analysis.}}
            \label{original}
        \end{subfigure}
        \hfill
        \begin{subfigure}[b]{0.475\textwidth}  
            \centering 
            \includegraphics[width=8.9cm,height=4.6cm,trim=2.4cm 8.5cm 0cm 8.3cm,clip=true]{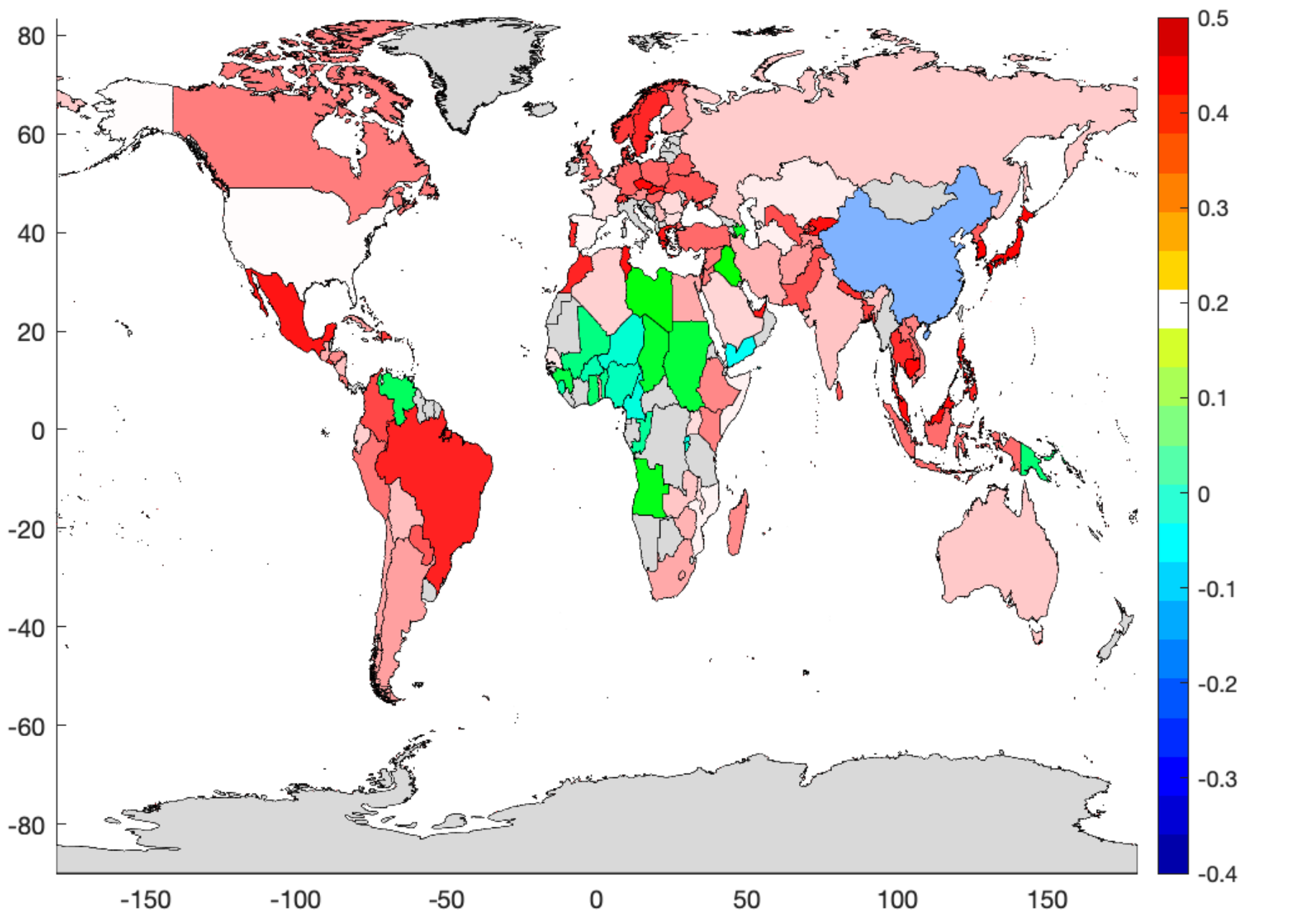}
            \caption{{\small Difference between the GENEPY indices computed starting from the original incidence matrix ${\bf M}$ and on its surrogate ${\hat{\bf M}}^{(MC)}$. Countries colored in grey are not considered in the analysis.}}
           \label{difference}
        \end{subfigure}
        \caption{Original  GENEPY values for countries for the year 2018 with products aggregated at the HS-2 level, and their comparison with their ${\widehat{\rm GENEPY}}^{(MC)}$ values.}
    \end{figure}

\noindent Similarly, Figs. \ref{original}-
\ref{difference}, which refer to products aggregated at the HS-2 level, show results similar to those obtained at the HS-4 level.

\vspace{-0.4cm}





\begin{figure}[H]
        \centering
        \begin{subfigure}[b]{0.475\textwidth}
            \centering 
        \includegraphics[scale=0.4,trim=1.5cm 8cm 0cm 8.5cm,clip=true]{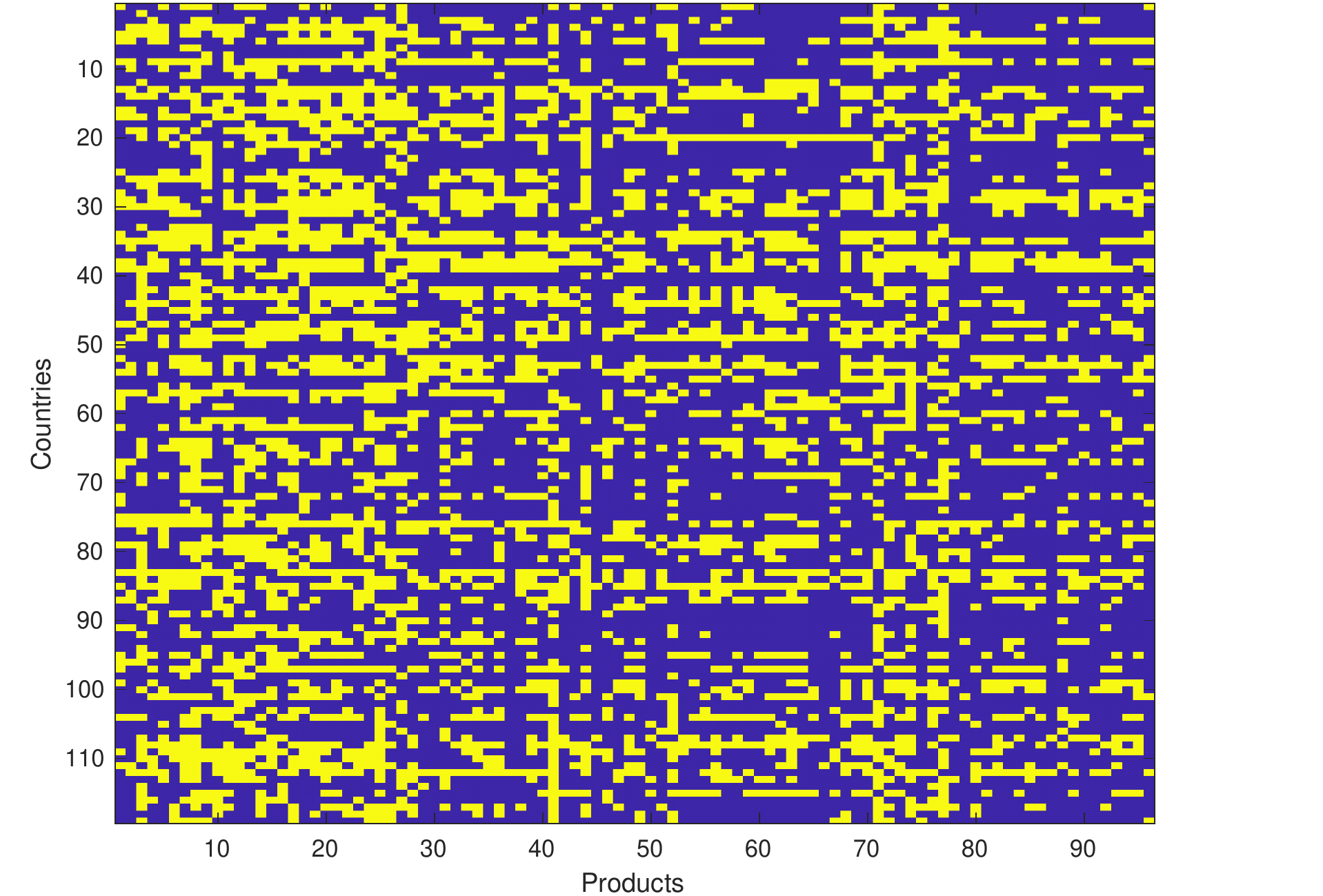}
            \caption{{\small Original incidence  matrix ${\bf M}$}.}
            \label{orighs2heat}
        \end{subfigure}
        \hfill
        \begin{subfigure}[b]{0.475\textwidth}  
            \centering 
        \includegraphics[scale=0.4,trim=1.5cm 8cm 0cm 8.5cm,clip=true]{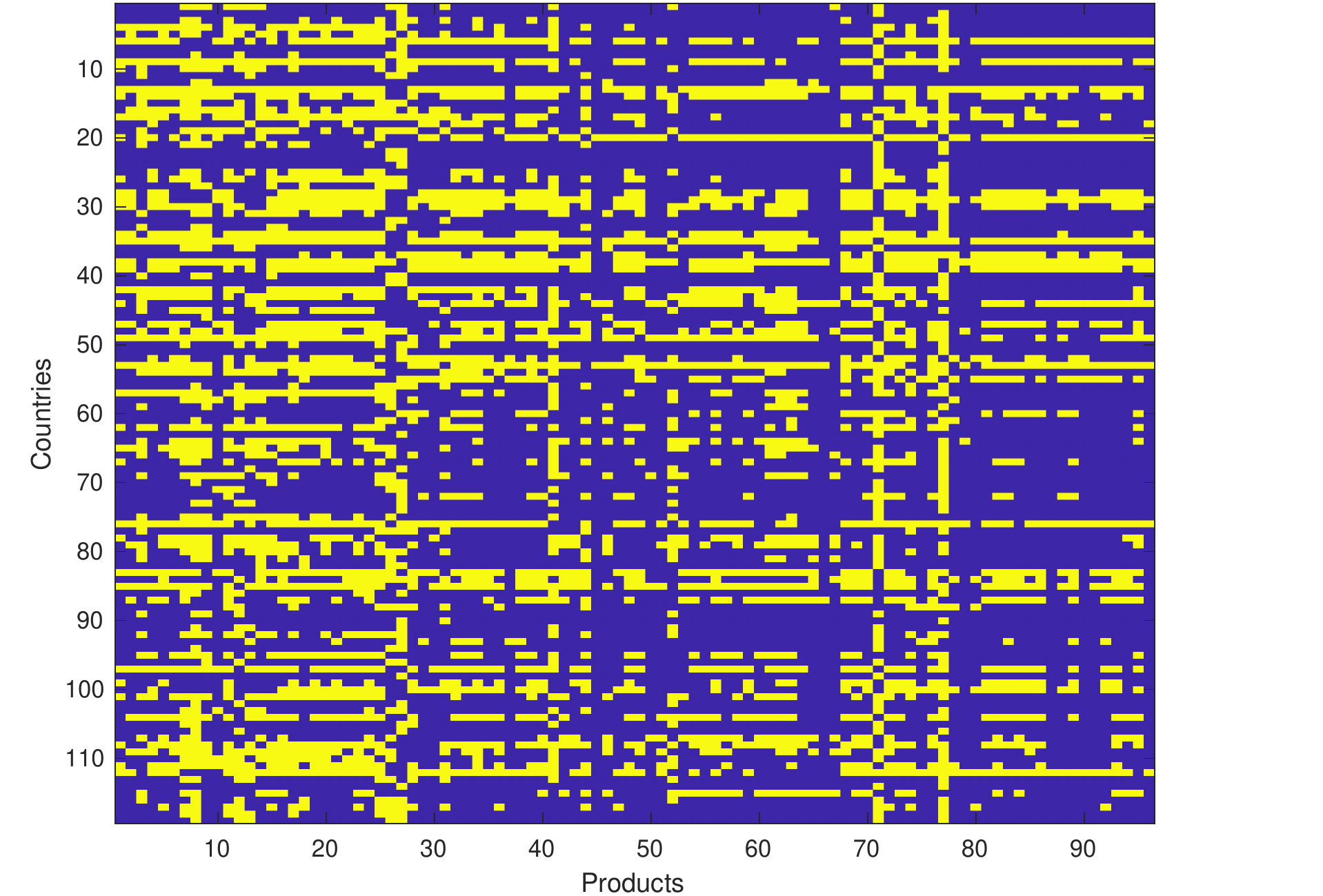}
            \caption{{\small Surrogate incidence matrix $\hat{\bf M}^{(MC)}$}.}
            \label{predhs2heat}
        \end{subfigure}
        \caption{Original versus surrogate incidence matrix for the year 2018 and the HS-2 level of product aggregation.}
    \end{figure}

\noindent Additionally, Figs. \ref{orighs2heat}-\ref{predhs2heat} report the original incidence matrix ${\bf M}$ as compared to its MC surrogate $\hat{\bf M}^{(MC)}$ obtained at the HS-2 level of product aggregation. Also in this case, the two matrices display similar but not identical entries. Thus, similar conclusions to the ones obtained for the HS-4 case apply. However, for the HS-2 case, the percentage of elements in which the two matrices differ is lower.

 \vspace{-0.2cm}

\noindent To conclude, Tab.      \ref{tab:deviations} reports the detail of the differences ${\rm GENEPY}-{\widehat{\rm GENEPY}^{(MC)}}$ for countries, for the year 2018 at the HS-2 level of aggregation. Such differences were already analysed in Fig. \ref{difference}. 
\begin{table}[H]
    \begin{minipage}{.5\linewidth}
      \centering
    \scalebox{0.6}{
\begin{tabular}{|l|l|}
\hline
GENEPY-${\widehat{\rm GENEPY}}^{(MC)}$ & \textbf{Country} \\ 
\hline 
-0.39257 & {\color{blue} Hong Kong} \\ 
-0.32742 & {\color{blue}China} \\ 
-0.11648 & {\color{blue}Italy} \\ 
0.012212 & {\color{teal} Iraq} \\ 
0.027759 & {\color{teal}Libyan Arab Jamahiriya} \\ 
0.029027 & {\color{teal}Angola} \\ 
0.045068 & {\color{teal}Azerbaijan} \\ 
0.047551 & {\color{teal}Chad} \\ 
0.054625 & {\color{teal}Sudan} \\ 
0.061237 & {\color{teal}Guinea} \\ 
0.063626 & {\color{teal}Congo} \\ 
0.064162 & {\color{teal}Venezuela} \\ 
0.07725 & {\color{teal}Ghana} \\ 
0.080608 & {\color{teal}Papua New Guinea} \\ 
0.081877 & {\color{teal}Mali} \\ 
0.083468 & {\color{teal}Benin} \\ 
0.08376 & {\color{teal}Congo} \\ 
0.084289 & {\color{teal}Sierra Leone} \\ 
0.089644 & {\color{teal}Burkina Faso} \\ 
0.089985 & {\color{teal}Nigeria} \\ 
0.090831 & {\color{teal}Burundi} \\ 
0.092596 & {\color{teal}Niger} \\ 
0.094417 & {\color{teal}Cameroon} \\ 
0.096485 & {\color{teal}Rwanda} \\ 
0.096501 & {\color{teal}Yemen} \\ 
0.097521 & {\color{red}United States} \\ 
0.09755 & {\color{red}South Sudan} \\ 
0.099289 & {\color{red}Spain} \\ 
0.10046 & {\color{red}Mozambique} \\ 
0.10293 & {\color{red}Somalia} \\ 
0.10471 & {\color{red}Turkmenistan} \\ 
0.10538 & {\color{red}Kazakhstan} \\ 
0.10658 & {\color{red}Senegal} \\ 
0.10809 & {\color{red}Ivory Coast} \\ 
0.10911 & {\color{red}France} \\ 
0.11116 & {\color{red}Bulgaria} \\ 
0.11269 &{\color{red}Malawi} \\ 
0.11666 & {\color{red}Uganda} \\ 
0.11966 & {\color{red}Romania} \\ 
0.12069 & {\color{red}Saudi Arabia} \\ 
0.12078 & {\color{red}Netherlands} \\ 
0.12121 & {\color{red}Algeria} \\ 
0.12298 & {\color{red}Ecuador} \\ 
0.12453 & {\color{red}Russia} \\ 
0.12549 & {\color{red}Australia} \\ 
0.12928 & {\color{red}Zambia} \\ 
0.13043 & {\color{red}India} \\ 
0.13344 & {\color{red}Nicaragua} \\ 
0.13778 & {\color{red}Bolivia} \\ 
0.14032 & {\color{red}Serbia} \\ 
0.14037 & {\color{red}Togo} \\ 
0.14076 & {\color{red}Iran Islamic Republic of} \\ 
0.14421 & {\color{red}Zimbabwe} \\ 
0.1452 & {\color{red}Belgium} \\ 
0.14975 & {\color{red}Guatemala} \\ 
0.15105 & {\color{red}Egypt} \\ 
0.1534 & {\color{red}South Africa} \\ 
0.15421 & {\color{red}Chile} \\ 
0.15709 & {\color{red}Cuba} \\ 
0.16116 & {\color{red}Argentina}
\\ 
\hline 
\end{tabular}}
\end{minipage}%
    \begin{minipage}{.5\linewidth}
\scalebox{0.6}{
\begin{tabular}{|l|l|}
\hline
GENEPY-${\widehat{\rm GENEPY}}^{(MC)}$ & \textbf{Country} \\ 
\hline \hline
0.16381 & {\color{red}Afghanistan} \\ 
0.16641 & {\color{red}Haiti} \\ 
0.16862 & {\color{red}Tajikistan} \\ 
0.16915 & {\color{red}Lebanon} \\ 
0.17253 & {\color{red}Austria }\\ 
0.17402 & {\color{red}Finland} \\ 
0.17561 & {\color{red}Honduras} \\ 
0.17755 & {\color{red}Madagascar} \\ 
0.17791 & {\color{red}Ethiopia} \\ 
0.18072 & {\color{red}Viet Nam} \\ 
0.18157 & {\color{red}Kenya} \\ 
0.19125 & {\color{red}Syrian Arab Republic} \\ 
0.19381 & {\color{red}Canada} \\ 
0.19577 & {\color{red}Jordan} \\ 
0.19605 & {\color{red}Costa Rica} \\ 
0.19724 & {\color{red}Peru} \\ 
0.19725 & {\color{red}Lao People's Democratic Republic} \\ 
0.19882 & {\color{red}United Kingdom} \\ 
0.19962 & {\color{red}Indonesia} \\ 
0.201 & {\color{red}Turkey} \\ 
0.20222 & {\color{red}Hungary} \\ 
0.20318 & {\color{red}Poland} \\ 
0.20356 & {\color{red}Germany} \\ 
0.20357 & {\color{red}Belarus}\\ 
0.2047 & {\color{red}Korea, Democratic People's Republic of} \\ 
0.20659 & {\color{red}Sri Lanka} \\ 
0.20667 & {\color{red}Paraguay} \\ 
0.20859 & {\color{red}Ukraine} \\ 
0.21088 & {\color{red}Pakistan} \\ 
0.21494 & {\color{red}Uzbekistan} \\ 
0.21626 & {\color{red}Israel} \\ 
0.21999 & {\color{red}Denmark} \\ 
0.2241 & {\color{red}Colombia} \\ 
0.22585 & {\color{red}Tanganyika} \\ 
0.22832 & {\color{red}Bangladesh} \\ 
0.2353 & {\color{red}Singapore} \\ 
0.23852 & {\color{red}Norway} \\ 
0.23875 & {\color{red}Switzerland} \\ 
0.23978 & {\color{red}Dominican Republic} \\ 
0.242 & {\color{red}Republic of the Union of Myanmar} \\ 
0.24262 & {\color{red}El Salvador} \\ 
0.24271 & {\color{red}Nepal} \\ 
0.25116 & {\color{red}Thailand} \\ 
0.2632 & {\color{red}Portugal} \\ 
0.26521 & {\color{red}Sweden} \\ 
0.26842 & {\color{red}Morocco} \\ 
0.27857 & {\color{red}Brazil} \\ 
0.28289 & {\color{red}Philippines} \\ 
0.28441 & {\color{red}Slovakia} \\ 
0.30354 & {\color{red}United Arab Emirates} \\ 
0.30412 & {\color{red}Greece} \\ 
0.3389 & {\color{red}Mexico} \\ 
0.35332 & {\color{red}Tunisia} \\ 
0.35893 & {\color{red}Kyrgyzstan} \\ 
0.37339 & {\color{red}Malaysia} \\ 
0.38045 & {\color{red}Czech Republic} \\ 
0.39382 & {\color{red}Cambodia} \\ 
0.50948 & {\color{red}Japan} \\ 
0.5384 & {\color{red}Korea, Republic of} \\ 
\hline 
\end{tabular}}
\end{minipage} 
\caption{Countries classified according to the deviations GENEPY$-{\widehat{\rm GENEPY}}^{(MC)}$ for the year 2018 at the HS-2 level. Countries colored in blue are countries for which GENEPY$<{\widehat{\rm GENEPY}}^{(MC)}$. Countries colored in green are countries for which GENEPY$\simeq{\widehat{\rm GENEPY}}^{(MC)}$. Countries colored in red are countries for which GENEPY$>{\widehat{\rm GENEPY}}^{(MC)}$.}\label{tab:deviations}
\end{table}

\section*{Results of the analysis for countries, for the years 2005 and 2014 and at the HS-2 level}

\begin{figure}[H]
        \centering
        \hspace{-0.6cm}
        \begin{subfigure}[b]{0.475\textwidth}
            \centering 
            \includegraphics[width=9.6cm,height=5cm,trim=2cm 8.5cm 0cm 8.5cm,clip=true]{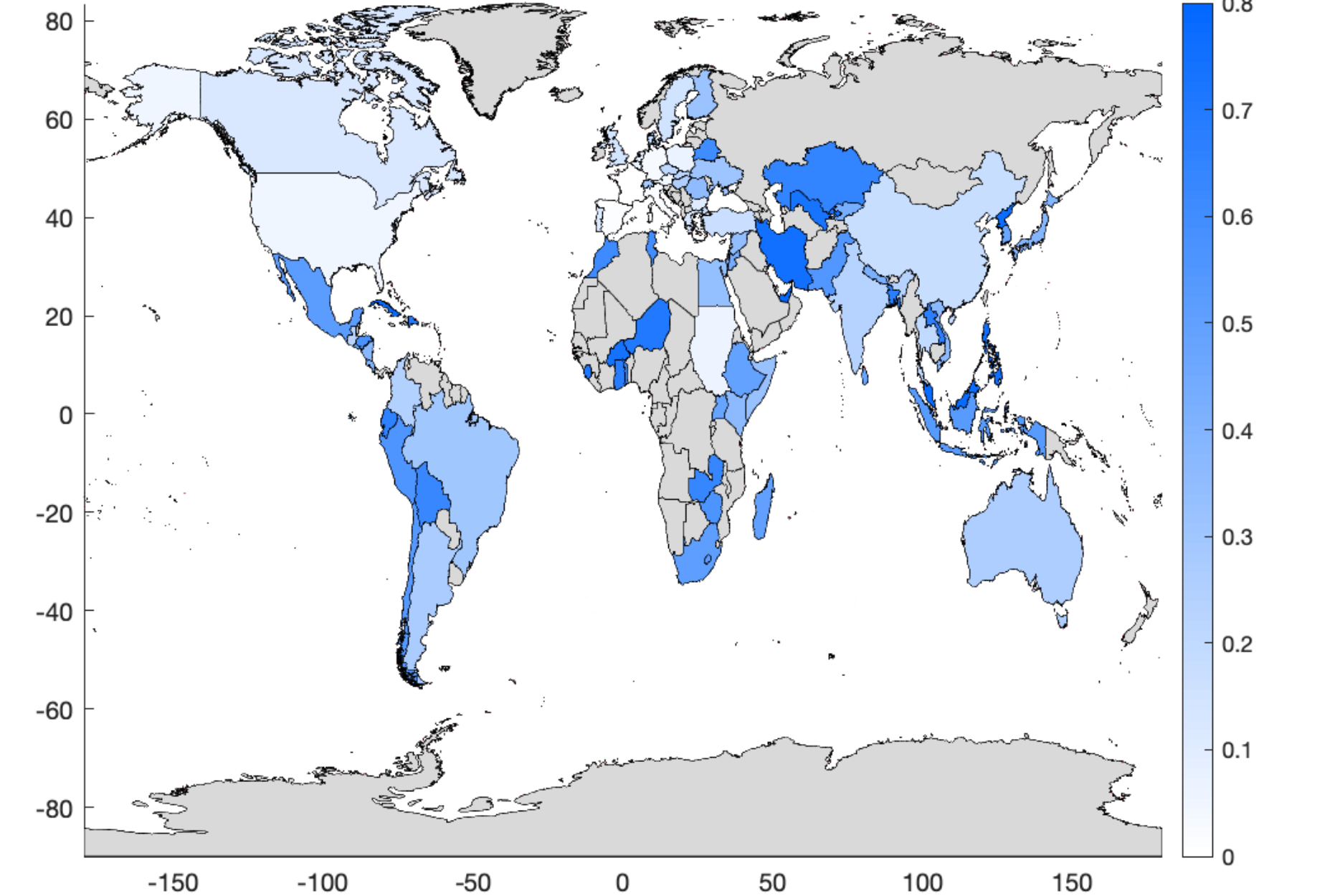}
            \caption{{\small False negative rate $fnr_c$}, reported proportionally to the shade of blue for the year 2005. Countries colored in grey are not considered in the analysis.}
            \label{fnegtesworld_2005}
        \end{subfigure}
        \hfill
        \hspace{0.6cm}
        \begin{subfigure}[b]{0.475\textwidth}   
            \centering 
            \hspace{2cm}
            \includegraphics[width=9.6cm,height=5cm,trim=2cm 8.5cm 0cm 8.5cm,clip=true]{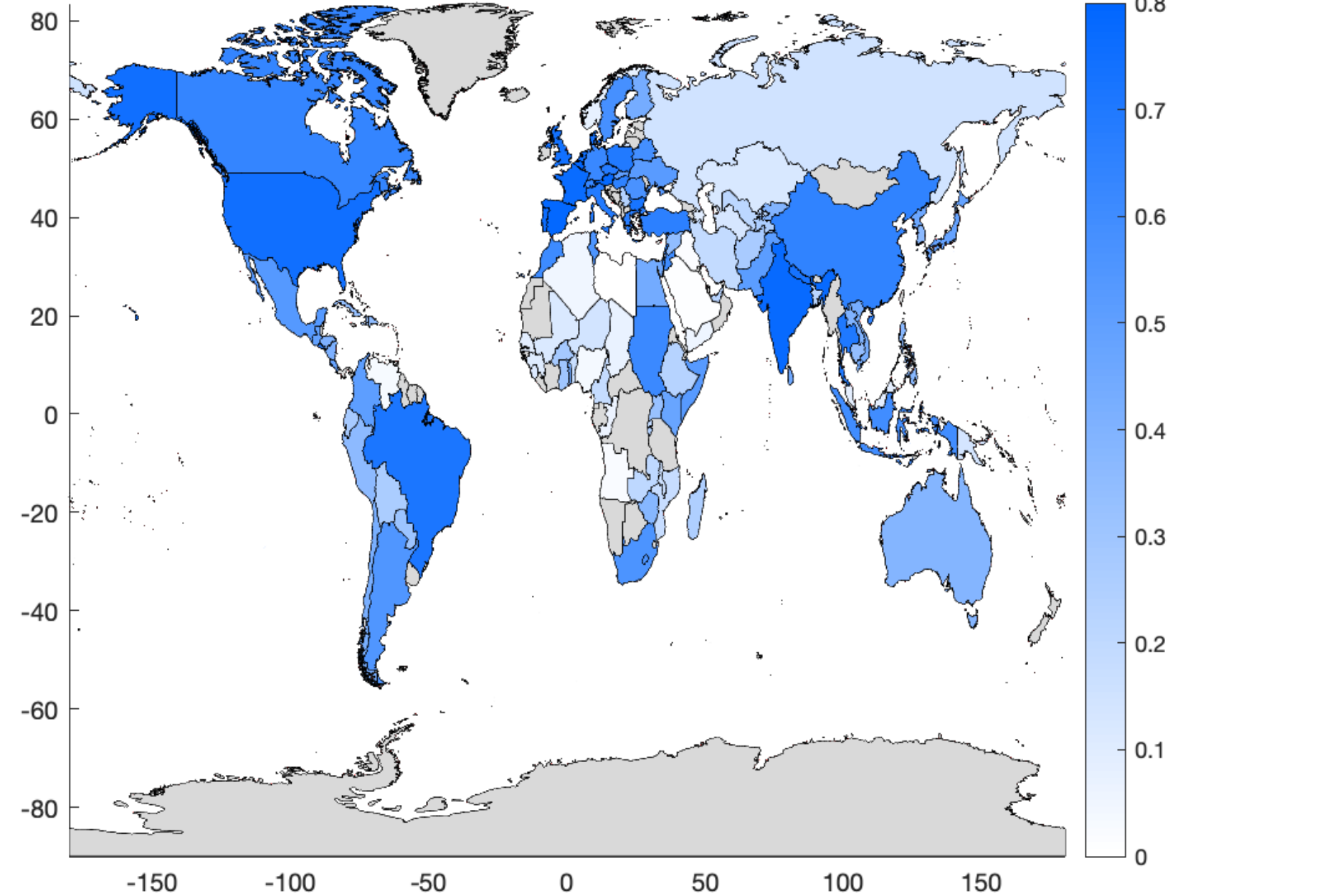}
            \caption{{\small False positive rate $fpr_c$}, reported proportionally to the shade of blue for the year 2005. Countries colored in grey are not considered in the analysis.}
            \label{fpostesworld_2005}
        \end{subfigure}
        \vskip\baselineskip
        \hspace{-0.6cm}
        \begin{subfigure}[b]{0.475\textwidth}   
            \centering 
            \includegraphics[width=9.6cm,height=5cm,trim=2cm 8.5cm 0cm 8.5cm,clip=true]{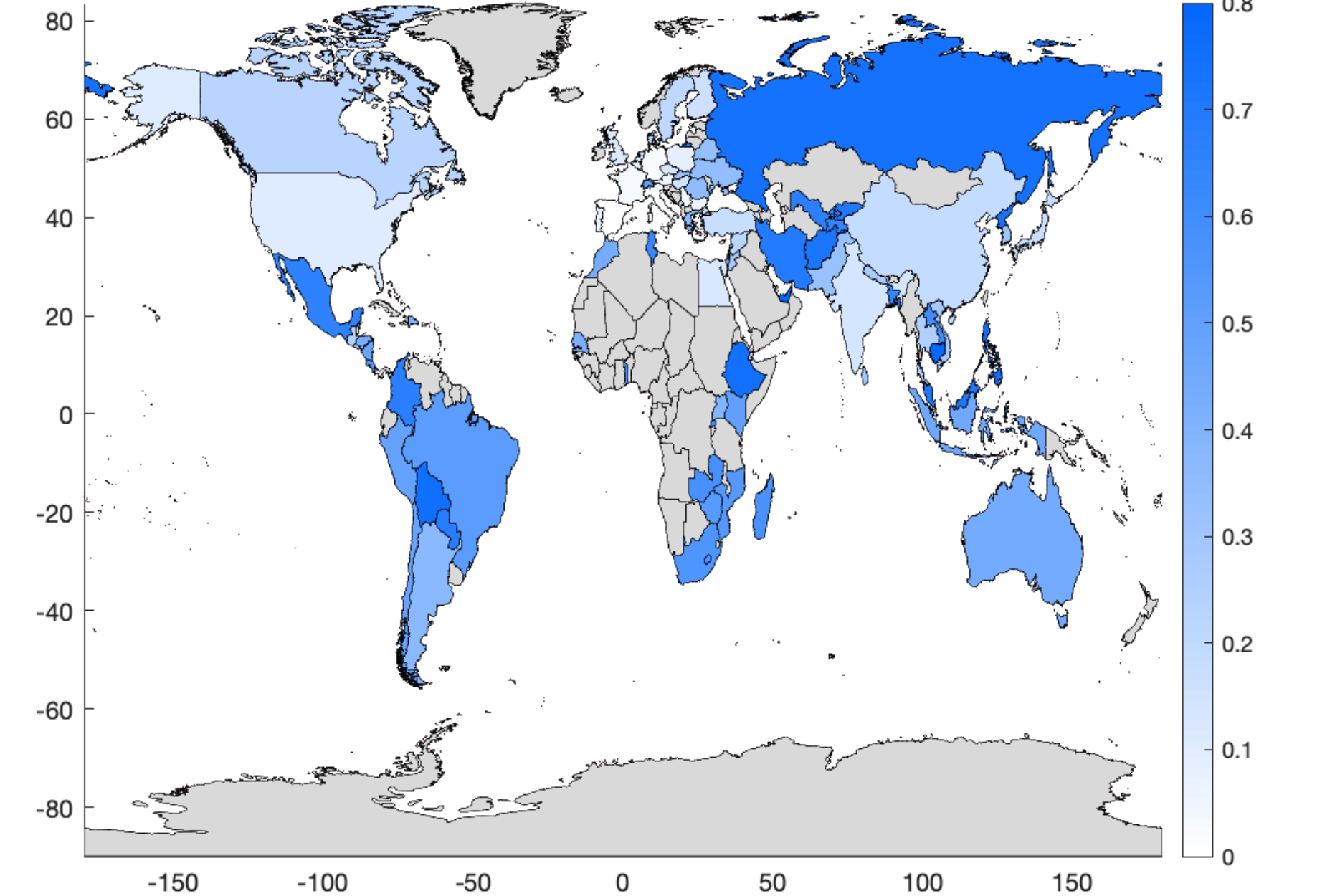}
            \caption{{\small False negative rate $fnr_c$}, reported proportionally to the shade of blue for the year 2014. Countries colored in grey are not considered in the analysis.}
            \label{fnegtesworld_2014}
        \end{subfigure}
        \hfill
        \hspace{0.6cm}
        \begin{subfigure}[b]{0.475\textwidth}   
            \centering 
            \hspace{2cm}
            \includegraphics[width=9.6cm,height=5cm,trim=2cm 8.5cm 0cm 8.5cm,clip=true]{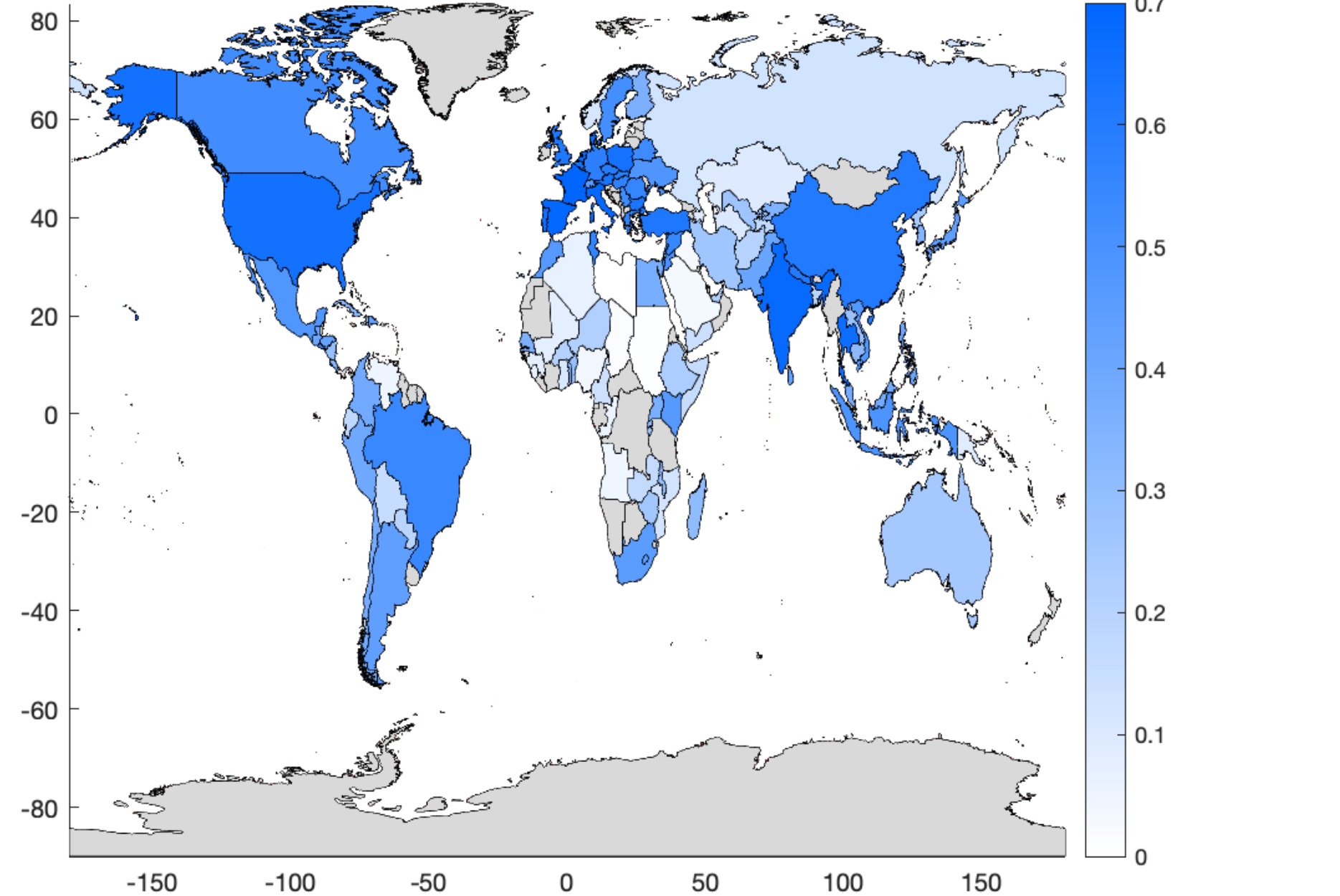}
            \caption{{\small False positive rate $fpr_c$}, reported proportionally to the shade of blue for the year 2014. Countries colored in grey are not considered in the analysis.}
            \label{fpostesworld_2014}
        \end{subfigure}
        \caption{False negative and false positive rates for countries, obtained by the  method reported in Step \ref{step:5} of Section \ref{ourMC} for the years 2005 and 2014, with products aggregated at the HS-2 level.}
        \label{fig:mat_products}
    \end{figure}

In the following, results similar to those obtained in the main text are reported for the years 2005 and 2014. In this case, in order to reduce the computational effort, the analysis was made at the HS-2 level of product aggregation. Fig. \ref{fig:mat_products} 
 reports false negative and false positive rates for the two years, whereas Tab. \ref{genep_vs_ft} reports Kendall correlation coefficients between the ranking produced using GENEPY against the rankings produced respectively by $fnr_{c,t}$ and $fpr_{c,t}$, for the years $t=2005$ and $t=2014$.

\begin{table}[H]
\centering
\begin{tabular}{|l|l|l|}
\hline
                            & \textbf{GENEPY$_t$ ($\tau_k$)} & \textbf{GENEPY$_t$ ($\text{$p$-value}$)} \\ \hline
\textbf{$fnr_{c,2005}$} & 0.2018                      & 0.0020                    \\ \hline
\textbf{$fpr_{c,2005}$} & 0.6212                     & 0.0000                     \\ \hline
\textbf{$fnr_{c,2014}$} & 0.1936                    & 0.0031                      \\ \hline
\textbf{$fpr_{c,2014}$} & 0.6165                    & 0.0000                      \\ \hline
\end{tabular}
\caption{$\tau_k$ and relative $p$-values for the rankings produced using GENEPY against the ranking produced respectively by $fnr_{c,t}$ and $fpr_{c,t}$, for the years $t=2005$ and $t=2014$, with products aggregated at the HS-2 level.}
\label{genep_vs_ft}
\end{table}

\subsection*{Application of the analysis to the products at the HS-2 level}
The same analysis made in the main text for the countries has been  repeated for the products, still referring to the year 2018. This is obtained simply by replacing at the beginning of the analysis the $\bf{RCA}$ matrix with its transpose. Notice that this analysis, as some of the analyses reported in this Supplemental, was made at the HS-2 level for computational time reasons. The results obtained at the HS-2 level, however, correlated at the 95\% with the ones obtained at the HS-4 level. 

\begin{figure}[H]
        \centering
        \hspace{-0.6cm}
        \begin{subfigure}[b]{0.475\textwidth}
            \centering 
            \includegraphics[scale=0.45,trim=3.5cm 8cm 0cm 8cm,clip=true]{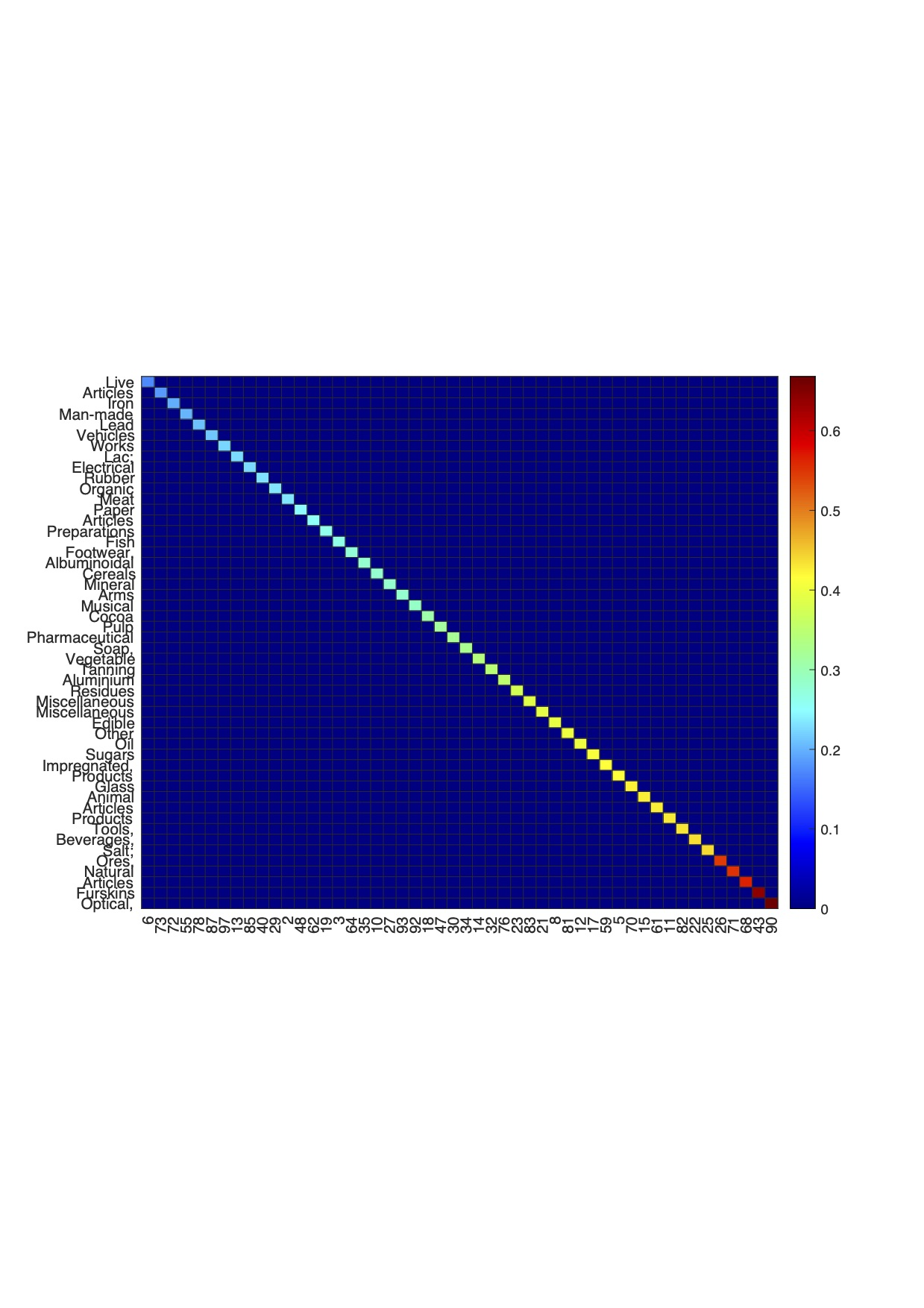}
            \caption{{\small False negative rate $fnr_p$} for the products, reported as a function of the color shade from blue (associated with the lowest value) to red (associated with the highest value).}
            \label{postesmat}
        \end{subfigure}
        \hfill
        \hspace{0.6cm}
        \begin{subfigure}[b]{0.475\textwidth}   
            \centering 
            \hspace{2cm}
            \includegraphics[scale=0.45,trim=3.5cm 8cm 0cm 8cm,clip=true]{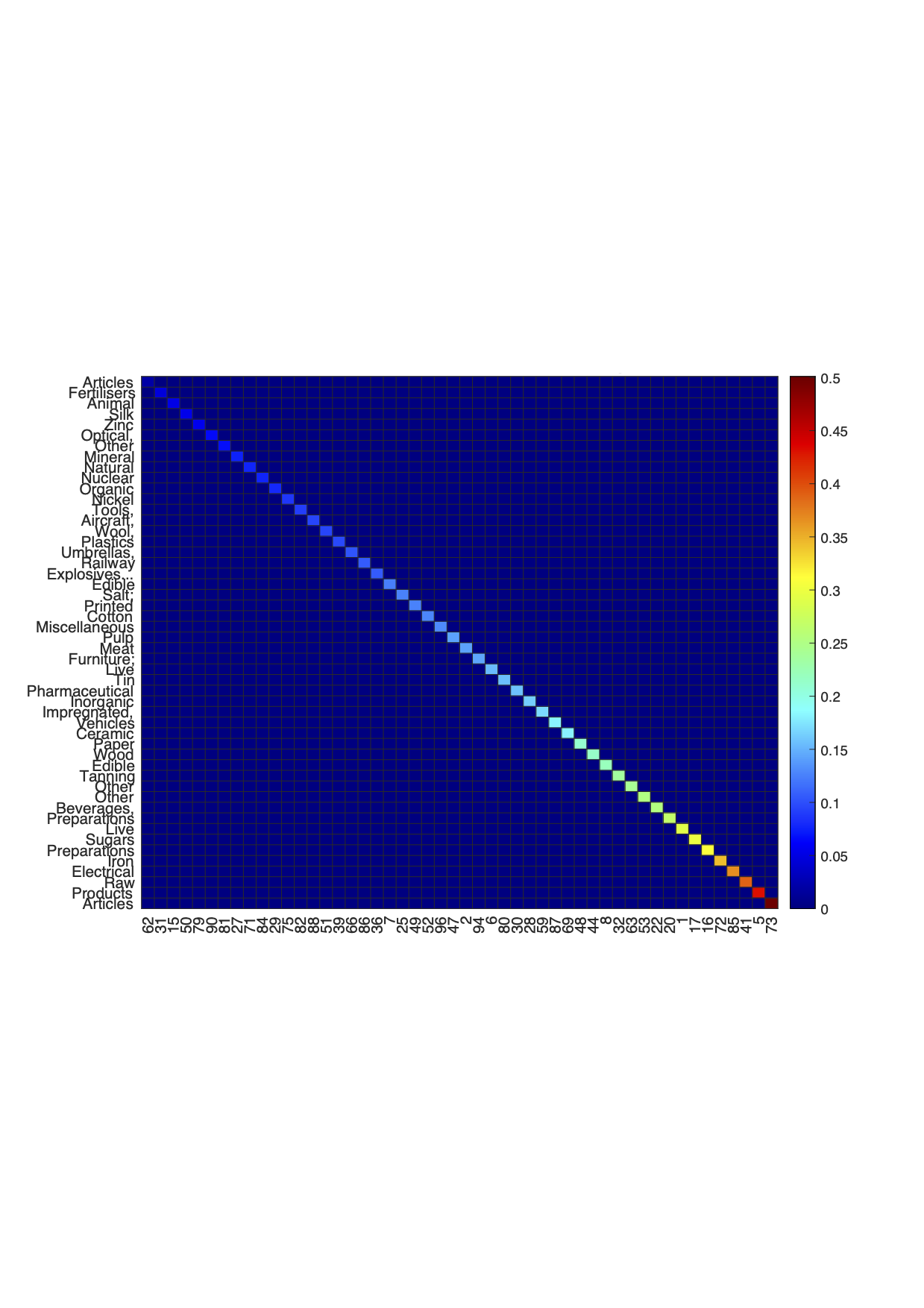}
            \caption{{\small False positive rate $fpr_p$} for the products, reported as a function of the color shade from blue (associated with the lowest value) to red (associated with the highest value).}
            \label{negtesmat}
        \end{subfigure}
        \vskip\baselineskip
        \hspace{-0.6cm}
        \begin{subfigure}[b]{0.475\textwidth}   
            \centering 
            \includegraphics[scale=0.45,trim=3.5cm 8cm 0cm 8cm,clip=true]{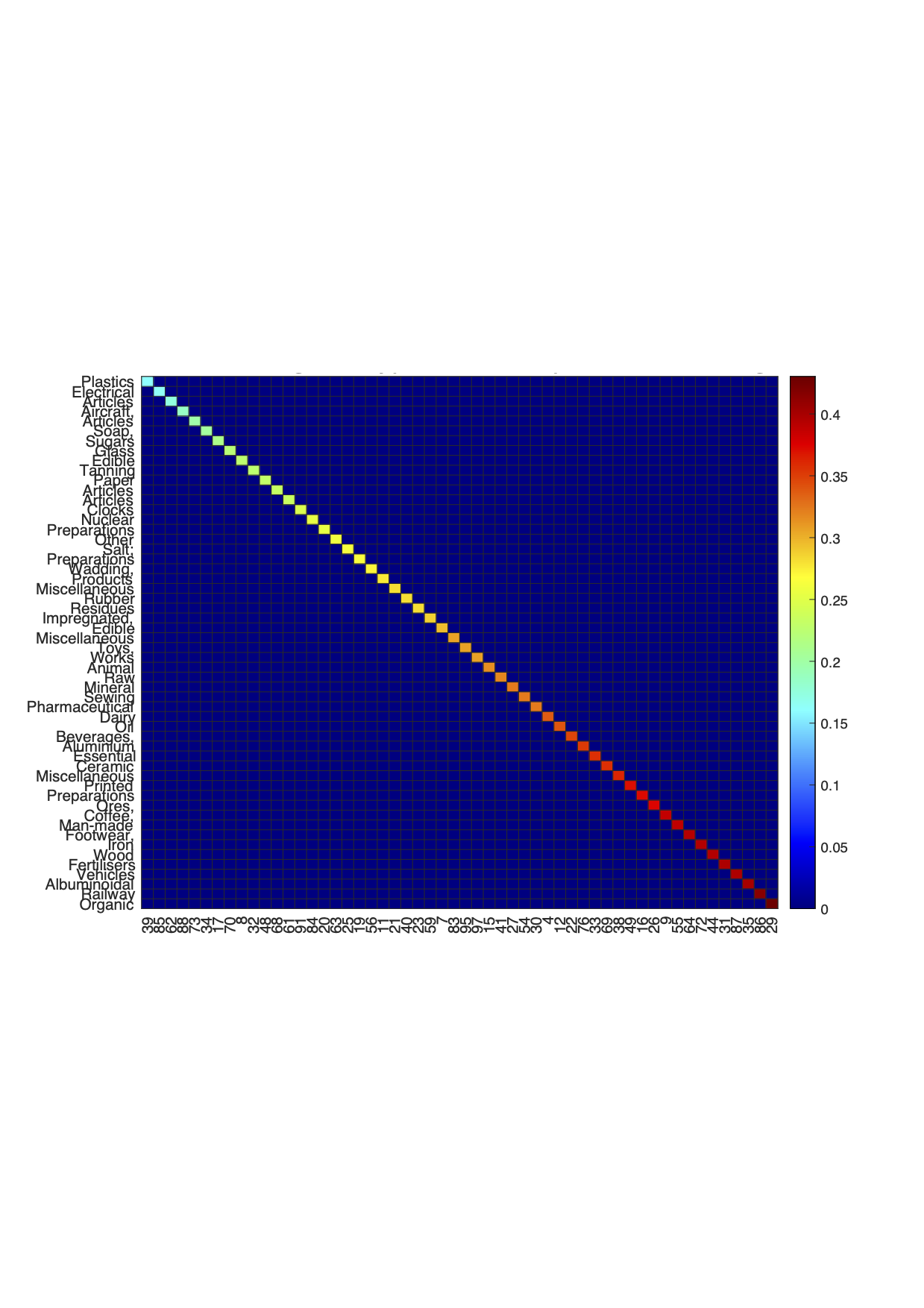}
            \caption{{\small False negative rate $fnr_p$ for the products having both $fnr_c<0.5$ and $fpr_c<0.5$. Values of $fnr_c$ are reported as a function of the color shade from blue (associated with the lowest value) to red (associated with the highest value).}}    
            \label{signpostesmat}
        \end{subfigure}
        \hfill
        \hspace{0.6cm}
        \begin{subfigure}[b]{0.475\textwidth}   
            \centering 
            \hspace{2cm}
            \includegraphics[scale=0.45,trim=3.5cm 8cm 0cm 8cm,clip=true]{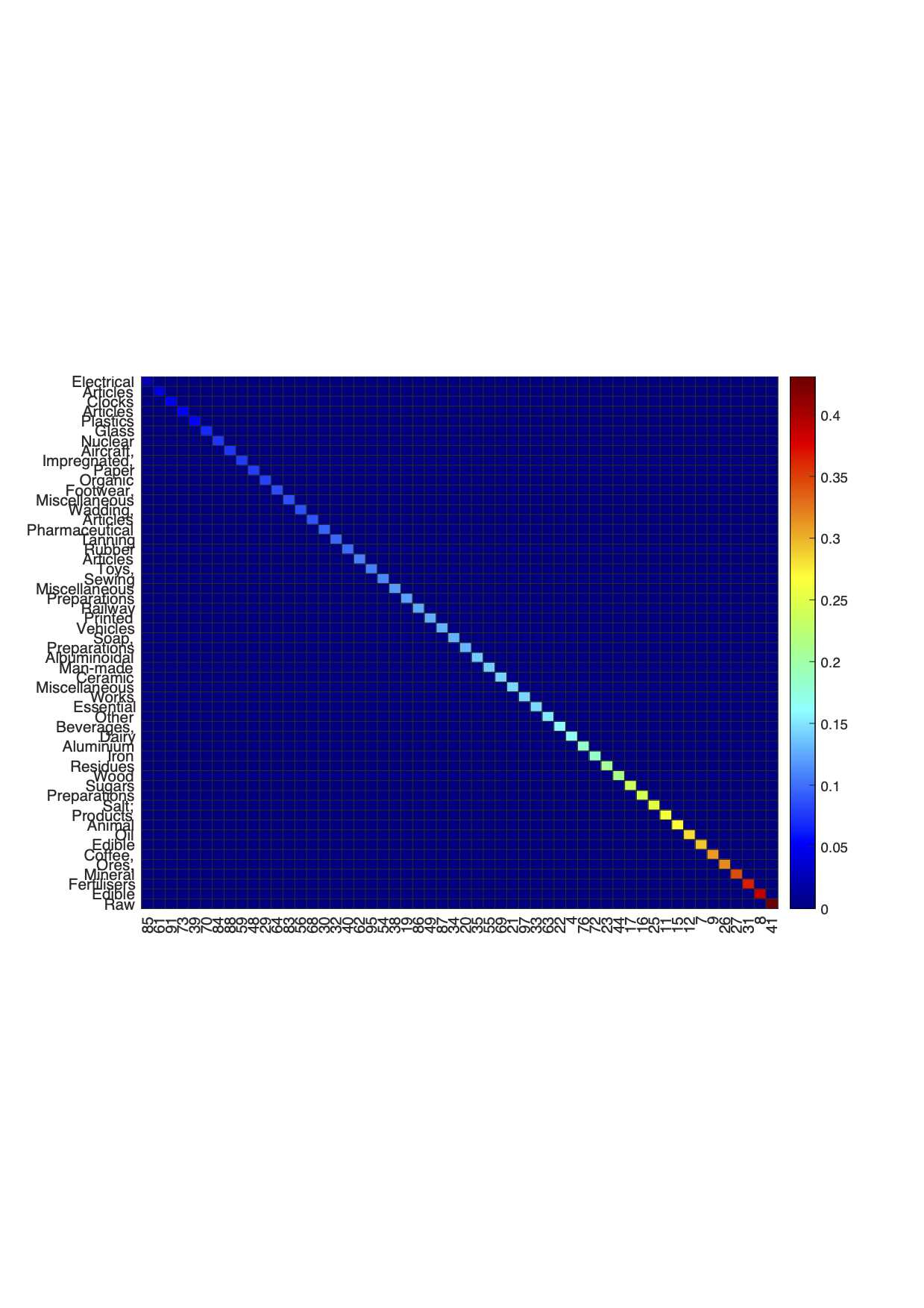}
            \caption{{\small False positive rate $fpr_p$ for the products having both $fnr_c<0.5$ and ${fpr}_c<0.5$. Values of $fpr_c$ are reported as a function of the color shade from blue (associated with the lowest value) to red (associated with the highest value).}}    
            \label{signnegtesmat}
        \end{subfigure}
        \caption{False negative and false positive rates for products, obtained by the  method reported in Step \ref{step:5} of Section \ref{ourMC} for the year 2018 and the HS-2 level.}
        \label{mat_products}
    \end{figure}

\noindent Figs. \ref{mat_products} displays respectively, on the main diagonal of each matrix reported, and for a subset of product codes,
\begin{itemize}
    \item the false negative rate $fnr_p$ for each product $p$ (Fig.\ref{postesmat});
    \item the false positive rate $fpr_p$ for each product $p$ (Fig.\ref{negtesmat});
    \item the false negative rate $fnr_p$, for the subset of products $p$ for which both $fnr_p$ and $fpr_p$ are lower than $0.5$ (Fig.\ref{signpostesmat});
    \item the false positive rate $fpr_p$, for the subset of products $p$ for which both $fnr_p$ abd $fpr_p$ are lower than $0.5$ (Fig.\ref{signnegtesmat}).
\end{itemize}
In all these cases, the products have been ordered increasingly with respect to the (either false positive or false negative) rate.

\noindent The correspondence between product codes and their names is reported in Tab. \ref{prodcodes}. The names are reported only for the HS-2 level 
for a better readability.

\begin{table}[H]
\label{prodcodes}
    \scalebox{0.6}{
    \begin{tabular}{|l|l|}
    \hline
\textbf{Row index} & \textbf{Product name} \\ 
\hline \hline
1 & Live animals \\ 
2 & Meat and edible meat offal \\ 
3 & Fish and crustaceans, molluscs and other aquatic invertebrates \\ 
4 & Dairy produce \\ 
5 & Products of animal origin, not elsewhere specified or included \\ 
6 & Live trees and other plants \\ 
7 & Edible vegetables and certain roots and tubers \\ 
8 & Edible fruit and nuts; \\ 
9 & Coffee, tea, mate and spices \\ 
10 & Cereals \\ 
11 & Products of the milling industry; \\ 
12 & Oil seeds and oleaginous fruits; \\ 
13 & Lac; gums, resins and other vegetable saps and extracts \\ 
14 & Vegetable plaiting materials \\ 
15 & Animal or vegetable fats and oils and their cleavage products \\ 
16 & Preparations of meat, of fish or of crustaceans \\ 
17 & Sugars and sugar confectionery \\ 
18 & Cocoa and cocoa preparations \\ 
19 & Preparations of cereals, flour, starch or milk; pastrycooks’ products \\ 
20 & Preparations of vegetables, fruit, nuts or other parts of plants \\ 
21 & Miscellaneous edible preparations \\ 
22 & Beverages, spirits and vinegar \\ 
23 & Residues and waste from the food industries; prepared animal fodder \\ 
24 & Tobacco and manufactured tobacco substitutes \\ 
25 & Salt; sulphur \\ 
26 & Ores, slag and ash \\ 
27 & Mineral fuels, mineral oils and products of their distillation; bituminous substances \\ 
28 & \pbox{20cm}{Inorganic chemicals; organic or inorganic compounds of precious metals,\\ of rare-earth metals, of radioactive elements or of isotopes} \\ 
29 & Organic chemicals \\ 
30 & Pharmaceutical products \\ 
31 & Fertilisers \\ 
32 & Tanning or dyeing extracts \\ 
33 & Essential oils and resinoids \\ 
34 & \pbox{20cm}{Soap, organic surface-active agents, washing preparations,\\
lubricating preparations, artificial waxes...} \\ 
35 & Albuminoidal substances \\ 
36 & Explosives, fireworks \\ 
37 & Photographic or cinematographic goods \\ 
38 & Miscellaneous chemical products \\ 
39 & Plastics and articles thereof \\ 
40 & Rubber and articles thereof \\ 
41 & Raw hides and skins \\ 
42 & Articles of leather; saddlery and harness \\ 
43 & Furskins and artificial fur \\ 
44 & Wood and articles of wood \\
45 & Cork and articles of cork \\ 
46 & Manufactures of straw, of esparto or of other plaiting materials \\ 
47 & Pulp of wood or of other fibrous cellulosic material \\ 
48 & Paper and paperboard
\\
\hline
\end{tabular}}
\scalebox{0.6}{
\begin{tabular}{|l|l|}
\hline
\textbf{Row index} & \textbf{Product name} \\ 
\hline \hline
49 & Printed books, newspapers, pictures and other products of the printing industry \\ 
50 & Silk \\ 
51 & Wool, fine or coarse animal hair \\ 
52 & Cotton \\ 
53 & Other vegetable textile fibres \\ 
54 & Sewing thread of man-made filaments \\ 
55 & Man-made staple fibres \\ 
56 & Wadding, felt and nonwovens \\ 
57 & Carpets and other textile floor coverings \\ 
58 & Special woven fabrics \\ 
59 & Impregnated, coated, covered or laminated textile fabrics \\ 
60 & Knitted or crocheted fabrics \\ 
61 & Articles of apparel and clothing accessories, knitted or crocheted \\ 
62 & Articles of apparel and clothing accessories, not knitted or crocheted \\ 
63 & Other made up textile articles \\ 
64 & Footwear, gaiters and the like; parts of such articles \\ 
65 & Headgear and parts thereof \\ 
66 & Umbrellas, sun umbrellas and similar articles \\ 
67 & Prepared feathers and down and articles made of feathers or of down \\ 
68 & Articles of stone, plaster, cement, asbestos, mica or similar materials \\ 
69 & Ceramic products \\ 
70 & Glass and glassware \\ 
71 & Natural or cultured pearls, precious or semi-precious stones, precious metals \\ 
72 & Iron and steel \\ 
73 & Articles of iron or steel \\ 
74 & Copper and articles thereof \\ 
75 & Nickel and articles thereof \\ 
76 & Aluminium and articles thereof \\ 
78 & Lead and articles thereof \\ 
79 & Zinc and articles thereof \\ 
80 & Tin and articles thereof \\ 
81 & Other base metals; cermets; articles thereof \\ 
82 & Tools, implements, cutlery, spoons and forks, of base metal; parts thereof of base metal \\ 
83 & Miscellaneous articles of base metal \\ 
84 & Nuclear reactors, boilers, machinery and mechanical appliances \\ 
85 & Electrical machinery and equipment and parts thereof \\ 
86 & Railway or tramway locomotives, rolling-stock and parts thereof \\ 
87 & Vehicles other than railway or tramway rolling-stock, and parts and accessories thereof \\ 
88 & Aircraft, spacecraft, and parts thereof \\ 
89 & Ships, boats and floating structures \\ 
90 & \pbox{20cm}{Optical, photographic, cinematographic, measuring, checking, precision,\\ medical or surgical instruments and apparatus; parts and accessories thereof} \\ 
91 & Clocks and watches and parts thereof \\ 
92 & Musical instruments; parts and accessories of such articles \\ 
93 & Arms and ammunition; parts and accessories thereof \\ 
94 & Furniture; bedding, mattresses, mattress supports, cushions and similar stuffed furnishings \\ 
95 & Toys, games and sports requisites; parts and accessories thereof \\ 
96 & Miscellaneous manufactured articles \\ 
97 & Works of art, collectors’ pieces, and antiques \\ 
\hline 
\end{tabular}
}
\caption{Product codes and corresponding names at the HS-2 level.}
\label{prodcodes}
\end{table}

\subsection*{Confusion matrix for products at the HS-4 and HS-2 levels}
\begin{figure}[H]
 \centering
 \includegraphics[width=100mm,scale=0.8]{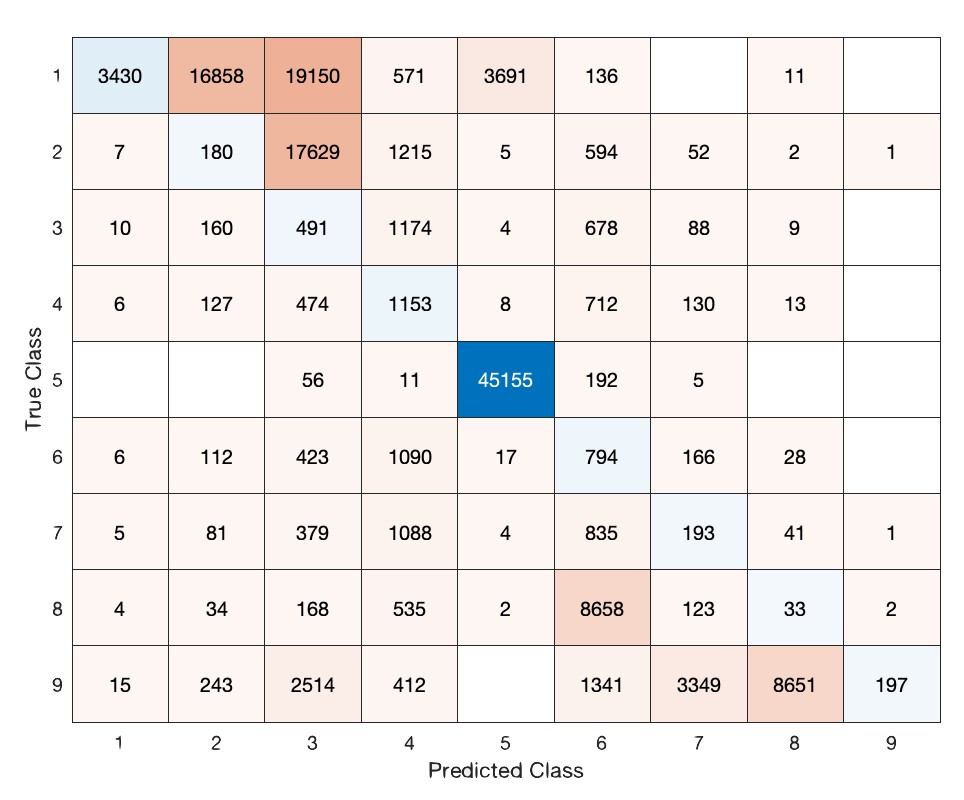}
            \vspace{0.5cm}
            \caption{Confusion matrix built from the normalized 2018 GENEPY values associated with the products at the HS-4 level, computed respectively based on the matrices ${\bf M}^\top$ and $({\hat{\bf M}}^\top)^{(MC)}$, then discretized in $8$ classes according to the percentiles of the respective GENEPY distributions.}
            \label{confusionproducts}
 \end{figure}

 Fig. \ref{confusionproducts} reports, for the HS-4 level of product aggregation, the confusion matrix built from the normalized GENEPY\footnote{Since the GENEPY and $\widehat{\rm GENEPY}^{(MC)}$ span different ranges, they have been rescaled to the same interval $[0,1]$.} values associated with the products for the year 2018, computed respectively based on the incidence matrix ${\bf M}^\top$ and the surrogate incidence matrix $({\hat{\bf M}}^\top)^{(MC)}$, then discretized in $8$ classes according to the percentiles of the respective GENEPY distributions. The class $1$ corresponds to the lowest GENEPY values, whereas the class $8$ corresponds to the highest GENEPY values. Notice that the whole procedure applied to countries was repeated here from scratch starting from the transpose of the ${\bf RCA}$ matrix, focusing in this case on the products. Very similar results ($\tau_k \simeq 0.96$) are obtained if instead, one computes the GENEPY for products using as input the transpose of the surrogate incidence matrix ${\bf{\hat{ M}}}^{(MC)}$ obtained in the application of MC to countries, described in the main text. In the figure, the true classes are the ones computed starting from the GENEPY applied to the incidence matrix ${\bf M}^\top$, whereas the predicted classes are the ones computed starting from the GENEPY applied to the surrogate incidence matrix $({\hat{\bf M}}^\top)^{(MC)}$.

 \noindent As evidenced by Fig. \ref{confusionproducts}, the GENEPY indices computed based on $\bf{M}^\top$ and $(\hat{M}^{T})^{(MC)}$ are quite similar, since the non-zero elements of the confusion matrix 
 are concentrated above all
 near its main diagonal (i.e., the number of classification errors is quite small far from the main diagonal). Some outliers emerged. The deviations regarded mainly the elements belonging to higher categories. This is somewhat reasonable since they refer to more complex products, which might be more complex to classify.\\
 In this case, the Kendall rank correlation coefficient between the GENEPY rankings computed based on ${\bf M}^\top$ and $(\hat{\bf M}^\top)^{(MC)}$ turned out to be $\tau_k \simeq 0.6$, with a $p$-value nearly equal to $0$.
 
\begin{figure}[H]
 \centering
 \includegraphics[scale=0.5,trim=2.5cm 8.7cm 0cm 8.7cm]{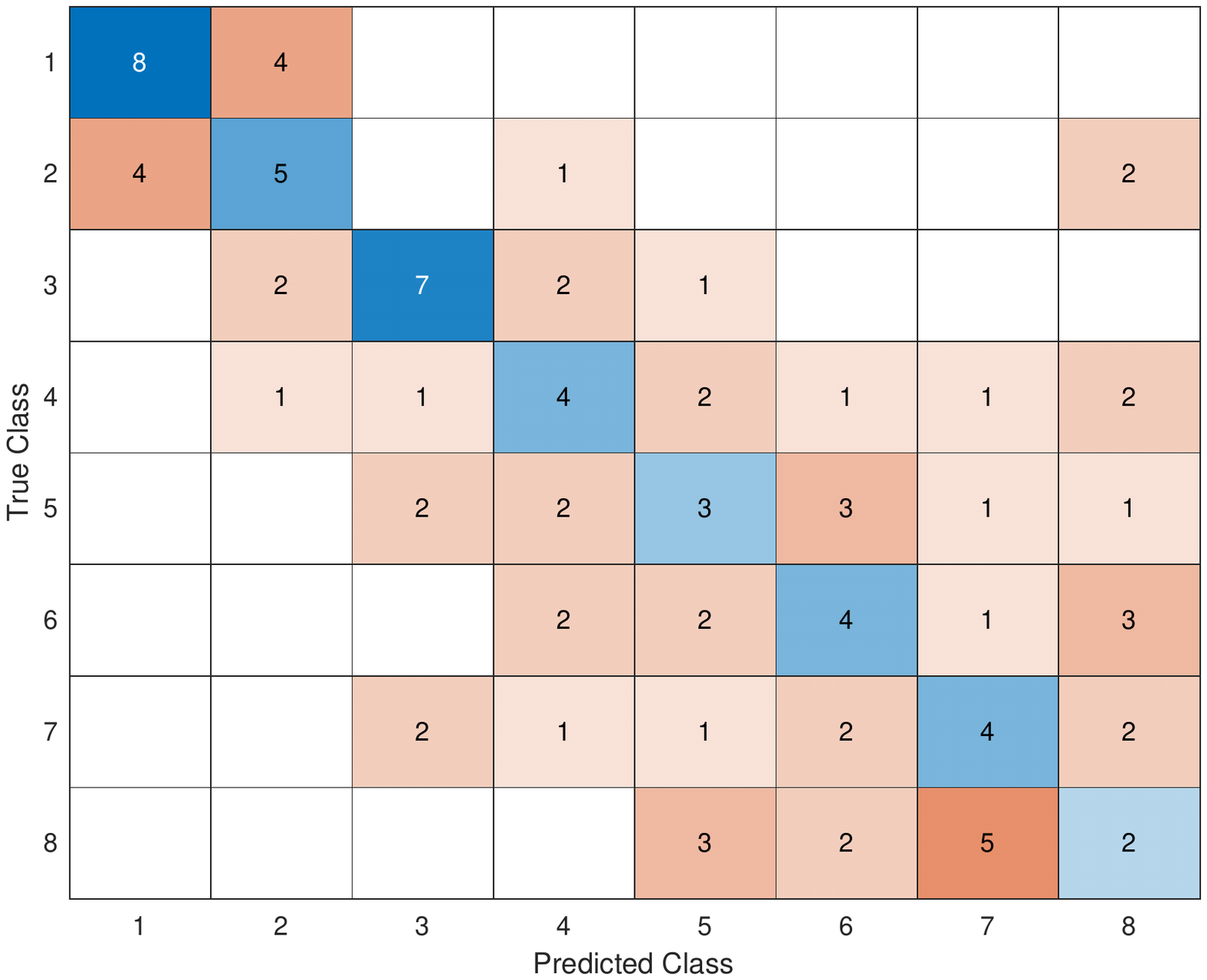}
            \vspace{0.5cm}
            \caption{Confusion matrix built from the normalized 2018 GENEPY values associated with the products at the HS-2 level, computed respectively based on the matrices ${\bf M}^\top$ and $({\hat{\bf M}}^\top)^{(MC)}$, then discretized in $8$ classes according to the percentiles of the respective GENEPY distributions.}
            \label{confusionproducts_hs2}
 \end{figure}

\noindent Similarly, we built up a confusion matrix also for the HS-2 case. This is reported in Fig. \ref{confusionproducts_hs2}.
\vspace{0.2cm}

\subsection*{Application of a variation of the analysis to countries, based on the entry-wise logarithm of the original ${\bf RCA}$ matrix for the year 2018, with products aggregated at the HS-4 level}

In the following, a variation of our analysis is applied to countries. In this variation, instead of discretizing the elements of the original ${\bf RCA}$ matrix, they are replaced by their natural logarithm\footnote{No issues arise when taking the natural logarithm, because no $0$ or negative entries are present in that matrix. Moreover, entries originally equal to $NaN$ are never included in the training, validation or test sets.}. Then, the rest of the proposed method is unchanged with respect to Section \ref{ourMC}. In the main text, the discretization method has been preferred, because it generates values more symmetrically distributed around $0$.

\begin{figure}[H]
        \centering
        \begin{subfigure}[b]{0.475\textwidth}
            \centering
          \includegraphics[width=9.6cm,height=5cm,trim=2cm 8.5cm 0cm 8.4cm,clip=true]{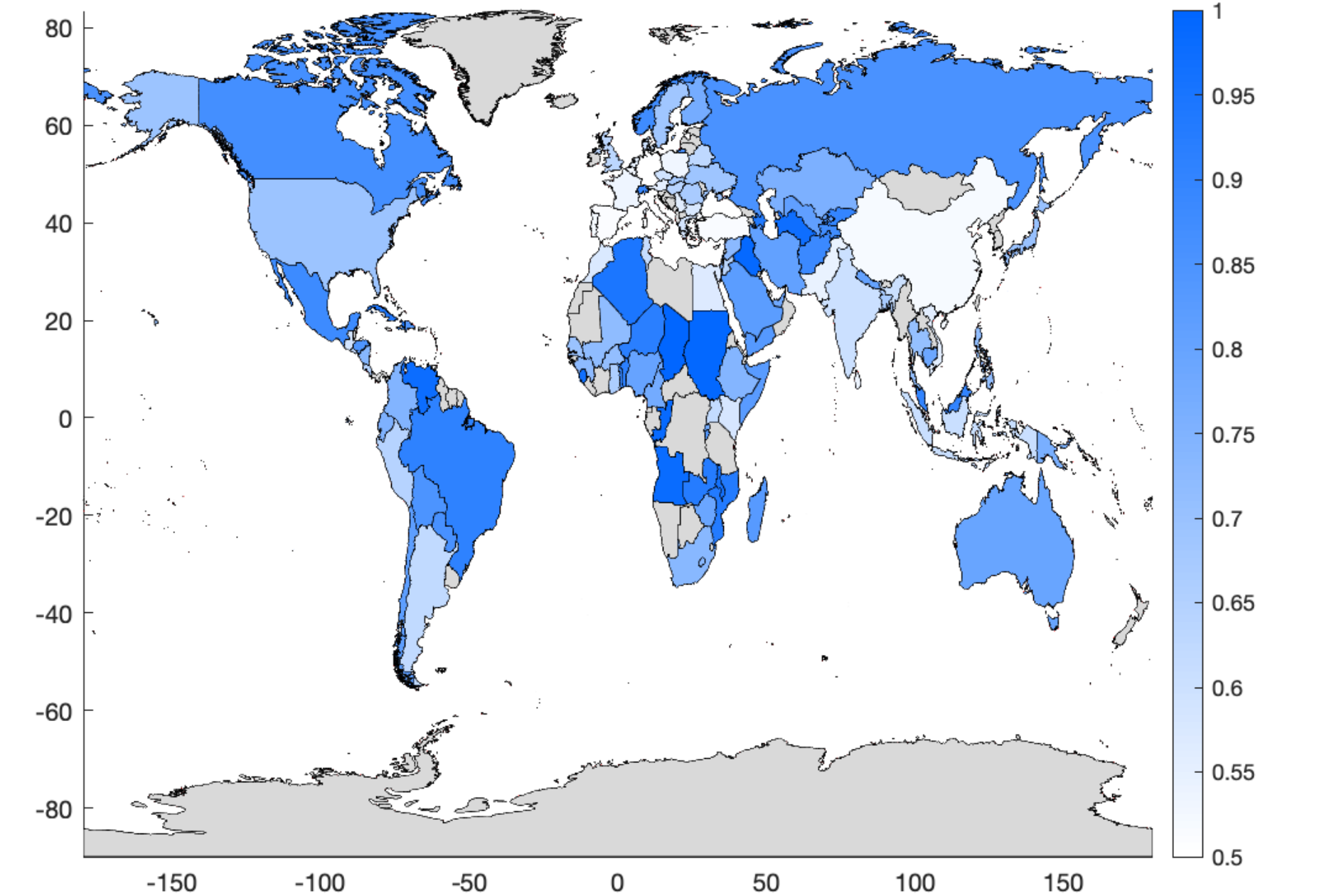}
            \caption{{\small False negative rate $fnr_c$, reported proportionally to the shade of blue for the year 2018 and the product aggregation level HS-4, for the case in which the entry-wise natural logarithm of the original ${\bf RCA}$ matrix is employed by the proposed method. Countries colored in grey are not considered in the analysis.}}
            \label{fnegths4_2018_log}
        \end{subfigure}
        \hfill
        \begin{subfigure}[b]{0.475\textwidth}  
            \centering 
          \includegraphics[width=9.6cm,height=5cm,trim=2cm 8.5cm 0cm 8.4cm,clip=true]{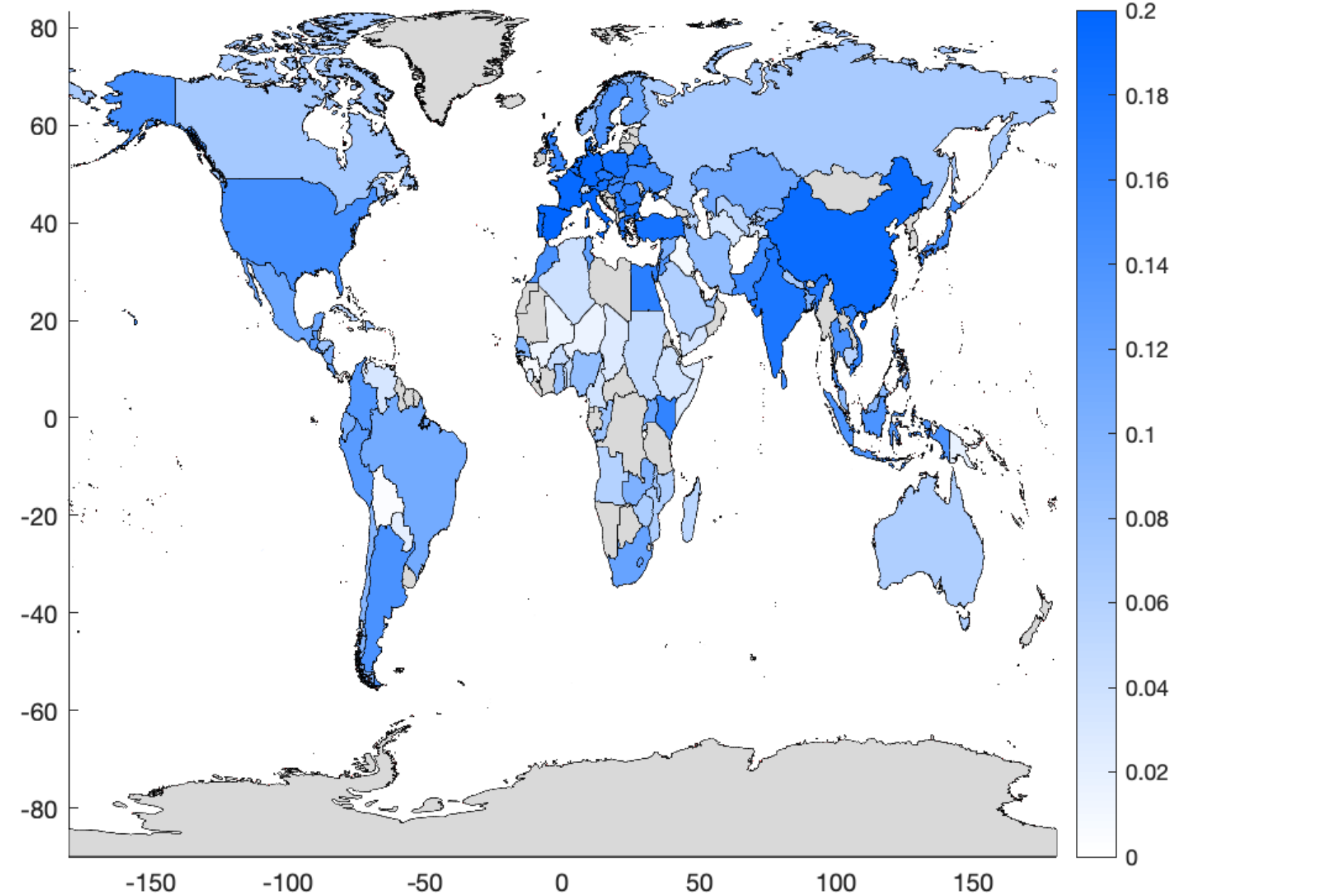}
            \caption{{\small False positive rate $fpr_c$}, reported proportionally to the shade of blue for the year 2018 and the product aggregation level HS-4, for the case in which the entry-wise natural logarithm of the original ${\bf RCA}$ matrix is employed by the proposed method. Countries colored in grey are not considered in the analysis.}
            \label{fposths4_2018_log}
        \end{subfigure}
        \caption{False negative and false positive rates for countries for the year 2018 and the product aggregation level HS-4, for the case in which the entry-wise natural logarithm of the original ${\bf RCA}$ matrix is employed by the proposed method.}
        \label{hs4_2018_log}
    \end{figure}

\noindent Fig. \ref{hs4_2018_log} reports, for this variation of analysis and for the product aggregation level HS-4, results similar to those obtained in Figs. \ref{postesworld} and \ref{negtesworld} for the original analysis and for the product aggregation level HS-2.

    \noindent Similarly, Tab. \ref{genephs4_vs_ft_log} finds, for this variation of analysis and for the product aggregation level HS-4, results similar to those obtained in Tab. \ref{genep_vs_ft} for the original analysis and for the product aggregation level HS-2.
    
\begin{table}[H]
\centering
\begin{tabular}{|l|l|l|}
\hline
                            & \textbf{GENEPY ($\tau_k$)} & \textbf{GENEPY ($\text{$p$-value}$)} \\ \hline
\textbf{$fnr_{c,log,hs-4}$} & -0.2569                      & 0.0000                    \\ \hline
\textbf{$fpr_{c,log,hs-4}$} & 0.5903                     & 0.0000                     \\ \hline
\end{tabular}
\caption{$\tau_k$ and relative $p$-values for the rankings produced using GENEPY against the ranking produced respectively by $fnr_{c,t}$ and $fpr_{c,t}$, for the product aggregation level HS-4, for the case in which the entry-wise natural logarithm of the original ${\bf RCA}$ matrix is employed by the original analysis.
}
\label{genephs4_vs_ft_log}
\end{table}

\begin{figure}[H]
        \centering
          \includegraphics[width=9.6cm,height=5cm,trim=2cm 8.5cm 0cm 8.5cm,clip=true]{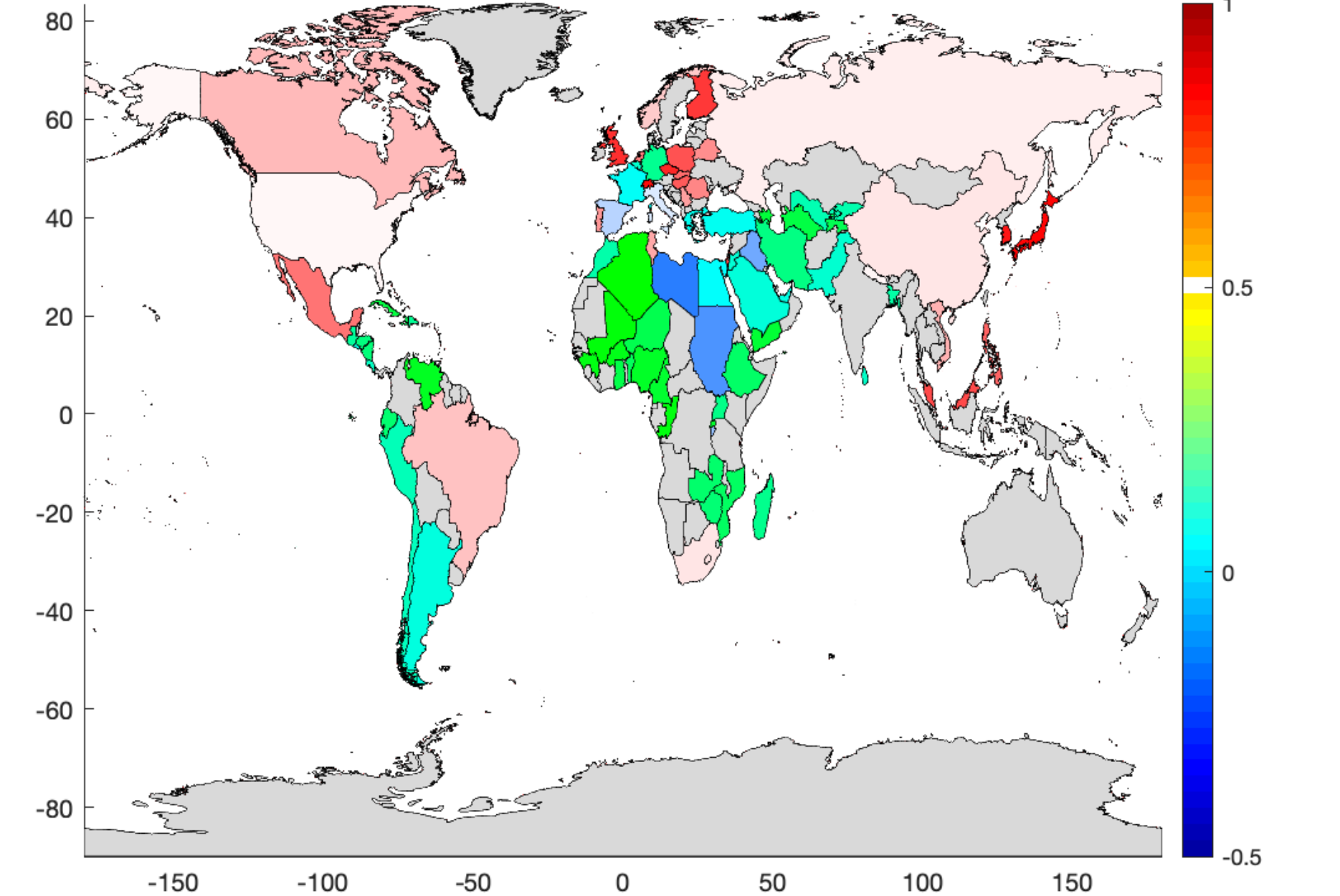}
            \caption{Difference between GENEPY and ${\widehat{\rm GENEPY}}^{(MC)}$ for the year 2018, with products aggregated at the HS-4 level, for the case in which the entry-wise natural logarithm of the original ${\bf RCA}$ matrix is employed by the proposed method.}
            \label{differenceHS4_log}
    \end{figure}

\noindent Finally, also Fig. 
\ref{differenceHS4_log}, which refers to the variation of analysis and to products aggregated at the HS-4 level, shows results similar to those obtained in Fig. 
\ref{difference}, which refers to the original analysis and to products aggregated at the HS-2 level.

\end{document}